\documentclass[]{book}

\usepackage{amsmath, amsthm, amssymb, deluxetable,chngcntr}
\usepackage{greybox}
\usepackage{mathptmx}       
\usepackage{helvet}              
\usepackage{courier}            
\usepackage{type1cm}         

\usepackage{makeidx}          
\usepackage{graphicx}         
\usepackage{multicol}           
\usepackage[bottom]{footmisc}  

\newcommand \mum{$\mu$m}

\counterwithout{section}{chapter}
\begin{document}

    \frontmatter

    \mainmatter

 \chapter*{~~~~~~~~Dusty Planetary Systems}\label{Debris_Disks}
\begin{center}
\large\bf{Amaya Moro-Mart\'{\i}n$^{a,b}$}\\

\normalsize{$^{a}$Centro de Astrobiolog\'{\i}a, INTA-CSIC, Spain.}\\
\normalsize{$^{b}$Department of Astrophysical Sciences, Princeton University, USA.}\\

\end{center}

Extensive photometric stellar surveys show that many main sequence stars show emission at infrared and longer wavelengths that is in excess of the stellar photosphere; this emission is thought to arise from circumstellar dust. The presence of dust disks is confirmed by spatially resolved imaging at infrared to millimeter wavelengths (tracing the dust thermal emission), and at optical to near infrared wavelengths (tracing the dust scattered light).  Because the expected lifetime of these dust particles is much shorter than the age of the stars ($>$ 10$^{7}$ yr), it is inferred that this solid material not primordial, i.e. the remaining from the placental cloud of gas and dust where the star was born, but instead is replenished by dust-producing planetesimals. These planetesimals are analogous to the asteroids, comets and Kuiper Belt objects (KBOs) in our Solar system that produce the interplanetary dust that gives rise to the zodiacal light (tracing the inner component of the Solar system debris disk). The presence of these "debris disks" around  stars with a wide range of masses, luminosities, and metallicities, with and without binary companions, is evidence that planetesimal formation is a robust process that can take place under a wide range of conditions.  

This chapter is divided in two parts. {\it Part~I} discusses how the study of the Solar system debris disk and the study of debris disks around other stars can help us learn about the formation, evolution and diversity of planetary systems by shedding light on the frequency and timing of planetesimal formation, the location and physical properties of the planetesimals, the presence of long-period planets, and the dynamical and collisional evolution of the system.  It first describes the interplanetary dust in the inner and outer Solar system and the evolution of dust production though the Solar system's history, followed by a summary of the properties of debris disks around other stars (their frequency, evolution, spatial structure and composition). 
{\it Part~II} reviews the physical processes that affect dust particles in the gas-free environment of a debris disk, like the Solar system's interplanetary space, and their effect on the dust particle size and spatial distribution; the discussion focuses on radiation and stellar wind forces, 
gravitational forces  in the presence of planets
and grain collisions. 

\clearpage

\begin{center}
\Large
{\bf Part I}\\
\huge 
{\bf Solar and Extra-Solar Debris Disks}
\end{center}
\normalsize

\section{Debris disks are evidence of the presence of extra-solar planetesimals}
\label{sec:planetesimals}

Circumstellar disks play a fundamental role in the formation of stars and planets  and the subsequent evolution of planetary systems. Pre-stellar disks form from the contraction and conservation of angular momentum of the densest regions of molecular clouds (consisting on gas and dust in a 100:1 mass ratio). The accretion of mass onto the forming star is regulated by mass and angular momentum transfer mechanisms within the disk, and requires the ejection of material and angular momentum by bipolar outflows.  
With time, the mass reservoir of the cloud gets depleted, the disk begins to dissipate from the inside out, and the transfer of mass onto the star weakens. Observations indicate that by 3 Myr, half of the disks show inner cavities 10s of AU in size, and the signatures of mass accretion onto the star weakens. Simultaneously, planet formation takes place by accretion of dust particles into larger and larger bodies, resulting in a few massive cores and a swarm of embryos. Because the formation of giant planets requires a gas disk to provide material for the formation of a gaseous envelope around a massive core, it needs to happen before the protoplanetary gas disk dissipates in approximately 6 Myr.  These massive planets might be responsible for carving the large inner cavities inferred to be present in young disks.  {\it Spitzer} Space Telescope surveys indicate that the majority of young stars (with stellar types later than B) are surrounded by gas-rich protoplanetary disks, indicating that most stars harbor the raw material required to form planetary systems. The formation of terrestrial planets and massive planets beyond the  ice line  is not limited by the presence of gas in the disk, and continues for approximately 100 Myr; a critical step in this process is the formation of planetesimals. 

\begin{figure}[]
\begin{center}
\includegraphics[width=0.75\textwidth]{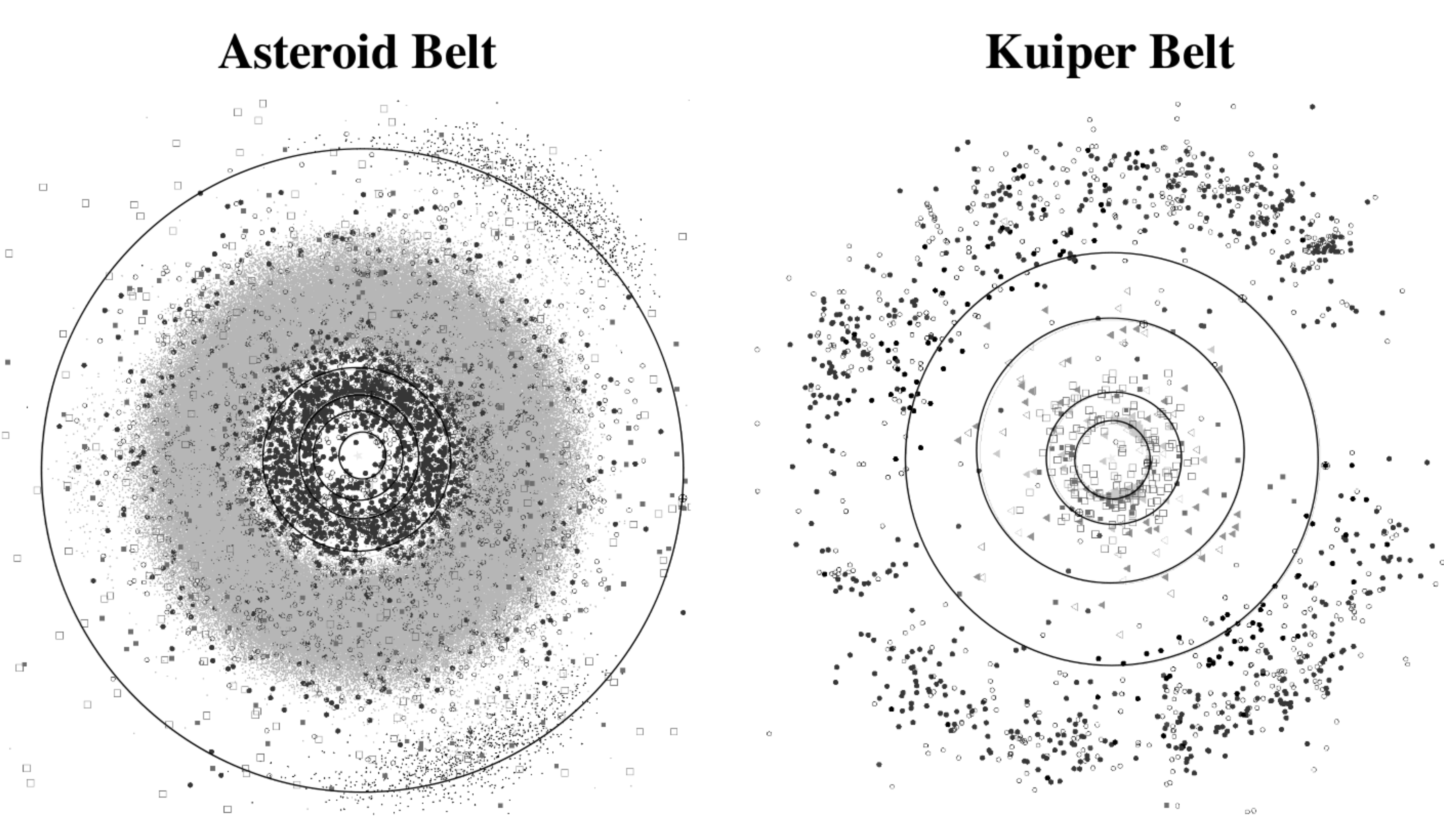}
\caption{Distribution of planetesimals in the Solar system. The  circles represent the orbits of the eight planets;  the outermost circles correspond to the orbits of Jupiter ({\it left}) and Neptune ({\it right}). Courtesy of G. Williams at the Minor Planet Center. 
}
\label{minorplanet}    
\end{center}
\end{figure}

\begin{figure}[]
\begin{center}
\includegraphics[width=0.75\textwidth]{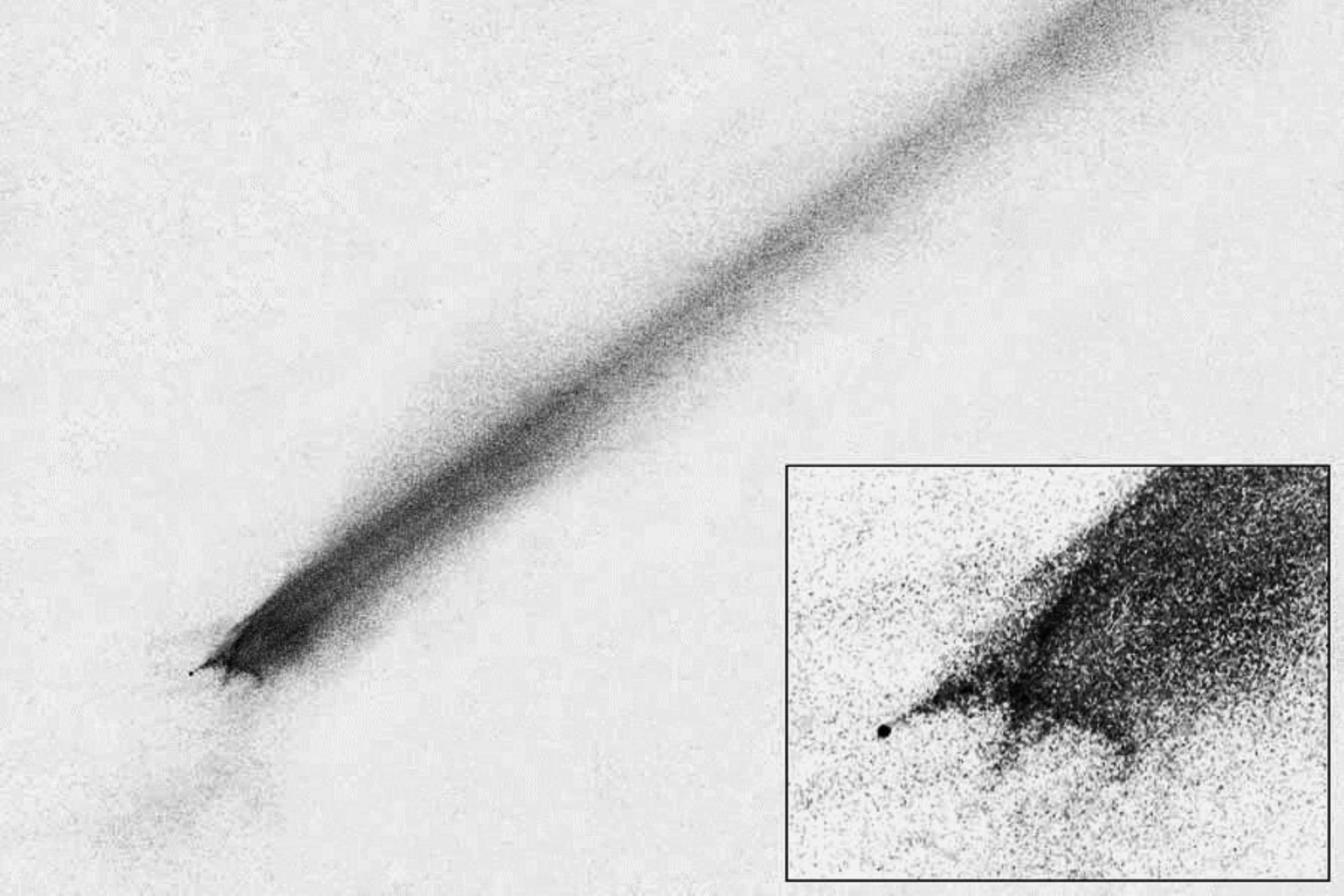}
\caption{Dust production by the inner belt asteroid P/2010A2 as seen by the {\it Hubble} Space Telescope Wide Field Camera 3.  Its comet-like morphology is due to a tail of millimetre-sized dust particles emanating from a 120 meter nucleus. Figure from Jewitt et al. (2010). 
}
\label{2010A2}    
\end{center}
\end{figure}

Observations with {\it Spitzer} show that there is evidence that {\it at least} 15\% of mature stars (10 Myr--10 Gyr) of a wide range of masses (0.5--3 M$_{Sun}$) harbor planetesimal belts with sizes of 10s--100s AU. This evidence comes from the presence of an infrared emission in excess of that expected from the stellar photosphere, thought to arise from a circumstellar dust disk. The reason why these dust disks are evidence of the presence of planetesimals is because the  lifetime of the dust grains is of the order of 0.01--1 Myr (see Sections \ref{sec:radiation_stellar_wind_effect_on_dynamics} and \ref{sec:coll_lifetime}), much shorter than the age of the star ($>$10 Myr); therefore, the origin of these dust grains cannot be primordial, i.e. from the cloud of gas and dust where the star was born, but must be the result of on-going dust production generated by planetesimals, like the asteroids, comets and Kupier Belt Objects (KBOs) in our Solar system (see Figures  \ref{minorplanet} and \ref{2010A2}). This is why these dust disks are known as {\it debris disks}. 

\begin{figure}[]
\begin{center}
\includegraphics[width=0.75\textwidth]{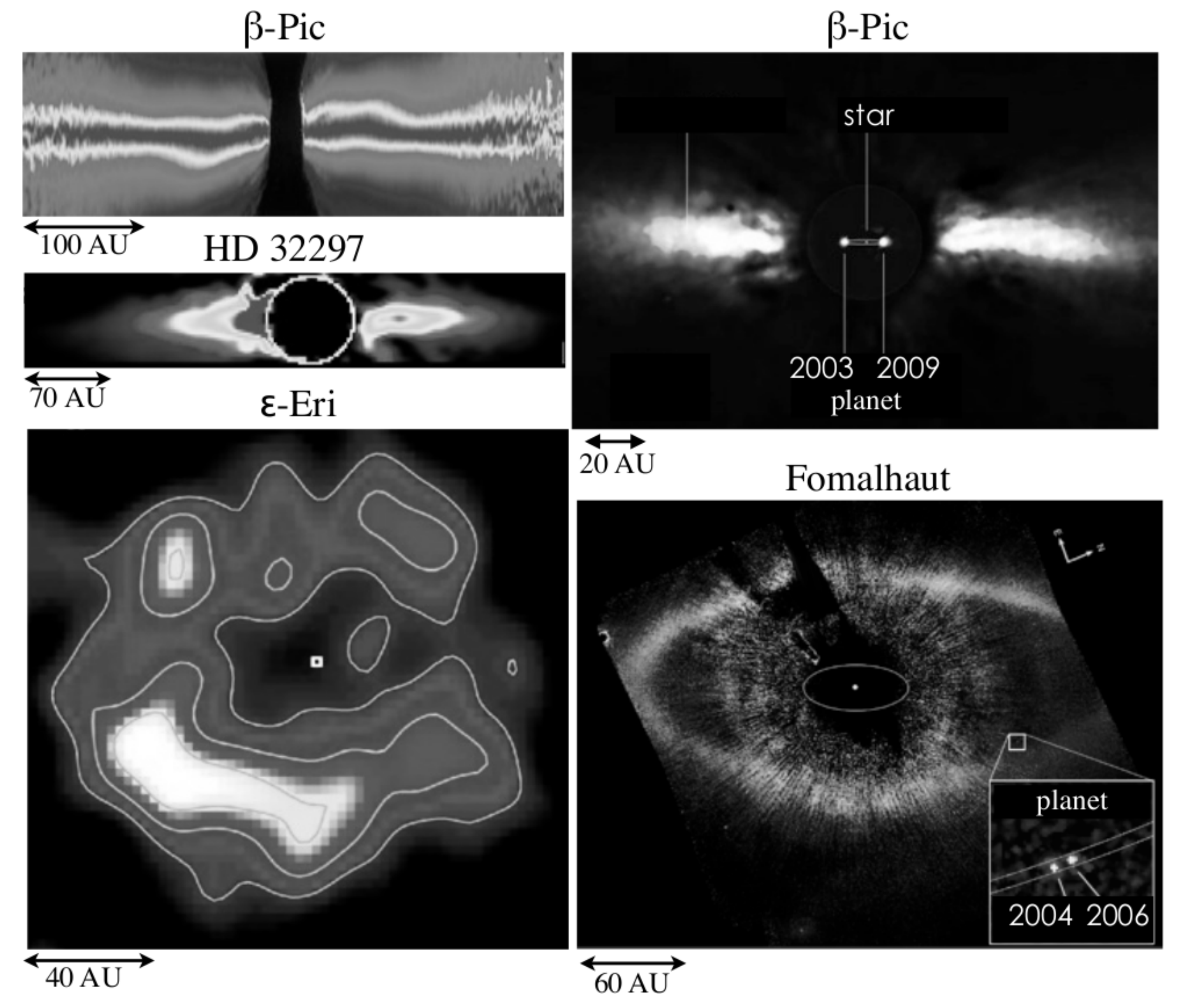}
\caption{({\it Left}): Spatially resolved images of nearby debris disks showing dust emission from 10s to 100s of AU with a wide diversity of complex features including inner gaps, warps, brightness asymmetries, offsets and clumply rings, some of which may be due to the presence of massive planets; from top to bottom: $\beta$-Pic (HST/STIS CCD coronography at 0.2--1 $\mu$m; Heap et al. 2000), HD 32297 (HST/NICMOS coronography at 1.1 $\mu$m; Schneider, Silverstone and Hines 2005) and $\epsilon$-Eri (JCMT/SCUBA at 850 $\mu$m; Greaves et al. 1998, 2005). ({\it Right}): Direct detection of planets in debris disks systems predicted to exist from the disk morphology; from top to bottom: $\beta$-Pic (VLT/NACO at 3.78 $\mu$m; Lagrange et al. 2010) and Fomalhaut (HST/ACS at 0.69--0.97 $\mu$m; Kalas et al. 2008). 
}
\label{structure_planets}    
\end{center}
\end{figure}

Debris disks are therefore evidence of the formation of planetesimals around other stars. The study of this population of extra-solar planetesimals can give us a more complete picture of the diversity of planetary systems, shedding light on their formation and dynamical histories. Even though these planetesimals will remain undetected in the foreseeable future, the study of their debris dust can  help us learn about some of the planetesimals' characteristics (e.g. location, composition and evolution). Debris disks can also help us learn about the planet population. As will be discussed in Sections \ref{structurecreatedbyplanets} and 
\ref{sec:gravitational_forces}, the structure of the debris disk is sensitive to planets with a wide range of masses and semi-major axes, and is independent of the system's age (see Figure \ref{structure_planets}). Therefore, the study of debris disk structure can serve as a planet-detection method, covering a parameter space complementary to that of radial velocity, transit and direct imaging techniques.  The goal of this chapter is to describe how debris disks can shed light on the formation, evolution an diversity of planetary systems, helping us place our Solar system into context.
	
\section{The Solar system debris disk}
\label{sec:solardd}

The Sun also harbors a disk of dust produced in the inner and outer Solar system by its population of minor bodies, like the asteroids, Kuiper Belt objects (KBOs) and comets (Figure \ref{minorplanet}). 

\subsection{Debris dust in the inner Solar system}
\label{sec:inner}

Figure \ref{2010A2}  shows an spectacular example of the production of dust by the asteroid P/2010A2 (from Jewitt et al. 2010). In this case, the dust is produced either by the rotational disruption of the asteroid (due to the spin-up produced by radiation torques), or by the collision with a meter-sized projectile, an event that must have taken place around February or March 2009.   Long before the technological development allowed detailed images like this one to be obtained, there were some phenomena, visible to the naked eye and that must have been noticed since the  beginning of humankind,  that hinted that our planetary system was a dusty one: the zodiacal light, the "shooting stars" and the comet tails. Their analysis with current instrumentation sheds light on the origin and properties of the Solar system's debris dust. 

\subsubsection{Zodiacal dust}
\label{zodiacaldust}
The zodiacal light is a glow that appears along the ecliptic and can be seen about an hour before and after sunset (in the Eastern and Western horizons, respectively).  The term "zodiacal" refers to its location along the ecliptic. It is caused by the scattering of solar photons by the dust grains in the inner Solar system. It appeared in ancient Egyptian art represented by a triangle inclined with respect to the horizon (worshiped as the war god Sopdu), and it was also  known to the Greek and the Romans, the Chaldeans and the Aztecs ({\it Codex Telleriano-Remensis}). The first explicit description in Europe appeared in 1661 (Childrey's {\it Britannia Raconica}) and the first scientific observations were done by Cassini in 1683, and correctly interpreted by de Duiliers a year later as produced by sunlight reflected from small particles orbiting the Sun. Scattered light observations of the zodical light can help determine the properties of the dust particles, revealing non-spherical, irregular or fluffy aggregates, 10--100 $\mu$m in size grain  (see Figure \ref{idp}), composed of a mixture of silicates and organic material with a low albedo of $\sim$ 0.08. Regarding their spatial distribution, the observed brightness profile of the scattered light follows $r^{-2.3}$  from 0.3--1 AU,  $r^{-2.5}$ out to 2.3 AU,  and  $r^{-2.37}$ out to the asteroid belt, corresponding to a number density distribution not too different from the $r^{-1}$ dependency expected in the case of dust dynamics  dominated by Poynting-Robertson drag (see Section \ref{sec:radiation_stellar_wind_effect_on_distribution}). It is also observed that the plane of symmetry of the zodiacal light is warped, as a result of gravitational perturbations by the planets (see Sections \ref{sec:secular_resonances} and \ref{sec:grav_effect_on_dynamics}).  The motion of the dust particles can be studied from the absorption lines in the scattered stellar light, revealing particles in elliptical prograde orbits that belong to two dust populations: one with a spherical distribution and a $r^{-2}$ number density radial profile, likely due to dust particles produced by long period comets, and another with a flattened low inclination distribution and a number density $r^{-1}$ dependency, likely due to asteroids or short period comets (see review in Levasseur-Regourd et al. 2001 and references therein). 

The heating of the dust particles by the Sun makes them emit at infrared and sub-milimeter wavelengths; this thermal emission dominates the night sky between 5--500 $\mu$m and has been mapped by the {\it IRAS}, {\it COBE}, {\it ISO} and {\it Spitzer} space telescopes. These observations indicate that the ratio of the dust-to-stellar luminosities, known as the fractional luminosity, is $f = L_{dust}/L_{\odot}$ $\sim$ 10$^{-8}$--10$^{-7}$ (Dermott et al. 2002); this is more than two orders of magnitude fainter than the extra-solar debris disks observed with {\it Spitzer}  (see Figure \ref{bryden} -- the difference being due to the limited sensitivity of the {\it Spitzer} observations). It is inferred that the particles dominating the thermal emission from the zodiacal cloud are rapidly-rotating grains 10--100 $\mu$m in size, with low albedos, located near 1 AU, and with an amorphous forsterite/olivine composition; there is also a population of smaller 1 $\mu$m-sized grains made of crystalline olivine and hydrous silicate that accounts for a weak silicate emission feature  at 10 $\mu$m (from the vibration of stretching Si-O bonds -- Reach et al. 2003).  The laboratory analysis of interplanetary dust particles (IPDs) collected in the Earth's stratosphere also reveal this mixture of amorphous and crystalline olivine and pyroxene, that has also been identified in the mid-infrared spectra of extrasolar debris disks (see Section \ref{sec:composition}). Regarding the spatial distribution, the thermal emission from the zodiacal cloud shows long, narrow arcs that coincide with the perihelion passage of some short-period comets (produced by  low albedo, porous, millimeter-sized dust particles), and broader dust bands  at low ecliptic latitudes,  thought to originate from the break-up of the asteroids that gave rise to the Themis, Koronis and Eos asteroidal families (Sykes \& Greenberg 1986; Dermott et al. 2002 argued that the formation of the Veritas family 8.3 Myr ago accounts for $\sim$25\% of the zodiacal thermal emission today). The gravitational perturbations exerted by the planets are also evident in the spatial distribution of the thermal emission: a ring of asteroidal dust particles are trapped in the exterior mean motion resonances (MMRs) with the Earth at around 1 AU, forming a ring-like structure with a 10\% number density enhancement on the Earth's wake that results from the resonance geometry of the  1:1 MMR (Dermott et al. 1994; Reach et al. 1995; see discussion in Section \ref{sec:grav_effect_on_dynamics}).

\subsubsection{Dust particles falling on Earth}
\label{dustfallinonearth}

"Shooting stars" are evidence that dust particles fall on Earth. This phenomenon, that must have been noticed since the beginning of humankind, is produced by the ionization of atmospheric atoms along the path of the incoming high-velocity dust grain.  The grain generally gets destroyed before it reaches the surface of the Earth because the impact of atmospheric molecules increases its temperature to $\sim$ 1000--2000 K, at which point the grain's atoms and molecules begin to ablate; another destruction process is the loss of the volatile glue that maintains the aggregate together. The trails can be used to trace back the orbits of the incoming dust particles, revealing that most of the sporadic meteors are generated by particles on prograde, low ecentric orbits, with small relative velocities with respect to the Earth of a few km/s. Meteors also occur in showers when the Earth crosses the dusty trail of an asteroid or a comet, the latter produced by the gradual released of dust during  ice sublimation or by the breakup of inactive comets. The origin of the dust can be traced back to the parent body because their orbital elements remain similar during the first 10$^4$ years (see review in McDonnell et al. 2001 and references therein). 

Some of the dust grains survive atmospheric entry and can be collected on the Earth's surface (from deep sea sediments and Greenland and Antarctic ice) in the form of micrometeorites with sizes in the 20 $\mu$m--1 mm range (Maurette et al. 1994). Because of the effect of atmospheric entry, particles larger than 50 $\mu$m are strongly affected and there is a selection effect against particles with high entry velocities, high densities and/or fragile structure. Micrometeorites can be collected in the Antarctic ice, e.g. at the bottom of the water well of the Amundsen-Scott South Pole Station, where the particles accumulate as the ice melts during the drilling of the well; a ton of this ice contains approximately 100 cosmic spherules larger than 50 $\mu$m, and about 500 micrometeorites in the 50-400 $\mu$m range (Maurette et al. 1994, 1996); the slope of the cumulative size distribution for particles larger than 200 $\mu$m is -5.2.   The estimated particle flux, compared to that found in the upper atmosphere in a similar size range, indicates that only about 4\% of the incoming dust grains survive atmospheric entry.  Micrometeorites  have also been collected in deep sea sediments (see Section \ref{earlysolarsystem}). The fate of the organic material on the dust particles that enter the Earth's atmosphere is particularly interesting for astrobiology; this matter may survive entry if it is in the form of complex compounds. (For a review see Jessberger et al. 2001 and references therein). 

\begin{figure}
\begin{center}
\includegraphics[width=0.70\textwidth]{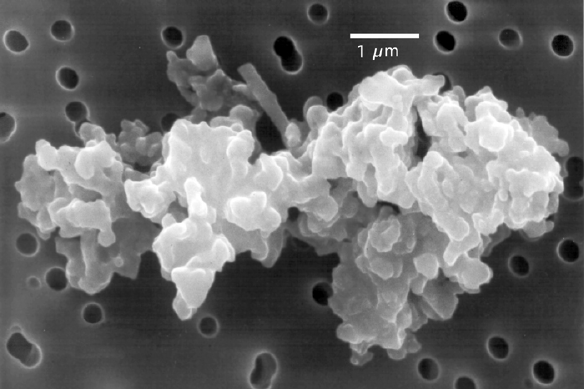}
\caption{Interplanetary dust particle (IDP) collected in the Earth's stratosphere.  Figure from D. E. Brownlee (University of Washington) and E. Jessberger (Institut f\"ur Planetologie, M\"unster). 
}
\label{idp}
\end{center}
\end{figure}

Dust particles can also be  collected from the Earth's stratosphere using aerogel collecting plates on high altitude flying aircraft.  These interplanteary dust particles (IDPs) are distinguished from high altitude terrestrial dust  (e.g. the dust produced by solid rockets) by their elemental and isotopic composition, and by the effect of their long journey in interplanetary space, the latter revealed  by He atoms implanted by the solar wind in the bubbles, voids and crystal defects of the grains, and by the radiation damage due to the impacts  with high-energy cosmic rays and stellar wind particles. The collected IDPs are on average 15 $\mu$m in size (with a size range of 5--25 $\mu$m determined by the collecting method and the terrestrial contaminants). Figure \ref{idp} shows a  typical IDP, formed by aggregates of thousands to millions of sub-micron grains, with about 40\% porosity and a bulk density of about 2 g/cm$^{-3}$  (this is consistent with the formation of the IDP as a random aggregate of similar-sized components). About 85\% of the IDPs are formed by aggregates of different minerals (chondritic composition), while the rest are aggregates of just a few minerals (non-chondritic composition). Chondritic IDPs tend to be dark (with low albedos of 0.05--0.15, due to the presence of carbon, sulfides and nanometer-sized metal grains), and their mineralogy is similar to that of the matrix of some carbonaceous chondrites and can be anhydrous (dominated by olivine and pyroxene) or hydrous; their structure ranges from porous (generally anhydrous silicates), to compact (generally hydrated silicates that have been subject to aqueous alteration). Because the chondritic IDPs are aggregates of thousands to millions of mineral grains, their overall elemental composition is similar to solar.  The entry velocities of the IDPs can be estimated from the amount of solar wind-implanted He that was lost during atmospheric entry (due to the heating of the dust particles). There are two populations of IDPs: (1) Low velocity grains ($<$ 14 km/s), likely associated with the low eccentricity and low inclination orbits of asteroidal dust; they tend to be compact, with average bulk densities of about 3 g/cm$^3$, and they commonly have a chodritic composition of low crystalline Fe-Mg hydrous silicates similar to that of carbonaceous chondrite meteorites. (2) High velocity grains ($>$ 18 km/s), with velocities similar to those expected from cometary dust; they have low bulk densities of about 1 g/cm$^3$, and a composition is similar to that found in comets like e.g. Hale-Bopp, with both crystalline and non-crystalline Fe-Mg anhydrous silicate minerals (pyroxenes and olivines) and GEMS. (For a review see Jessberger et al. (2001) and references therein). 

\subsubsection{In situ dust detections in the inner Solar system}
\label{innerdustinsitu}

Another of the earliest observations of debris dust in the Solar system are  the sightings of comet tails. Records of comets appear as early as in Chinese oracle bones, and in many cultures they have been considered bad omens. Comets are now treasured as the most pristine planetesimals in the Solar system, harboring invaluable information about our planetary system's formation and evolution. Comets are one of the main sources of dust; because sublimation drives the cometary activity,  it is assumed that their dust production generally  takes place in the inner Solar system. However, there are isolated flare-ups that also produce dust at large heliocentric distances. The dust production rate of comets is difficult to estimate because the cometary activity is not steady (e.g. comet Holmes had a massive mass-loss in 2007 that made it become 10$^6$ times brighter in a day --  Li et al. 2010).  A study by Nesvorny et al.  (2010) concluded  that about 85\% of the dust in the inner Solar system is produced by Jupiter-family comets and $<$10\% by long-period comets (a result that is model-dependent). 

Cometary dust particles have been studied {\it in situ} in the case of comets Halley, Tempel 1 and Wild 2 using instruments on-board the {\it VeGa} 1, {\it VeGa} 2, {\it Giotto}, {\it Deep Impact} and {\it Stardust} spacecrafts. {\it Deep Impact} sent a copper-core impactor into comet Tempel 1 to study the ejected cloud of sub-surface material (observed from ground-based telescopes and from the flyby section of the spacecraft). It found that the surface of the comet was more dusty than expected, depleted of ice that is present at depths of about 1 m (A'Hearn 2008); ground-based observations of the dust cloud released by the impact revealed the presence of forsterite (Mg2SiO4) and enstatite (MgSiO3). Comet Wild 2 was the target of the {\it Stardust} mission, that was able to return cometary dust particles to Earth for detailed laboratory analysis. Their study reveals that even though Wild 2 is thought to have originated in the cold environment of the Kuiper Belt,  it contains aggregates where highly refractory and volatile components are found close together in the micron-scale, indicating that the comet has aggregated dust particles that have formed at a very wide range of heliocentric distances:  there is pre-solar material identified from the isotopic evidence and that has suffered little alteration since it was accreted; there are large 5 $\mu$m Fe-poor forsterite crystals that are likely formed from the vaporization, melting or heating of material above 1000 K; there are calcium-aluminium inclusions (CAIs -- the oldest solids of the Solar system) thought to have condensed at very high temperatures during a brief period of time, and that must have been transported from the inner Solar system (where CAIs formed) to the region where the comet formed; and there are also amorphous glassy grains some of which have an interstellar origin. The returned samples contained a broad range of minerals, each of which has been found in primitive meteorites but never such a rich diversity in the same object. This  heterogeneous mixture is far from being in chemical or mineralogical equilibrium (Brownlee et al., 2006, McKeegan et al. 2006, Zolensky et al., 2006a, Sandford et al., 2006). 

Dust particles have also been detected {\it in situ} at different heliocentric distances by the spacecrafts  {\it HEOS} (1 AU), {\it Hitten} (1 AU),  {\it Helios} (0.3--1AU), {\it Galileo} (0.7--5 AU), {\it Pioneer} 8 and 9 (0.75--1.08 AU), {\it Ulysses} (1.3-2.3 AU), {\it Cassini} and {\it Pioneer} 10 and 11. Impact velocities   $>$ 1 km/s result in the vaporization and ionization of material from the target and the impactor, that is separated by an applied electric field and produces an electronic signal that depends on the mass, velocity and chemical composition of the impacting grain. One of the challenges is to calibrate the detectors in the laboratory using impactors with characteristics similar to that of interplanetary particles.  Another challenge is that these {\it  in situ} measurements are limited to the trajectory of the spacecraft, so the interpretation of the impact data depends on assumptions regarding the dust spatial distribution. These measurements are also biased against particles at high inclinations because they spend little time near the ecliptic where most of the experiments take place  (see review by Gr\"un et al. 2001). 

The current dust production rate in the inner Solar system is of the order of 10$^4$ kg/s; the relative contribution of the different sources is still under debate and has likely changed with time. As was mentioned above, Nesvorny et al. (2010) argued that the splitting of Jupiter-family comets accounts for 85\% of the dust in the inner Solar system with $<$10\% produced by Oort Cloud long-period comets, while a previous study by Dermott et al. (1994) concluded that the asteroids contribute to at least 33\%; other contributors of dust to the inner Solar system could include Halley Type comets and KBOs. 

\subsection{Debris dust in the outer Solar system}
\label{sec:outer}

As discussed in  Section \ref{zodiacaldust}, the dust in the inner Solar system ("zodiacal dust") has been studied via remote observations of its scattered light and thermal emission. The situation is very different for the dust in the outer Solar system, for which no remote detections have been achieved so far: at optical wavelengths,  its emission is dwarfed by the much brighter zodiacal light in the foreground, while at infrared and longer wavelengths, its emission is also confounded by  the  thermal emission of the zodiacal dust in the foreground and emission from the galactic dust (known as the "cirrus") in the background.  The cosmic microwave  background radiation provides a very uniform source against which emission from the Kuiper Belt (KB) might  be detected in the future. The derived upper limit on the total mass of dust in the outer Solar system is about 10$^{-5}$ M$_{\oplus}$ or 10$^{20}$ kg, a thousand times the mass inferred from the {\it Voyager} 1 detection experiments (Backman, Dasgupta \& Stencel 2005; Jewitt \& Luu 2000; Moro-Mart\'in \& Malhotra 2003). Dynamical modeling of the dust produced in the Kuiper Belt, together with recent in situ detections by the {\it New Horizons} spacecraft, indicates that the fractional luminosity of the Kuiper Belt dust, L$_{dust}$/L$_{Sun}$, is $\sim$10$^{-7}$; this is below the detection limit {\it Herschel}/PACS, meaning that Kuiper Belt dust disk analogs around other stars could not be detected with present-day instrumentation in space (Vitense et al. 2012). 

{\it Pioneer} 10 and 11 detected dust out to 18 AU and 13 AU, respectively, with detectors that consisted on pressurized gas cells with metallic walls 25-50 $\mu$m thick, that  would lose pressure when penetrated by a projectile (with the problem that  the usable area of the detector diminished with time). For the detectable mass range (10$^{-12}$ to 10$^{-11}$ kg for impact velocities of 20 km/s), it was found that the flux was nearly constant between Jupiter and 18 AU (Humes 1980).  Even though Pioneer 10 and 11 did not sample the KB region directly (because the former failed at 18 AU and the latter was turned off at 13 AU), dynamical models indicate and that the KB was likely the source of the dust detected beyond 10 AU (Landgraf et al. 2002). Therefore, the detection of Kuiper Belt dust preceded the discovery of the Kuiper Belt itself by Jewitt \& Luu (1993), but at the time the origin of the dust went unrecognized. The dynamical models also require a significant contribution from comets (including short-period Oort cloud and Jupiter-family comets) to account for the flat number density distribution of the dust.  

{\it Voyager} detected dust in the 30--60 AU Kuiper Belt region with a number density of  $n \sim$ 2$\times$10$^{-8}$ m$^{-3}$. The data remains poorly calibrated because the detections were done indirectly by measuring the pulses in the conductivity of the medium adjacent to the spacecraft, caused by plasma generated from the vaporization of the impactor; it is thought that smallest dust particles detected were $s \sim$2 $\mu$m in size (Gurnett et al. 1997). Because the size of the dust particles follow a power-law distribution (see Section \ref{sec:coll_effect_on_dust_size}), a reasonable approximation is that most impactors were of this minimum size.  Adopting a  KB vertical height of $H \sim$ 10 AU, this would corresponds to an optical depth of $\tau \sim \pi s^2 H n \sim 4\times 10^{-7}$ (Jewitt \& Luu 2000), about two orders of magnitude smaller than that of the extra-solar debris disks detected by {\it Spitzer} (Figure \ref{bryden}, where the fractional luminosity can be approximated by the optical depth, $f = L_{dust}/L_{star} \sim \tau$ -- see discussion in Section \ref{mutualcollisions}); this is comparable to the normal optical depth of the zodiacal dust, $\tau \sim$ 10$^{-8}$--10$^{-7}$ (see Section \ref{zodiacaldust}).   Additional evidence of dust-producing collisions in the KB are the impact craters imaged by the {\it Deep Impact} mission on the surface of the comet Tempe1, thought to be created during the time the comet belonged to the KB. 

The dust production rate estimates in the outer Solar system are in the range (0.2--5)$\times$10$^4$ kg/s (from Voyager and Pioneer data, respectively; Jewitt \& Luu 2000; Landgraf et al. 2002). Dust production rates from theoretical models are in the range of (0.1--1)$\times$10$^4$ kg/s, from the erosion of KBO surfaces by the flux of interstellar grains (Yamamoto \& Mukai 1998), to (1--300)$\times$10$^6$, from mutual grain-grain collisions (Stern 1996). For comparison, the dust production rate in the inner Solar system is of the order of 10$^4$ kg/s (see Section \ref{innerdustinsitu}).  

{\it Cassini} also detected 17 dust impact events between the orbits of Jupiter and Saturn (however, the orientation of the spacecraft was not optimized to maximize the number of detections). The particles are inferred to have sizes in the sub-micron to micron range, and are found on bound and unbound orbits. Particles on bound orbits have low eccentricities and low inclinations; the shape of the impact signals indicated that the grains are irregular and have high porosity, as expected from a cometary origin; particles on unbound orbits are inferred to have sizes of $\sim$ 0.4 $\mu$m, in agreement with an interstellar origin (Altobelli et al. 2007). 

\subsection{Evolution of the dust production rate in the Solar system}
\label{earlysolarsystem}

\begin{figure}[]
\begin{center}
\includegraphics[width=0.80\textwidth]{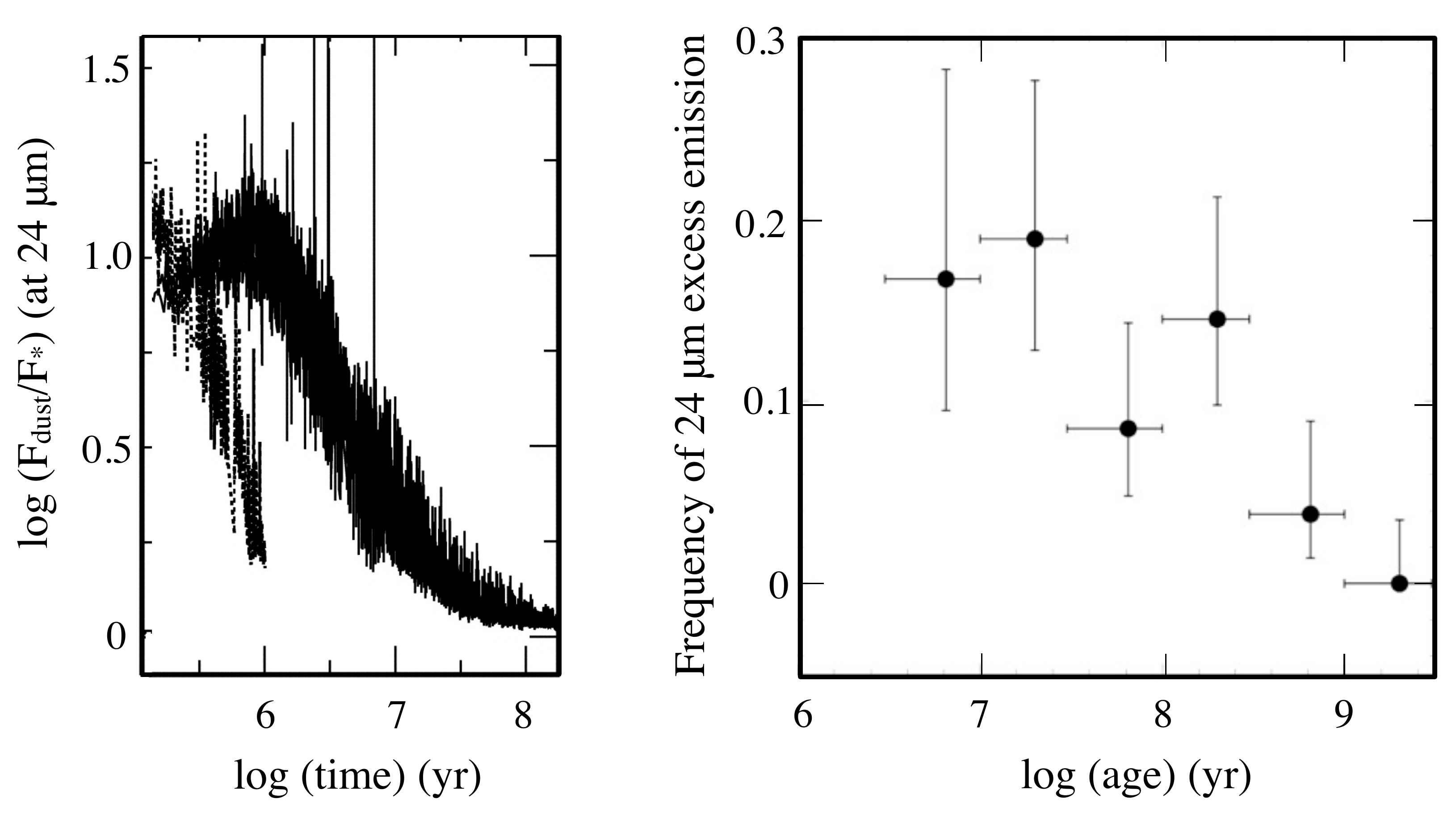}
\caption{({\it Left}): Evolution of the 24 $\mu$m excess as a function of time for two planetesimal disks extending from 0.68 to 1.32 AU (dashed line) and 0.4 to 2 AU (solid line). The central star is solar type. Excess emission decreases as planetesimals grow into Mars-sized or larger objects and collisions become increasingly rare. From Kenyon and Bromley (2005).  ({\it Right}): Fraction of stars in a sample of 309 FGK stars with detectable 24 $\mu$m excess plotted as a function of age. Each bin spans a factor of 3 in age. The vertical bars are Poisson errors. From Meyer et al. (2008). 
}
\label{24evol}    
\end{center}
\end{figure}

The dust production rate in the Solar system has changed significantly with time. It is thought that the Solar system was significantly more dusty in the past because the asteroid and the Kuiper belts were more densely populated.  Evidence for a massive primordial KB is the existence of KBOs larger than 200 km, which formation by pairwise accretion must have required a number density of objects about two orders of magnitude higher than today. Evidence for a massive primordial asteroid belt (AB) comes from the minimum mass solar nebula, showing a strong depletion in the AB region unlikely to be primordial. The Solar system then became progressively less dusty as the planetesimal belts eroded away by mutual planetesimal collisions. Evidence of collisional evolution comes from the modeling and observation of the size distribution of the asteroids and KBOs (see discussion in Section \ref{sec:coll_effect_on_dust_size} and Figure \ref{bottke_bernstein}).  This collisional evolution likely resulted in the production of large quantities of dust, as it can be seen in the left panel of Figure \ref{24evol} from a model by Kenyon and Bromley (2005). They found that in a planetesimal belt, Pluto-sized bodies $\sim$1000 km in size excite the eccentricities of the more abundant 1--10 km sized planetesimals, triggering a collisional cascade that produces dust and changes the planetesimal size distribution. Because  the dust production rate is proportional to the number of collisions, and this is proportional to  the square of the number of planetesimals, as the planetesimals erode and grind down to dust, the  dust production rate decreases and the expected thermal emission from the dust slowly decays with time as $1/t$ (see discussion Section \ref{sec:coll_effect_on_dust_evolution}). This decay is punctuated by large spikes that are due to large collisions happening stochastically (left panel of Figure \ref{24evol}).  Examples of stochastic events in the recent history of the Solar system are the fragmentation of the asteroids giving rise to the asteroidal families and the dust bands. 

\begin{figure}[]
\begin{center}
\includegraphics[width=0.95\textwidth]{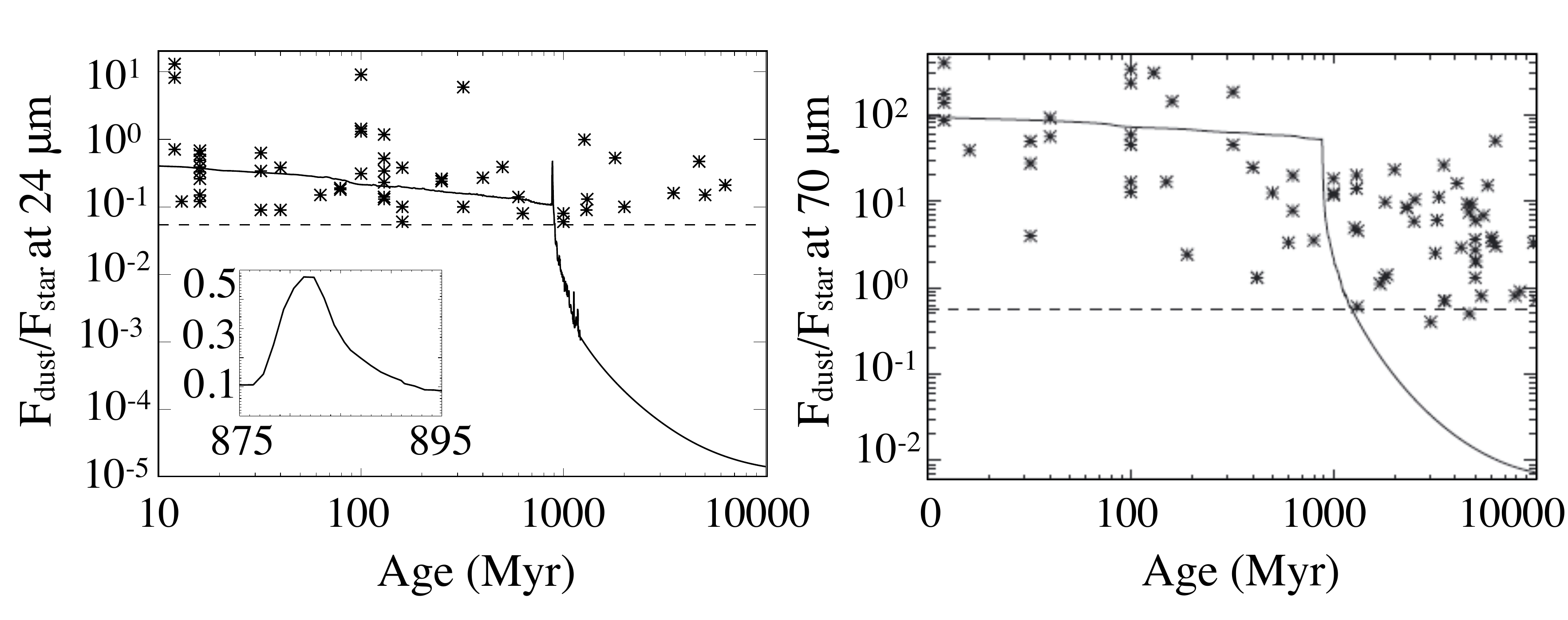}
\caption{Excess ratio ($F_{dust}/F_{star}$) versus time at 24 $\mu$m ({\it left}) and at 70 $\mu$m ({\it right}). The asterisks correspond to {\it Spitzer}/MIPS observations of FGK stars and the dashed lines are the observational limits. The solid lines correspond to a model of the dust production in the Solar system, assuming blackbody grains with a power-law size distribution of $n(s)ds \propto s^{-3.5}ds$; the sharp decrease is due to the drastic planetesimal depletion that took place at the time of the Late Heavy Bombardment.  From Booth et al. (2009). 
}
\label{booth}    
\end{center}
\end{figure}

A major change in the dust production rate is expected to have occurred in the early Solar system at the time of the Late Heavy Bombardment (LHB), a period in which a large number of impact craters in the Moon and the terrestrial planets were created (with an impact rate at Earth approximately 20000 $\times$ the current value).  Because this heavy bombardment deleted all record of previous impacts, it is not clear whether the LHB was a single event, the tail of a very heavy bombardment process, or the last of a series of multiple cataclysm (but see Chapman, Cohen \& Grinspoon 2007).  This event, dated  from lunar samples of impact melt rocks, happened during a very narrow interval of time 3.8 to 4.1 Gyr ago ($\sim$600 Myr after the formation of the terrestrial planets).  Thereafter, the impact rate decreased exponentially with a time constant  ranging from 10--100 Myr (Chyba, 1990). Strom et al. (2005) compared the impact cratering record and inferred crater size distribution on the Moon, Mars, Venus and Mercury to the size distribution of different asteroidal populations, and found that the LHB lasted  $\sim$20--200 Myr, that the source of the impactors was the main AB, and that the mechanism was size independent. The most likely scenario is that the orbital migration of the giant planets caused a resonance sweeping of the  AB and as a result many of the asteroidal orbits became unstable, causing a large scale  ejection of bodies into planet-crossing orbits (explaining the observed cratering record); the orbital migration of the planets also caused a major depletion  of the KB as Neptune migrated outward; it is estimated that $\sim$ 90\% of the KBOs were lost (Gomes et al. 2005). The LHB was probably a single event in the history of the Solar system that would have been accompanied by a high rate of collisions and dust production (see Figure \ref{booth}). After the LHB, there must have  been a sharp decrease in the dust production rate due to the drastic depletion of planetesimal (Booth et al. 2009). 

Some record of the interplanetary dust flux falling on the Earth can be found in the sedimentary rocks at the sea floor. As mentioned in Section \ref{dustfallinonearth},  stellar wind He atoms are implanted in the voids, bubbles and crystal defects of interplanetary dust particles. A significant fraction of the He is lost during atmospheric entry as the particle heats up, but some He survives and under special conditions can remained trapped for millions of years in sedimentary rocks  at the sea floor (Farley 1995).  Extraterrestrial material is associated with high $^3$He/$^4$He ratios, as opposed to that of terrestrial origin; the enhanced isotope ratio observed in some of the cores extracted from the sedimentary rock at the sea floor can be accounted for if approximately 0.5\% of the total mass in interplanetary dust suffered little He loss upon atmospheric entry. The concentration of  extraterrestrial material in the sedimentary rock depends on the flux of interplanetary dust particles and on the sedimentation rate; for pelagic clays, the concentration is large because they accumulate slowly. The record of interplanetary dust flux analyzed this way spans a timescale of 100 Myr (Farley 1995; see summary by Zolensky et al. 2006b), and because of the effects of atmospheric entry, it favors asteroidal dust in low eccentric orbits. It is found that:  (1) The flux increased by a factor of 5 between  36.5 and 34 Myr ago, coinciding with deposition of Ir, shocked quatz and spinnel (associated with major impacts); the sedimentation rate at Earth is well known during that time, but the origin of this enhanced interplanetary dust flux, whether due to asteroidal collisions or to increased cometary activity, is still uncertain. (2) The flux increased by a factor of 4 between 8.2$\pm$0.1  and 6.7 Myr ago; this is likely associated with the collisional cascade that resulted from the the most recent asteroid break up, the one that gave rise to the formation of the Veritas asteriodal family 8.3$\pm$0.5 Myr ago, thought to be caused by the breakup of a 150 km size C-type asteroid rich in hydrated minerals; at that time, it probably constituted the dominant source of (water rich) interplanetary dust at Earth (Dermott et al. 2002 argued that still contributes to about 25\% of the zodiacal dust today, but see Nesvorny et al. 2010).

\section{Extra-solar debris disks}
\label{sec:extrasolardd}

The extrapolation of radial velocity studies indicate that $\sim$17--19\% of solar-type stars harbor giant planets within 20 AU (Marcy et al. 2005).  A natural question arises whether stars also harbor planetesimals, thought to be the building-blocks of planets.  Long before extra-solar planets were discovered, it was inferred that the answer to this question was yes: dust-producing planetesimals had to be responsible for the infrared excesses observed around many mature stars (see discussion in Section \ref{sec:planetesimals}). These dust disks, first discovered by {\it IRAS} (Aumann et al. 1984) and later studied by {\it ISO} (e.g. Habing et al. 2001, Decin et al. 2003), were extensively surveyed by {\it Spitzer}. The goal of the {\it Sptizer} surveys was to characterize the frequency and properties of debris disks around  stars of different spectral types, ages and environment; taking advantage of the unprecedented sensitivity of the {\it Spitzer/MIPS} (24 $\mu$m and 70 $\mu$m) and {\it Spitzer/IRS} (5--35 $\mu$m) instruments (Rieke et al. 2004; Houck et al. 2004), more than 700 stars were surveyed and hundreds of debris disks were identified. The following is a brief summary of the results from the {\it Spitzer} debris disk surveys. 

\subsection{Debris disk frequency}
\label{frequency}
\begin{figure}[]
\begin{center}
\includegraphics[width=0.7\textwidth]{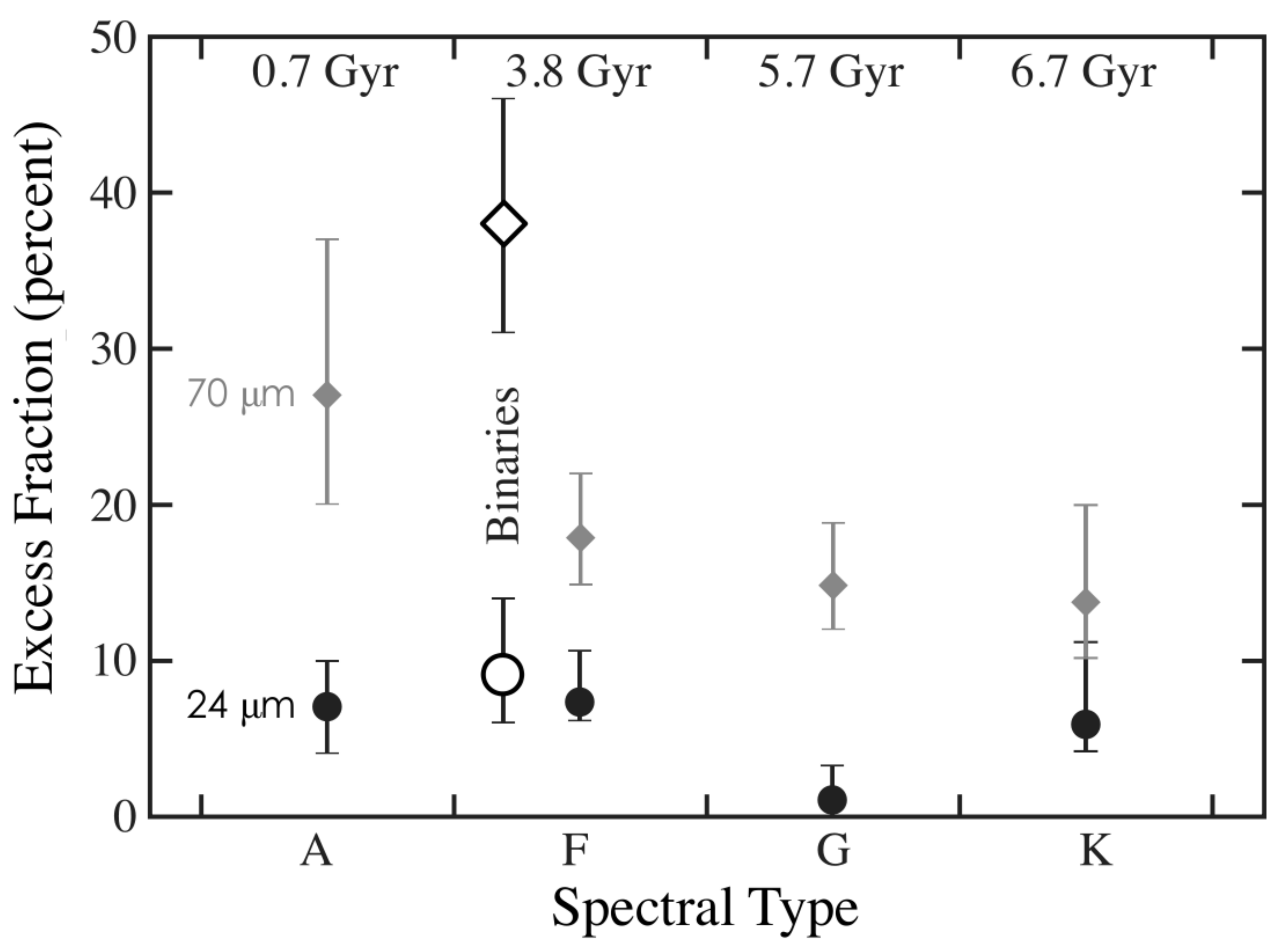}
\caption{The percentage of stars showing excess dust emission as a function of stellar type for ages $>$ 600 Myr (the mean ages within each type are shown at the top); the vertical bars correspond to 1-$\sigma$ Gaussian errors; solid black symbols correspond to single stars with excess emission at 24 $\mu$m (tracing warmer dust), while the solid grey diamonds correspond to excesses at  70 $\mu$m (tracing colder dust); empty symbols are for binary systems. Figure from Trilling et al. (2008). A different  survey of 328 solar-type FGK stars (30 Myr--3 Gyr) found that the frequency of 24 $\mu$m excess is 14.7\% at $<$ 300 Myr and  2\% at $>$300 Myr, while at 70 $\mu$m, the excess  rates are 6--10\% and are fairly independent of age (Meyer et al. 2008; Hillenbrand et al. 2008; Carpenter et al. 2009). 
}
\label{trilling}    
\end{center}
\end{figure}

Most of the debris disks detected by {\it Spitzer} are found around mature main-sequence stars A to K2 type, with stellar luminosities ranging from 0.3 to 3 L$_{\odot}$.  Figure \ref{trilling} summarizes the frequency of debris disks at two different wavelengths in a combined sample of 350 AFGK stars older than 600 Myr (from Trilling et al. 2008). From Wien's law, emission peaks at  24  $\mu$m and 70 $\mu$m would correspond to characteristic dust temperatures of 153 K and 52 K, respectively; assuming 1 L$_\odot$, blackbody grains would adopt that temperature if located at 3 AU (24 $\mu$m) and 28 AU (70 $\mu$m), while a 10 $\mu$m  grain in the intermediate size regime would adopt that temperature if located at 5 AU (24 $\mu$m) and 75 AU (70 $\mu$m -- see Equations \ref{tgrainbb} and \ref{tgrainis}).   Figure \ref{trilling}  shows that debris disks are more common around  A-type stars  than around solar-type FGK stars (the younger age of the A-star sample may bias this result because,  as discussed in Section \ref{ddevolution}, debris disk are common around young stars). It is also found that the frequency of debris disks is significantly higher for solar-type stars than for old M stars. There is a prominent debris disk around the M star AU Mic, but this star is relatively young ($\sim$ 12 Myr). In a survey of 64 old M dwarfs, none of the stars were found to have excesses (Gautier et al. 2007). This may be an observational bias because the peak emission of these colder disks would be at $\lambda$ $>$ 70 $\mu$m, i.e. beyond the wavelength where {\it Spitzer/MIPS} was most sensitive. On-going debris disks surveys with {\it Herschel} are increasing the number of disk detection rates made by {\it Spitzer} and most of the new detected debris disks are found around cold late-type stars (Eiroa et al. in prep. and Kennedy et al. in prep; preliminary results from the {\it Herschel}/DEBRIS survey show debris disk frequencies of 26\%, 24\%, 19\%, 9.5\% and 1.3\% around spectra types A, F, G, K and M respectively - Kennedy et al. in prep.). 

There is also evidence of the presence of planetesimals around white dwarfs. Some of these evolved stars show infrared excesses and high levels of pollutants (elements other than the expected pure H and He), thought to arise from tidally disrupted planetesimals. Section \ref{sec:composition} discusses how the atmospheric abundances of these white dwarfs provide a  unique opportunity to study the elemental composition of the disrupted planetesimals. The presence of planetesimals around stars with  a very wide range of spectral types, from M-type to the progenitors of white dwarfs -- with several orders of magnitude difference in stellar luminosities -- implies that planetesimal formation is a robust process that can take place under a wide range of conditions. 

\begin{figure}[]
\begin{center}
\includegraphics[width=0.55\textwidth]{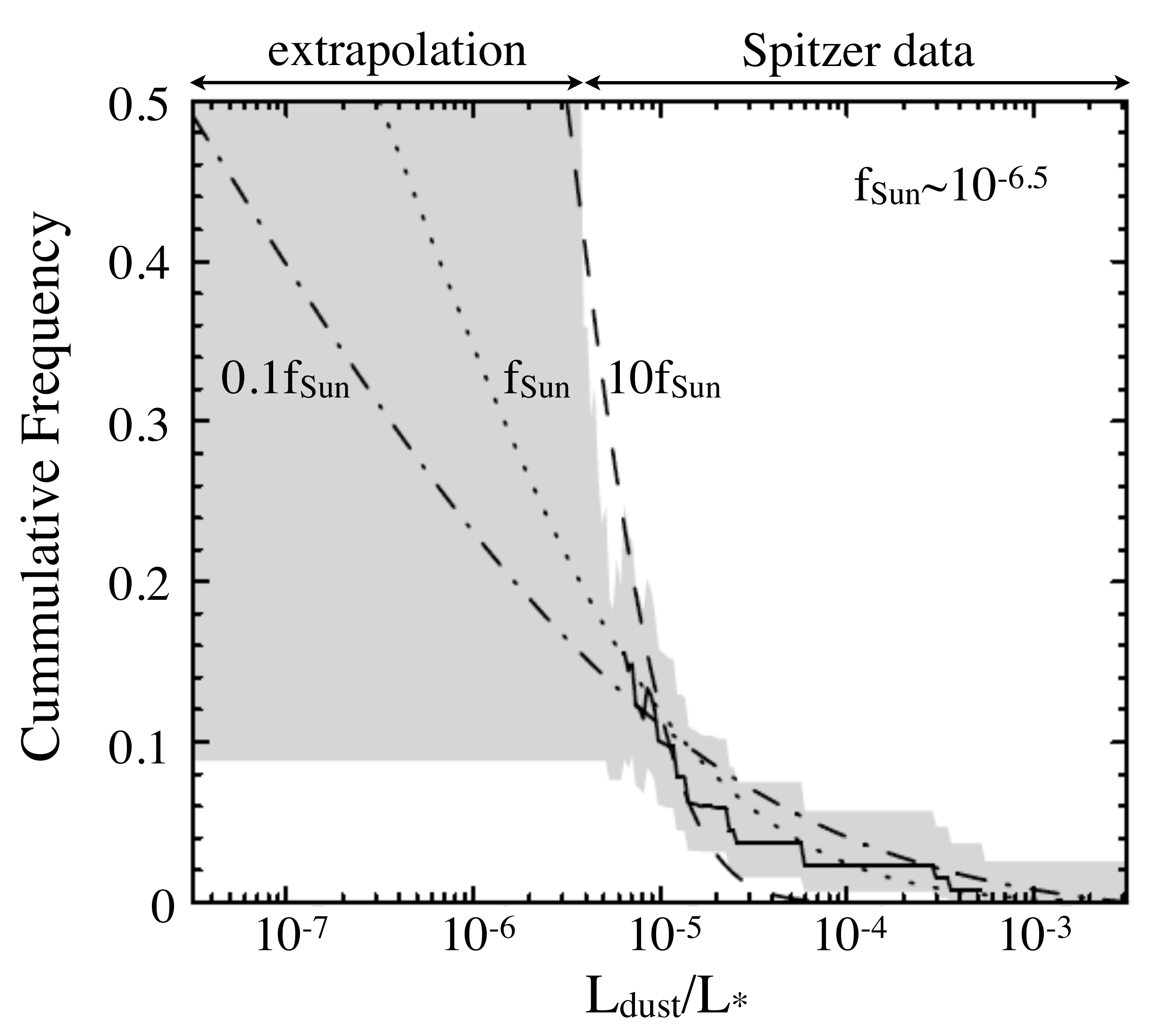}
\caption{Disk detection cumulative frequency as a function of dust fractional luminosity ($f = L_{dust}/L_*$). The grey region corresponds to Spitzer data (right) and an estimate based on extrapolation of these observations (left). The lines correspond to  theoretical debris disk distributions  assuming a gaussian distribution of debris disk fractional luminosities ($f$) and considering an average fractional luminosity of:
$f_{ave}$ = $f_{Sun} \sim10^{-6.5}$, i.e. similar to that of the Solar system's debris disk (dotted line); 
$f_{ave}$ = 10$f_{Sun}$ (dashed line); 
$f_{ave}$ = 0.1$f_{Sun}$ (dot-dashed line). 
Figure from Bryden et al. (2006). 
}
\label{bryden}    
\end{center}
\end{figure}

Figure \ref{bryden}  shows the disk detection frequency from {\it Spitzer} surveys, indicating that there is a steep increase with decreasing fractional luminosity ($f = L_{dust}/L_*$; from Bryden et al. 2006) .  Due to the limited sensitivity of the {\it Spitzer} debris disks surveys, the detected fractional luminosities are generally  $f \gtrsim 10^{-5}$; this is larger than those inferred for the Solar system's debris disk today : $f \sim 10^{-8}-10^{-7}$ for the inner Solar system and $f  \sim 10^{-7}-10^{-6}$ for the outer Solar system (although the latter is only an estimate because its emission is overwhelmed by the zodiacal dust foreground -- see Section \ref{zodiacaldust} and \ref{sec:outer}). Figure \ref{bryden}  compares the observations to theoretical debris disk distributions: assuming a gaussian distribution of debris disk luminosities and extrapolating from  {\it Spitzer} observations, Bryden et al. (2006) concluded the fractional luminosity of an average debris disk around a solar-type stars could be between 0.1--10 $\times$ that of the Solar system debris disk. In other words,  the observations are consistent  with debris disks at the Solar system level being common (but they would have been too faint to be detected by {\it Spitzer}). On-going debris disks surveys with {\it Herschel} (e.g. DUNES and DEBRIS -- Eiroa et al. in preparation, Kennedy et al. in preparation) are now probing the frequency of disk detections for fainter disks.  However, recent estimates of the KB dust disk emission by Vitense et al. (2012) indicate that, with a fractional luminosity of $f \sim10^{-7}$ peaking at 40--50 $\mu$m, the dust emission of a KB dust disk analog would be less than 1\% the stellar photosphere, still below the {\it Herschel}/PACS detection limits. This means that the detection of KB dust disk analogs still await for more sensitive far-infrared space instrumentation (e.g. {\it SPICA}/SAFARI). 

\subsection{Debris disk evolution}
\label{ddevolution}

\begin{figure}[]
\begin{center}
\includegraphics[width=0.95\textwidth]{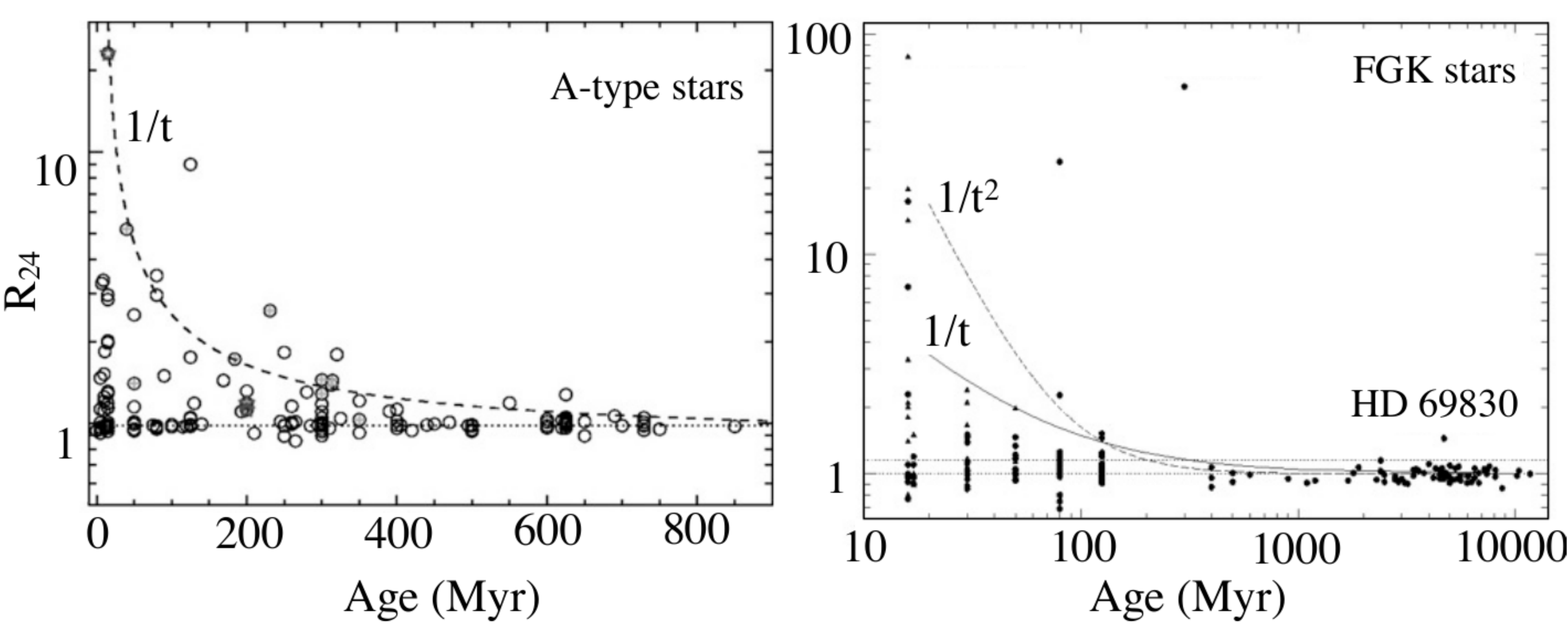}
\caption{Dust emission divided by the expected stellar emission at 24 $\mu$m as a function of stellar age, for A-type stars ({\it left}) and FGK stars ({\it right}),  from Su et al. (2006) and Siegler et al. (2007), respectively. The main features are a 1/t overall decay and a large variability for a given stellar age. 
}
\label{diskevolution}    
\end{center}
\end{figure}

The study of the frequency and properties of debris disks around stars of different ages can shed light on the evolution of debris disks with time. The main challenge in this case is that the ages of main sequence stars, in particular those that are not in clusters, are difficult to determine. As mentioned in Section \ref{earlysolarsystem} and the left panel of Figure \ref{24evol} shows, collisional models predict that the steady erosion of planetesimals will naturally lead to a decrease in the dust production rate; this slow decay will be punctuated by short-term episodes of increased activity  triggered by large collisional events that can make the disk look an order of magnitude brighter (see the more detailed discussion in Section \ref{sec:coll_effect_on_dust_evolution}). These models agree broadly with the observations derived from the {\it Spitzer} surveys, as the ones shown in Figurel \ref{diskevolution} for A-type and FGK (solar-type) stars.   It is found that the frequency and fractional luminosities  ($f=L_{dust}/L_*$) of debris disks around FGK stars  with ages in the range 0.01--1 Gyr declines in a timescale of 100--400 Myr, but there is no clear evidence of a decline in the 1--10 Gyr  age range (Trilling et al. 2008).  This indicates that different physical processes might be dominating the evolution of the dust around the younger and the older systems. A possible scenario is that, at young ages, stochastic dust production  due to individual collisions is more prominent, while at older ages, the steady grinding down of planetesimals dominates.  The  relative importance of these two processes is still under discussion. 

The {\it Spizer} surveys also showed that the evolution of dust around both A-type and FGK stars proceeds differently in the inner and outer regions, with the warmer dust (dominating the emission at 24 $\mu$m) declines faster than the colder dust (seen at 70 $\mu$m). This indicates that the clearing of the disk in the inner regions is more efficient, as would be expected from the shorter dynamical timescales. 

\begin{figure}[]
\begin{center}
\includegraphics[width=0.95\textwidth]{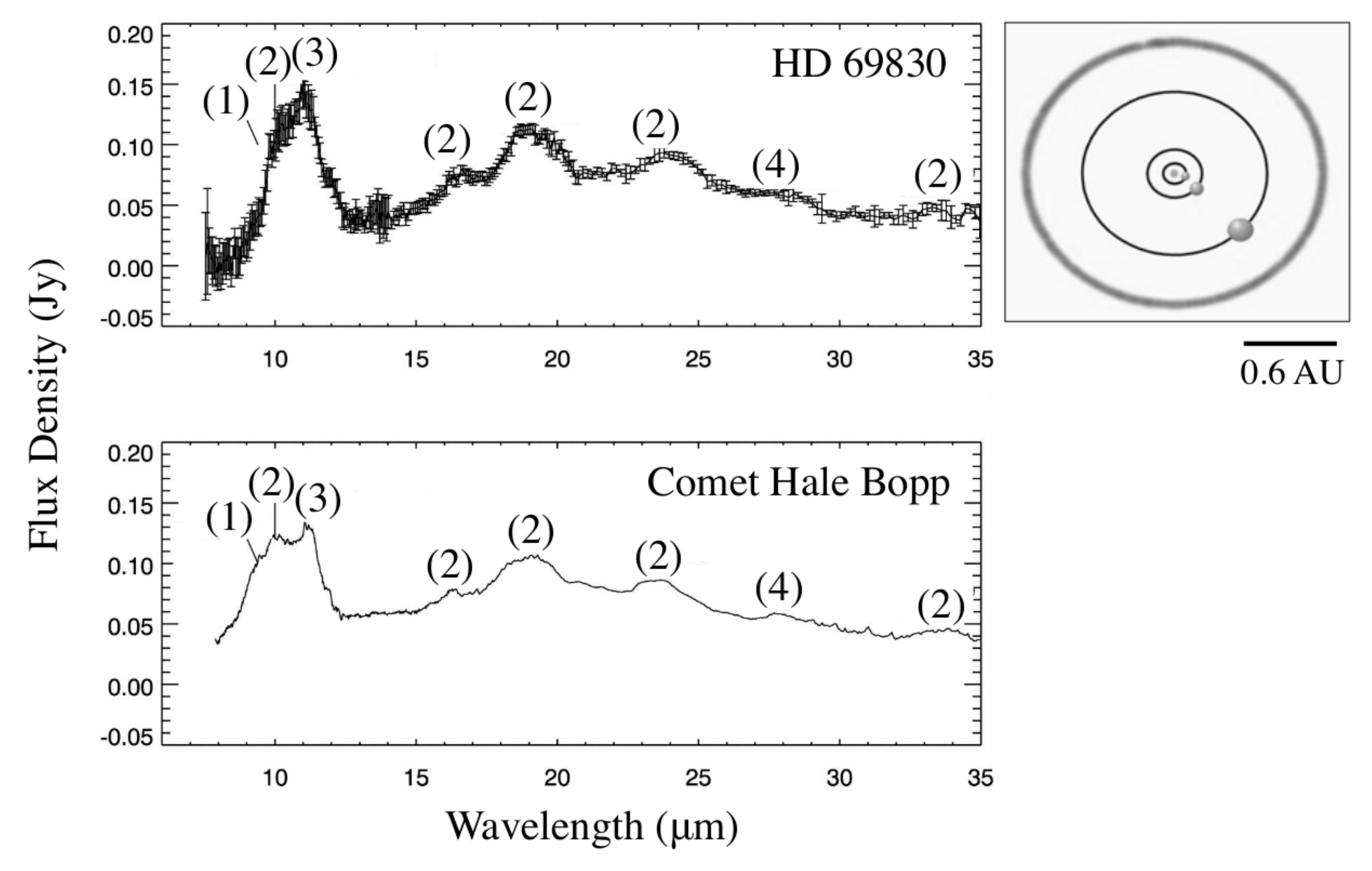}
\caption{({\it Top left}): Spectrum of the dust emission around HD 69830. ({\it Bottom left}): Spectrum of comet Hale-Bopp normalized to a blackbody temperature of 400 K. The spectral features labeled are: (1) amorphous olivine, (2) crystalline olivine, (3) crystalline olivine (forsterite) and (4) crystalline pyroxene (from Beichman et al. 2005). ({\it Right}): Further analysis showed that the best fit is to a highly processed low carbon P- or D-type asteroid belt located at $\sim$1 AU, outside the orbit of the three Neptune-like planets in the system (Lisse et al. 2007);  the planets are located at 0.0785 AU  ($\geqslant$ 10.2 M$_{\oplus}$),  0.186 AU  ($\geqslant$ 11.8 M$_{\oplus}$) and 0.63 AU ($\geqslant$ 18.1 M$_{\oplus}$).  
}
\label{hd69830}    
\end{center}
\end{figure}

Regarding the issue of steady state vs. stochastic dust production, some systems show evidence that transient events dominate the dust production. This is the case of HD 69830, a star that shows no excess emission at 70 $\mu$m, but shows strong excess at 24 $\mu$m (HD 69830 is one of the outliers in the right panel of Figure \ref{diskevolution}), with prominent spectral features in the {\it Spizter/IRS} wavelength range (see Figure \ref{hd69830} -- Beichman et al. 2005). The spectral features are  indicative of the presence of large quantities of small warm grains.  Because these small grains have very short lifetimes (see discussion in Sections \ref{sec:coll_lifetime}), it is inferred that the level of dust production is too high to be sustained for the age of the system (because the planetesimals would have not survived the inferred erosion rate). These led to Wyatt et al. (2007) conclude that the high rate of dust production in HD 69830, as well as in a few other systems, is transient.  HD 69830 is particularly interesting because it harbors three Neptune-like planets inside 1 AU (and the dust is inferred to be located near the 2:1 and 5:2 MMRs of the outermost planet), so there is the possibility that this transient event is triggered by gravitational perturbations from the planets.  The planet-debris disk connection will be discussed in Section \ref{correlation}. 

The duration of the dust production events (expected to be short if stochastic collisions dominate, and long otherwise) is critical to estimate what percentage of stars show evidence of dust production throughout their lives. And this is an important question to address because terrestrial planet formation is expected to result in the production of large quantities of dust in these regions (observable at 24 $\mu$m -- see the left panel  of Figure \ref{24evol}), so the percentage of stars showing excess emission at these wavelengths can shed light on the frequency of planet formation.  The right panel of Figure \ref{24evol} shows the evolution of the 24 $\mu$m emission as a function of stellar age. If the dust-producing events are very long-lived, the stars that show dust excesses in one age bin will also show dust excesses at later times, and this may result in that $<$ 20\% of the FGK stars in this survey show evidence of planetesimal formation near the terrestrial planet region.  On the other hand, if the dust-producing events are shorter than the age bins in the figure, the stars showing excesses in one age bin are not the same as the stars showing excesses at other age bins, and this might result in that $>$ 60\%  of these stars show evidence of planetesimal formation (assuming that each star only has one epoch of high dust production). An additional caveat is that most of these observations are spatially unresolved, and therefore it is not evident where the 24 $\mu$m emission is coming from in the disk; there is the possibility that the steady erosion of planetesimals in the KB-like region could be contributing to the 24 $\mu$m excess emission, in which case the interpretation in terms of the percentage of stars showing evidence of planetesimal formation near the terrestrial planet region would change.  Surveys of spatially resolved disks will help clarify this issue. 

As it was discussed in Section \ref{earlysolarsystem}, there is evidence that the migration of the giant planets in the early Solar system had an important effect on the evolution of its debris disk:  the drastic planetesimal clearing that took place at the time of the LHB, thought to be triggered by the migration of the planets, would have been associated with a sharp decreased in the dust production rate in both the AB and the KB (see solid line in Figure \ref{booth}).  Because  the extrapolation of radial velocity studies indicate that $\sim$17--19\% of solar-type stars harbor giant planets within 20 AU (Marcy et al. 2005), and the presence of hot Jupiters and planets locked in resonances are evidence that planet migration has taken place in some of these planetary systems, a natural question to ask is whether  these drastic planetesimal clearing events are common in other system. The interest is that frequency and the timing of these LHB-type of events can have important implications for the habitability of these systems.  Figure \ref{booth} shows together the expected evolution of dust in the Solar system at two different wavelengths, and the {\it Spitzer} debris disks observations.  A statistical study by Booth et al. (2009) concluded that less than 12\% of solar-type star suffered drastic planetesimal clearing events similar to that during the LHB. This is a preliminary result because, as  explained above, the {\it Spitzer} surveys were limited in sensitivity, so this issue needs to be revisited in the future using deeper surveys. In any case, this result might not be surprising because there is no evidence of a positive correlation between the presence of debris disks and the presence of giant planets (Greaves et al. 2004; Moro-Mart\'in et al. 2007a; Bryden et al. 2009; K{\'o}sp{\'a}l et al. 2009), so there is no reason to expect that the debris disks observed so far should show evidence of planetesimal depletion due to planet migration.  The planet--disk correlation will be discussed in Section \ref{correlation}. 

\subsection{Debris disk structure and inferred planetesimal location}
\label{sec:ddstructure}

Because most of the debris disks observed so far are located at distances $>$ 10 pc from the Sun, the limited spatial resolution of the instruments leaves them spatially unresolved. Before the launch of the {\it Herschel} infrared telescope (3.5 m in diameter, compared to 0.85 cm for {\it Spitzer}), only a few dozen of the closest and the largest debris disks (out of the hundreds known) were spatially resolved. The sizes of these disks range from 10s of AU up to 1000 AU, and imply that the dust-producing planetesimals are located on spatial scales similar to the KBOs in our Solar system  (extending out to $\sim$ 50 AU for the classical KB, and out to $\sim$ 1000 AU for the scattered KB). The location of the dust in the disks that are spatially unresolved can still be constrained or inferred. Because different wavelengths trace different dust temperatures and distances from the central star, the study of the debris disks SED can shed light on the radial distribution of the dust: for example, a disk with a central cavity where the warmer dust is absent would have an SED showing a depletion in the mid-IR wavelength region. In fact, the  SEDs of the unresolved debris disks show evidence of central cavities, with characteristic dust temperatures in the range of 50--150 K, corresponding to dust located in the 10s--100 AU range. This result may be biased because of the limited sensitivity and wavelength coverage of the {\it Spitzer} observations.  On-going debris disks surveys with {\it Herschel} are quickly increasing the number of resolved disks and/or providing better constraints for their outer radii with the help of more sensitive longer wavelength observations. 

\subsubsection{Inner gaps}
\label{innergaps}

\begin{figure}[]
\begin{center}
\includegraphics[width=0.6\textwidth]{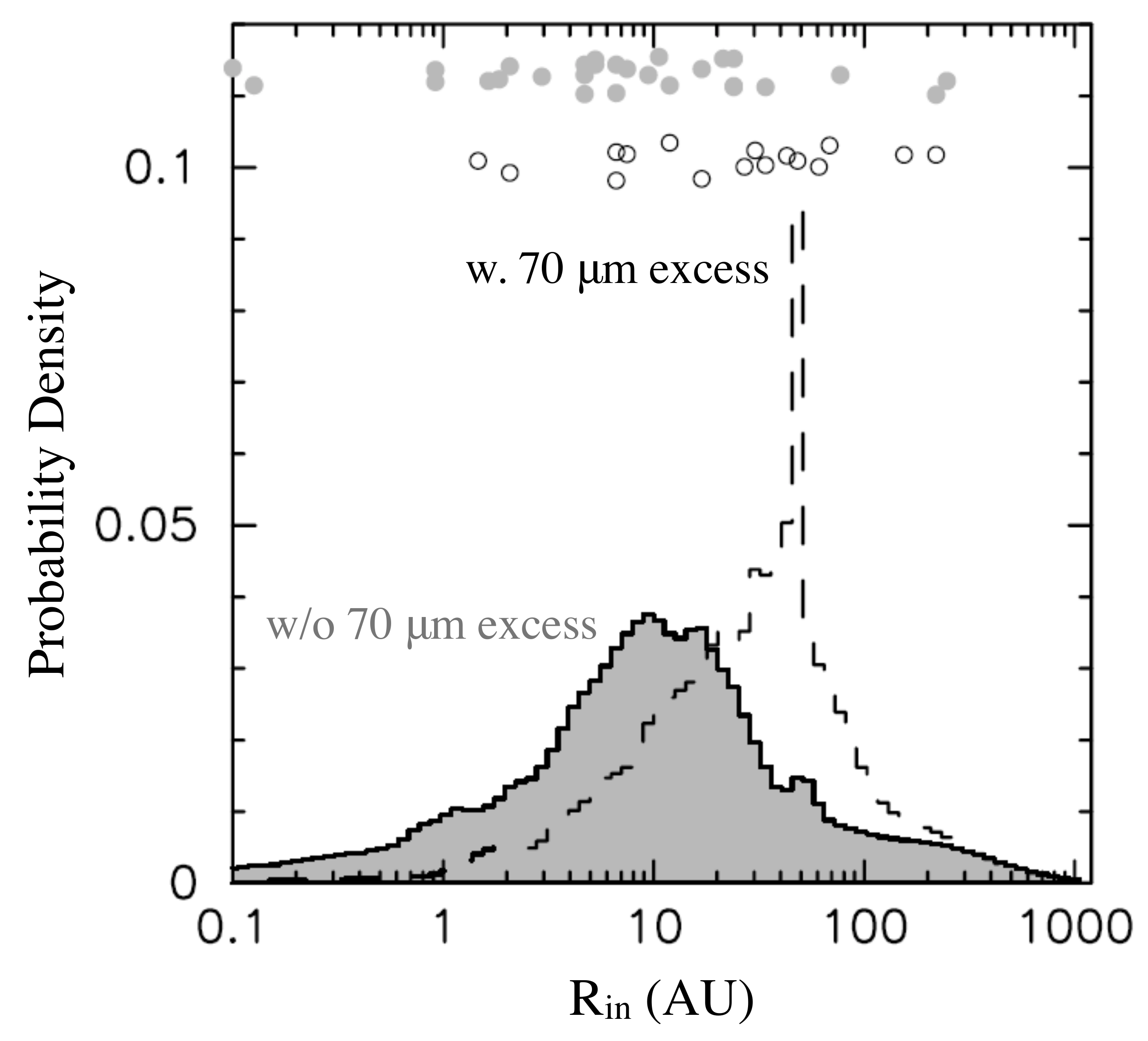}
\caption{Probability distribution for disk inner radii based on the analysis of the {\it Spitzer}/IRS spectra (12--35 $\mu$m) of 44 debris disks around FGK stars. The dashed and grey histograms correspond to sources with and without 70 $\mu$m excess, respectively (the best fit parameters are the open and grey circles on top). Typical disk inner radius are $\sim$40 AU and $\sim$10 AU for disks with and without 70 $\mu$m excess, respectively, indicating that most of the debris disks observed are KB-like. From Carpenter et al. (2009). 
}
\label{carpenter2}    
\end{center}
\end{figure}

From a {\it Spitzer} survey of 328 FGK stars at 24 and 70 $\mu$m, it was found that about 2/3 of the debris disks SEDs could be fitted with a single temperature blackbody ($T <$ 45--85 K) consistent with a ring-like configuration, while the rest would require either multiple rings or a continuous distribution of dust out to tens of AU (Hillenbrand et al. 2008).    Detailed analysis of the excess spectra from 12--35 $\mu$m of 44 of these stars showed that the characteristic dust temperature in these disks range from 60--180 K, that a cold component is needed to account for the 70 $\mu$m excess, and that inner disk cavities are common; the inner radii of these cavities are  $\sim$ 40 AU for the disks with 70 $\mu$m excess and $\sim$ 10 AU for the disks without 70 $\mu$m excess (see Figure \ref{carpenter2} from Carpenter et al. 2009). Because these dust disks are in a regime where the dynamics of the dust particles are mostly controlled by collisions (see discussion in \ref{sec:coll_effect_on_spatial_dist}), and therefore the dust traces the location of the planetesimals, these results indicate that most of the planetesimals inferred to exist around mature FGK (solar-type) stars are KB-like (in the sense that they have large inner cavities -- the inner radius of the KB is $\sim$ 35 AU). 

Inner cavities are also common around more massive stars: a {\it Spitzer} survey of 52 A-type and and late B-type stars known to have debris disks showed that the majority of the disks (39/52)  can be be fitted with a single-temperature blackbody with a median temperature of 190 K, corresponding to a characteristic distance of 10 AU, while the rest (13/52) are better fitted by extended disks without cavities (Morales et al. 2009). 

The presence of inner cavities has been confirmed by spatially resolved observations of nearby debris disks in both scattered light, as e.g. in HR 4796A, Fomalhaut and HD 139664, and in thermal millimeter and sub-millimeter emission, as e.g. in $\epsilon$-Eri, Vega and $\eta$ Corvi (with inner cavities of 50 AU, 80 AU and 100 AU, respectively). 

\subsubsection{Degeneracy of the SED analysis}
\label{seddegeneracy}

The analysis of the debris disks SEDs in terms of the dust location depends on assumptions on how efficiently the grains absorb and reemit the stellar radiation, because it is this balance that determines their equilibrium temperature. This in turn depends on the grain size and composition, which ideally can be constrained through the modeling of the solid state features in the debris disks spectra. The issue is that  most debris disks observed so far do not have any features in the 5--35 $\mu$m wavelength range covered by {\it Spitzer/IRS} (see in Section \ref{sec:composition}) where they show a smooth blackbody continuum (Chen et al. 2006, Beichman et al. 2006, Carpenter et al. 2009, Morales et al. 2009). It is generally assumed that this is because the grains have sizes $\gtrsim$10~$\mu$m (a grain size commonly adopted in the SED analysis).  Regarding the grain composition, a common assumption is that  they are made of ``astronomical silicates'' (i.e. silicates with  optical constants from Weingartner and Draine 2001). However,  the laboratory analysis of dust particles from the Solar system (IDPs and {\it Stardust} returned samples -- Section \ref{sec:inner}), and the analysis of debris disks with spectral features, like HD 69830 (Figure \ref{hd69830} and Section \ref{sec:composition}), indicate that this assumption might be too simplistic. Another caveat is that the debris disks SEDs are generally not constrained at the long wavelength range and most of them have only upper limits beyond 70 $\mu$m. As a result, the cold dust remains undetected and the outer disk radii unconstrained. On-going observations with {\it Herschel}, with increased sensitivity at longer wavelenghts,  are now contributing to characterize the cold dust. In fact, {\it Herschel} has detected a "new class" of very cold ($\sim$20 K) and faint ($f \sim 10^{-6}$) debris disks that only show dust emission beyond 100 or 160 $\mu$m (Eiroa et al. 2011). 

\begin{figure}[]
\begin{center}
\includegraphics[width=1.0\textwidth]{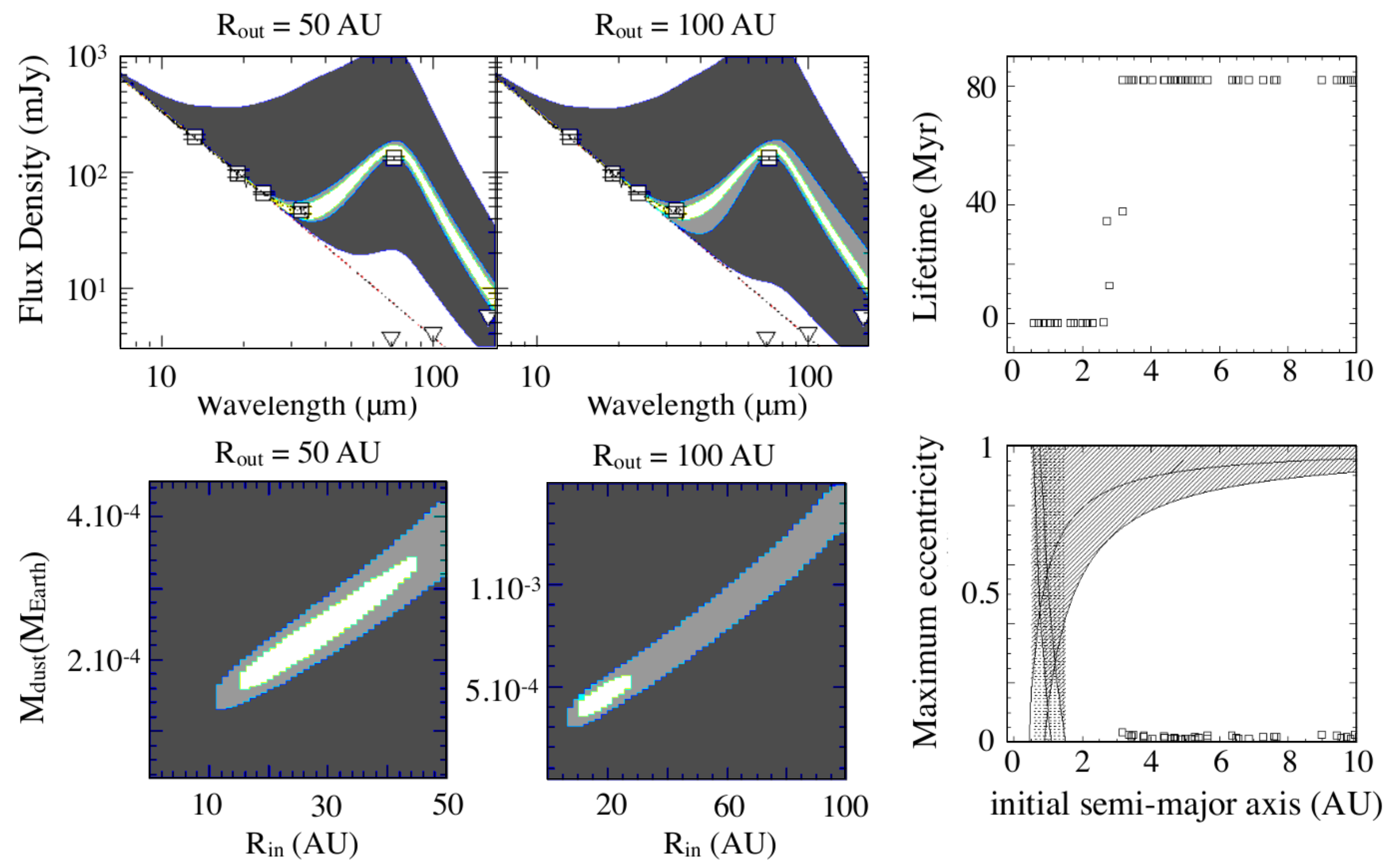}
\caption{({\it Top left and center}): Observed and modeled SEDs for HD~82943. The dotted line is the stellar photosphere. The {\it Spitzer} observations are represented by squares with 1-$\sigma$ error bars. The triangles are Herschel 5-$\sigma$/1hr sensitivity limits. The colored regions are formed by SED models of the star+disk emission, where the disk is composed of  10 $\mu$m size particles with optical properties typical of astronomical silicates, a total dust mass M$_{dust}$ and extends from R$_{in}$ to R$_{out}$ with a constant surface density. R$_{in}$ and M$_{dust}$ are the  free parameters; R$_{out}$ is kept fixed at  50 AU (left) and 100 AU (center).  The colors represent the goodness of the fit: for the white inner region, $\chi$$^{2}$ probability P($\chi$$^{2}$ $\mid$ $\nu$) $<$ 0.683;  $\it{light~grey}$ for P($\chi$$^{2}$ $\mid$ $\nu$) $>$ 0.683; and $\it{dark~grey}$ for P($\chi$$^{2}$ $\mid$ $\nu$) $>$ 0.9973, i.e. models that are excluded with 3-$\sigma$ certainty. ({\it Bottom left and center}): Parameter space of the modeled SEDs in the top panels showing the degeneracy of the SED analysis. In this case, the SED can rule out the presence of small grains, so these models assume a single grain radius of 10 $\mu$m. Adopting a disk outer radius of 50 AU, the best SED fits require the inner disk radius  to be 16 AU $\leq$ Rin $\leq$44 AU, while if adopting a disk outer radius of 100 AU, the best fit will be for 12 AU $\leq$ Rin $\leq$ 26 AU.  ({\it Right}): Results from the dynamical simulation (lasting 82 Myr) of 500 test particles in the HD 82943 planetary system, where the planets b and c have masses of 1.46 $M_{Jup}$ and 1.73 $M_{Jup}$,  semimajor axes of 0.75 AU and 1.19 AU, and eccentricities of 0.45 and 0.27, respectively. 
({\it Top right}): test particle's lifetimes. ({\it Bottom right}): allowed parameter space for the planetesimals' orbital elements, where the shaded areas indicate regions where test particle's orbits are unstable due to planet-crossing (striped area) or overlapping first order mean motion resonances (dotted area); the squared symbols show the maximum eccentricity attained by test particles on initially circular orbits.  The test particle orbits
are stable beyond $\sim$3 AU, with maximum eccentricities always $<$ 0.1. Long-lived, dust-producing planetesimals could therefore be located anywhere beyond 3 AU. From Moro-Mart\'in et al. (2010).}
\label{hd82943}    
\end{center}
\end{figure}

Figure \ref{hd82943} illustrates the degeneracy in the SED analysis of the debris disk around the planet-bearing star HD 82943. The top left and center panels show the observed and modeled SEDs, while the allowed parameter space is shown right below.  If the system is known to harbor planets (like in this case), an additional constraint on the dust location can be obtained from dynamical simulations that study the effect of the planetary perturbations on the stability of the planetesimals' orbits, and that can identify the regions where the planetesimals could be stable and long-lived (see right panels of Figure  \ref{hd82943}). Figure \ref{hd38529} shows a similar dynamical analysis used to constrain the dust location in the HD 38529 planetary system. In this case, it is found that the effect of the secular perturbations is very long-ranged (extending to 55 AU compared to the 3.74 AU semimajor axis  of the the outermost planet), and constrains the dust-producing planetesimals to the 20--50 AU region (resembling the KB). 

\begin{figure}[]
\begin{center}
\includegraphics[width=0.9\textwidth]{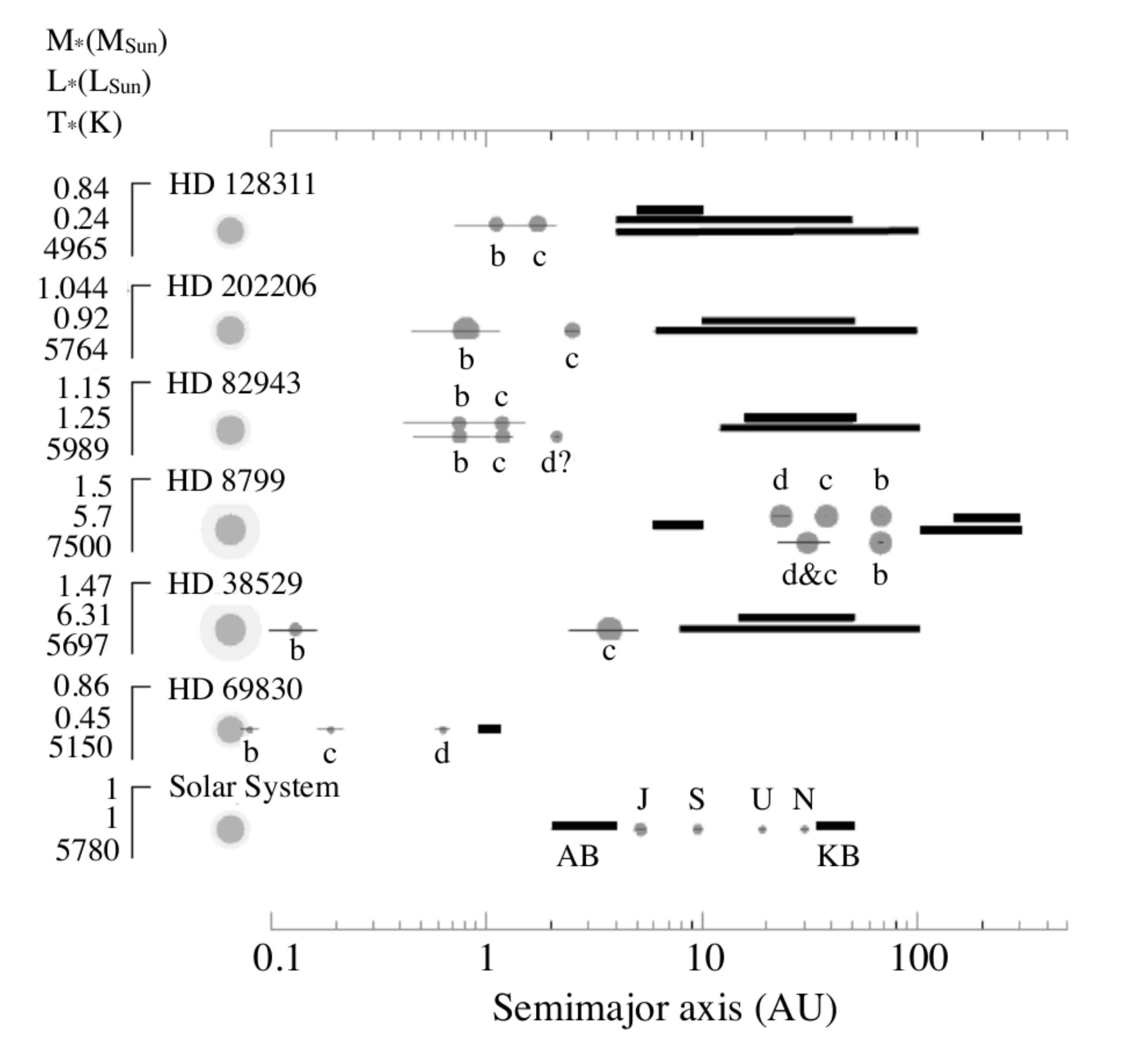}
\caption{Schematic representation of seven planetary systems known to harbor multiple planets and dust-producing planetesimals. The stellar mass, luminosity and effective temperature are labeled to the left. The sizes of the dark grey circles are proportional to the cube root of the stellar and planetary masses, while the sizes of the light  grey circles are proportional to the stellar luminosities. The thin lines extend from periastron to apoastron. For a given system, there is a range of planetary configurations that can fit the observations . The inferred location of the dust-producing planetesimals are represented by the thick black lines. Each line corresponds to a possible solution, showing the degeneracy of the problem and the need for spatially resolved observations. From Moro-Mart\'in et al. (2010).}
\label{bonita}    
\end{center}
\end{figure}

Figure \ref{bonita} illustrates how the study of debris disks can help us learn about the diversity of planetary systems. The figure shows the possible planet and planetesimal configurations of the systems known to date to harbor both multiple planets and debris dust (from Moro-Mart\'in et al. 2010). For most of the stars, the study is based on a combined SED and dynamical analysis (similar to that on Figure \ref{hd82943}).  The solutions are degenerate; to set tighter constraints on the planetesimal  location, there is the need to obtain spatially resolved images and/or accurate photometric points from the mid-infrared to the submillimeter, so the inner and outer radius of the disk can be better determined. Observations with {\it Herschel}, {\it JWST} and {\it ALMA} will be very valuable for this purpose


\subsubsection{Other structural features revealed by spatially resolved observations}
\label{otherstructuralfeatures}

The few dozen debris disks that have been spatially resolved so far show a rich diversity of structural features (Figure \ref{structure_planets}); Kalas et al. (2006) identified two basic architectures, either narrow belts about 20--30 AU wide and with well-defined outer boundaries (e.g. HR 4796A, Fomalhaut, and HD 139664), or wide belts with sensitivity limited edges implying widths $>$ 50 AU (HD 32297, $\beta$-Pic, AU Mic, HD 107146, and HD 53143). Structural 
features include clumpy rings (like in AU-Mic, $\beta$-Pic, $\epsilon$-Eri and Fomalhaut), sharp inner edges (Fomalhaut), brightness asymmetries ($\beta$-Pic, AU-Mic, HR 4796, HD 32297, Fomalhaut and Vega), offsets of the dust disk center with respect to the central star (Fomalhaut and $\epsilon$-Eri, with offsets of 15 AU and 6.6 AU--16.6 AU, respectively), warps of the disk plane (like $\beta$-Pic and AU-Mic) and spirals (HD 141569 --  Wyatt et al. 1999, Heap et al. 2000, Clampin et al. 2003, Holland et al. 2003, Stapelfeldt et al., 2004, Kalas et al. 2005, Greaves et al. 2005, Schneider et al. 2005, Krist et al. 2005). Some of these features have also been observed in the Solar system debris disk: the zodiacal cloud shows a warp in its plane of symmetry and an asymmetric ring near the Earth's orbit, and dynamical models of the KB dust disk predict an inner cavity around 10 AU, and an asymmetric ring outside the orbit of Neptune (see Figure \ref{KBdust} and discussion in Sections \ref{zodiacaldust} and \ref{sec:gravitational_forces}). 

\begin{figure}[]
\begin{center}
\includegraphics[width=1\textwidth]{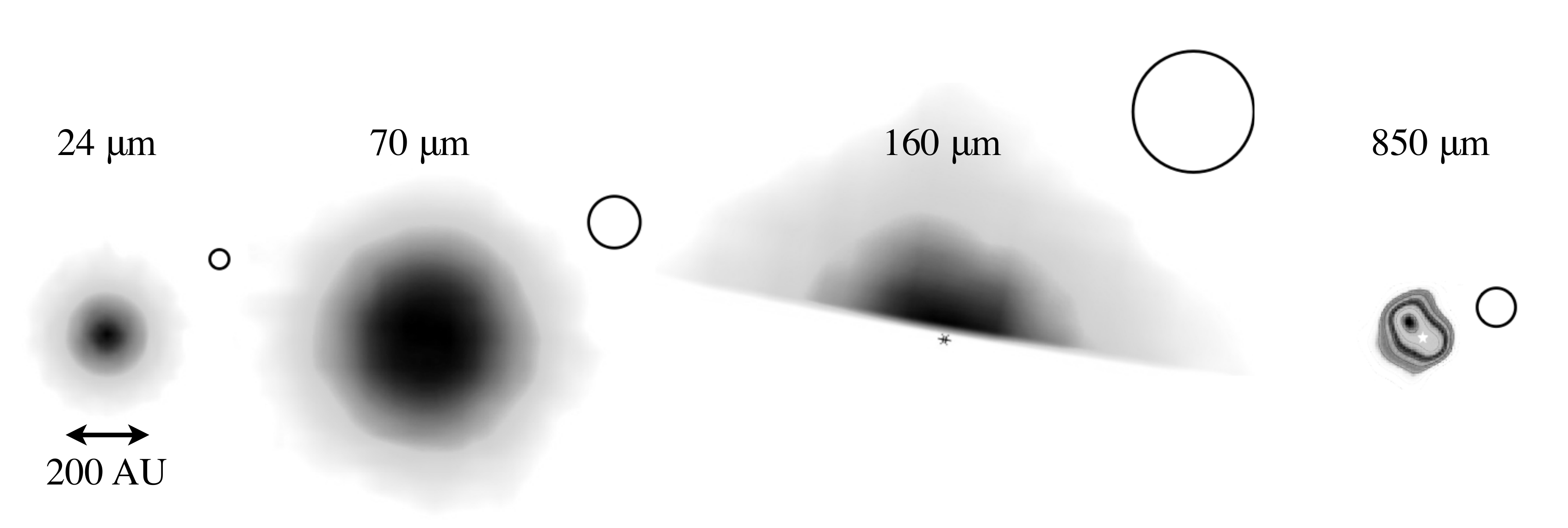}
\caption{Spatially resolved images of Vega from Spitzer/MIPS at 24, 70, 160 $\mu$m (Su et al. 2005) and from JCMT/SCUBA at 850 $\mu$m (Holland et al. 1998). All images are in the same scale. The instrument beam sizes (shown in white circles) indicate that the wide radial extent of the MIPS disk images compared to the SCUBA disk image is not a consequence of the instrumental PSF but due to a different spatial location of the particles traced by the two instruments. This indicates the need for spatially resolved observations at different wavelengths.  The sub-mm emission is thought to arise from large dust particles on bound orbits that originate from a planetesimal belt analogous to the KB, while the MIPS emission is though to correspond to smaller particles on unbound orbits produced by collisions in the planetesimal belt traced by the sub-mm observations; these particles are blown away by radiation pressure to distances much larger than the location of the parent bodies; this scenario would explain not only the wider extent of the MIPS disk but also its uniform distribution, in contrast with the clumply and more compact sub-mm disk (Su et al. 2005; cf. to M{\"u}ller, L{\"o}hne \& Krivov 2010, that interpret the observations with a collisional cascade in steady state). 
}
\label{vega}    
\end{center}
\end{figure}

These spatially resolved observations show that the disks look different at different wavelengths  because at a given wavelength the thermal emission is dominated by a particular grain  size, and grains of different sizes have different dynamical evolutions that result in different structural features. As it is discussed in more detail in {\it Part II}, large particles that dominate the emission at longer wavelengths show more structural features because they interact more weakly with the stellar radiation field and therefore their dynamical evolution is slow; in this case, the particles trace the location of their parent planetesimals, or if planets are present, they can also be subject to resonant trapping.  On the other hand,  small grains that dominate the emission at shorter wavelengths, interact more strongly with radiation, which results in a more uniform and extended disk. This can clearly be observed in the case of Vega (Figure \ref{vega}). Section \ref{seddegeneracy} discussed the need for spatially resolved observations to break the degeneracy in the debris disks SED analysis; the above discussion indicates that these resolved observations need to be taken at multi-wavelengths. 

\subsubsection{Debris disk structure can unveil the presence of planets}
\label{structurecreatedbyplanets}

Secular perturbations by Saturn are responsible for the creation of the inner edge of the asteroid belt around 2 AU, while other secular perturbations  are thought to account for the offset of the zodiacal cloud center with respect to the Sun, the inclination of the cloud with respect to the ecliptic and the cloud warp;  in addition, the trapping of dust particles in MMRs with the Earth is responsible for the asymmetric ring of dust along the Earth's orbit (Section \ref{zodiacaldust}). In light of these observations, a natural question arises: are the structural features observed in debris disks the result of gravitational perturbations with unseen planets? Clumpy rings have been explained as dust and/or dust-producing planetesimals trapped in MMRs with a planet; warps can be the result of secular perturbations of a planet in an orbit inclined with respect to the planetesimal/dust disk; and spirals, offsets and brightness assymmetires might also be the result of secular perturbations, in this case of an eccentric planet that forces an eccentricity on the planetesimals and the dust.  Even though the origin of individual features is still under discussion\footnote{It is possible that some of these disk features are created by mechanism other than planets. For example,  clumps could trace the location of a recent planetesimal collision, instead of the location of dust-producing planetesimals or dust particles trapped in MMRs with a planet. Azimuthal asymmetries and spirals could be created by binary companions or close stellar flybys, but the later involves fine tuning to create the perturbations without destroying the disk. Brightness asymmetries on the outermost edge of the disk could also be created by sandblasting from interstellar grains.  The interaction of the dust with remnant gas, stellar wind or magnetic fields could also be responsible for some of the structure. Even though it is possible that some of the disk features are created by mechanism other than planets, it seems unlikely that non-planet mechanism can account for all the debris disks structure observed.}, and the models require further refinements -- e.g. in the inclusion of dust collisional processes (Stark and Kuchner 2009) and the effects of gas drag -- the complexity of these features, in particular the azimuthal asymmetries, indicate that planets likely play a role in the creation of structure in the debris disks.  

The Fomalhaut case illustrates this idea: a massive planet was predicted to exist to account for both, the sharp inner edge observed in the dust disk (created by gravitational scattering), and the offset of the dust disk center (created by secular perturbations -- Wyatt et al. 1999, Kalas et al. 2005, Quillen et al. 2006). Later-on, follow-up observations were able to directly image a planet candidate around the predicted semi-major axis, where the planet eccentricity could be constrained from the offset of the dust disk (Kalas et al. 2008). Because the eccentricity of the dust particles at the edge of the chaotic region where the MMRs overlap depends on the planet mass, the radial distribution of the dust near the inner edge of the dust disk can set limits on the mass of the planet (Chiang et al. 2009). This is particularly interesting because in some systems a dynamical constraint for the planet mass can be compared to the mass estimate  based on the observed luminosity, with the advantage that the former is independent of the planet age (which in most cases is difficult to estimate) and the initial conditions of the planet evolutionary model.  In these systems, a dynamical constraint can be used to test and calibrate current evolutionary models of giant planets; this is important because, in most cases, dynamical constrains are not available and the planet masses need to be derived from evolutionary models alone. 

Another example of a planet successfully predicted to exist based on the debris disk structure is $\beta$-Pic b (Mouillet et al. 1997; Lagrange et al. 2010): with an estimated mass of $\sim$9 M$_{Jup}$, this planet located at 8--14 AU can account for some of the asymmetries observed in the debris disk, including its inner warp. This system is particularly relevant because the planet has a relative short orbit that will allow to constrain the mass with the radial velocity technique, enabling the much needed calibration of the planet evolutionary models at young ages ($\sim$ 10 Myr).

The connection between planets and inner cavities is under debate. The gravitational scattering of dust particles by planets is a very efficient process that can create a dust depleted region inside the orbit of the planet (see discussion in Sections \ref{sec:gravitational_scattering} and \ref{sec:grav_effect_on_dynamics} and Figures \ref{KBdust}, \ref{efficiency} and \ref{radialprofile}).  But some of the observed inner cavities might be the result of grain-grain collisions rather than gravitational scattering with a planet; however, this latter scenario assumes that the parent bodies have an inner edge to their spatial distribution which may require planets to be present to keep the planetesimals confined.  Core accretion models of planet formation predict the formation of inner cavities because the planets form  faster closer to the star, depleting planetesimals from the inner disk regions. 

Most of the structural features discussed in Section \ref{sec:gravitational_forces} depend on the mass and orbit of the planet. The case of the Solar system illustrates that the structure is sensitive to small planets (like the Earth) and to planets located far from the star (like Neptune). This opens the possibility of using the study of the dust disk structure as a  detection technique for planets of a wide range of masses and semi-major axes. This method  is complementary to radial velocity and transit surveys (limited to planets relatively close to the star), and to direct imaging (limited to young and massive planets). 


\subsection{Planet-debris disk relation}
\label{correlation}

In the core accretion models of planet formation, planetesimals are the building blocks of planets; in these models, giant planet formation requires the presence of a protoplanetary disk rich in planetesimals, which would favor a positive correlation between the presence of giant planets and debris disks. However, as it was discussed in Section \ref{earlysolarsystem}, the migration of giant planets can lead to drastic planetesimal clearing events, as the one expected to have occurred during the LHB in the early Solar system; this would favor an anti-correlation between giant planets and debris disks.

{\it Spitzer} debris disks surveys do not find any sign of positive or negative correlation between the presence of giant planets identified in radial velocity surveys and the presence of debris disks (Moro-Mart\'in et al. 2007a; Bryden et al. 2009; K{\'o}sp{\'a}l et al. 2009), i.e. there is no apparent difference between the incidence rate of debris disks around stars with and without known planetary companions. 

\begin{figure}[]
\begin{center}
\includegraphics[width=0.9\textwidth]{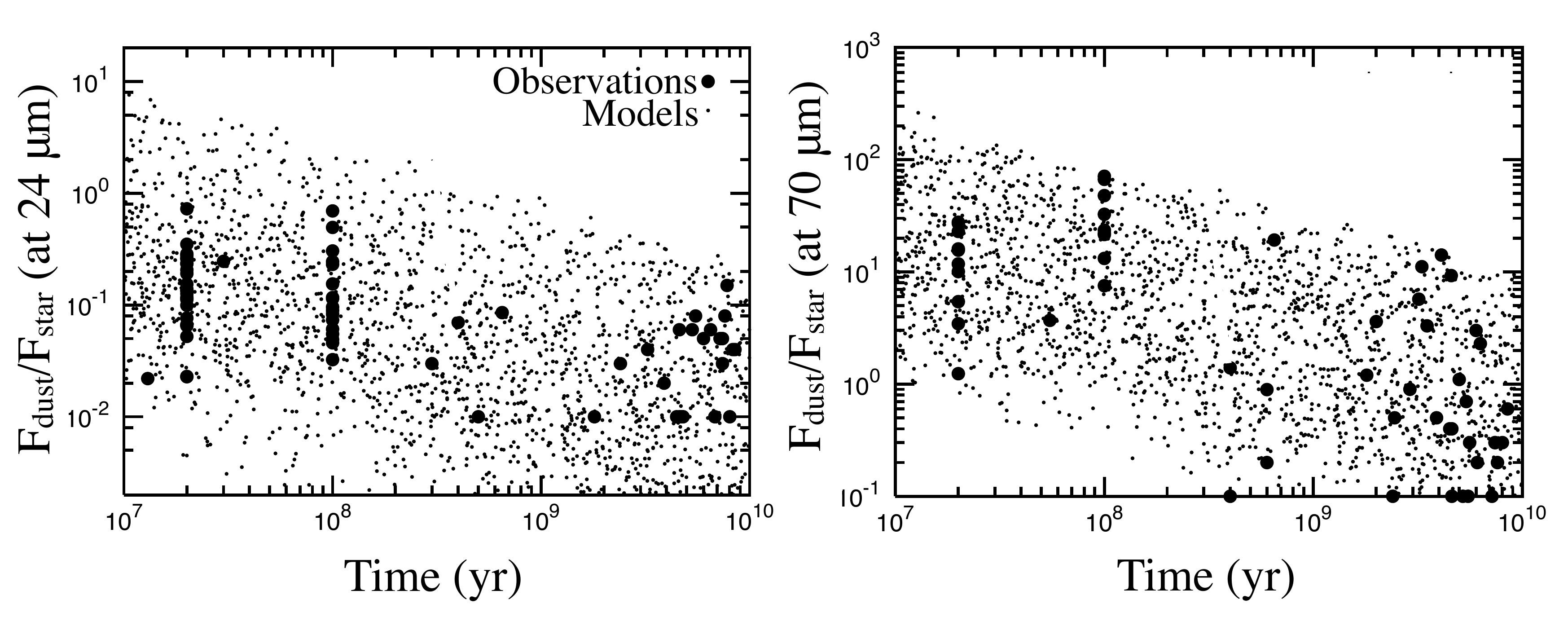}
\caption{Ratio of the excess dust flux to the stellar flux  at 24 $\mu$m ({\it left}) and at 70 $\mu$m ({\it right}) as a function of stellar age. The collisional cascade models (small dots) can reproduce the range of ratios observed by {\it Spitzer} (thick dots) by varying the disk location and disk masses.  The models predict an overall decay of the excess ratio with time. Figure from Lohne et al. (2008). 
} 
\label{lohne}
\end{center}
\end{figure}

A scenario that could account for this lack of correlation is the following.  {\it Spitzer} observations are sensitivity limited and can only detect  the brighter disks; based on the observed distribution of fractional luminosities, it seems likely that debris disks at the Solar system level are very common, i.e. that many stars harbor planetesimals (see discussion in Section \ref{frequency}). Debris disks are found around stars with a wide range of spectra-types, indicating that the planetesimal formation process is  very robust and can take place under a wide range of conditions. This is also in agreement with the observation that the presence of debris disks is not correlated with high stellar metallicities (Greaves et al. 2006). Giant planets, on the other hand, are strongly correlated with high stellar metallicities (because their formation may require the presence of a large surface density of solids in the disk, so that the planet can grow a core sufficiently large to accrete an atmosphere before the  gas disk disappears -- Fischer \& Valenti 2005). All this indicates that the conditions required to form planetesimals are more easily met than those to form giant planets, planetesimals are  more common, and massive planets may not be required to produce the debris dust. In a possible scenario for the production of debris at Gyr ages, even in a disk that is too low in solids to form a giant planet, a large 1000 km size planetesimals can stir up smaller planetesimals (0.1--10 km in size) along  their orbits, starting a collisional cascade that can produce dust excess emission over the relevant range of ages. Results from numerical models exploring this scenario are shown in Figures \ref{lohne}. 

The planet-debris disk relation will be revisited with the {\it Herschel} DEBRIS and DUNES surveys,  sensitive to lower-mass debris disks and to disks around colder stars. 
 
The increasing population of super-Earths discovered by the radial velocity surveys will soon allow to test whether there is a correlation between the presence of low-mass planets ($<$ 10 M$_{Earth}$) and the presence of debris disks. Contrarily to high-mass planets, recent results from the radial velocity surveys indicate that  there is no correlation between high metalicities and low-mass planets (Mayor et al. 2011). This might indicate that the conditions to form low-mass planets are more easily met than those to form high-mass planets and, therefore, there could be a correlation between low-mass planets and debris disks. At the time of the {\it Spitzer} planet-debris disk correlation studies, little was known about the frequency of low-mass planets and only a handful of these objects were known. Now that many more low-mass planet systems have been identified, and given that the {\it Herschel} surveys include low-mass planet host stars, it has became possible to explore whether a low-mass planet-debris correlation exists, as hinted by the metallicity studies. 

\begin{figure}[]
\begin{center}
\includegraphics[width=0.5\textwidth]{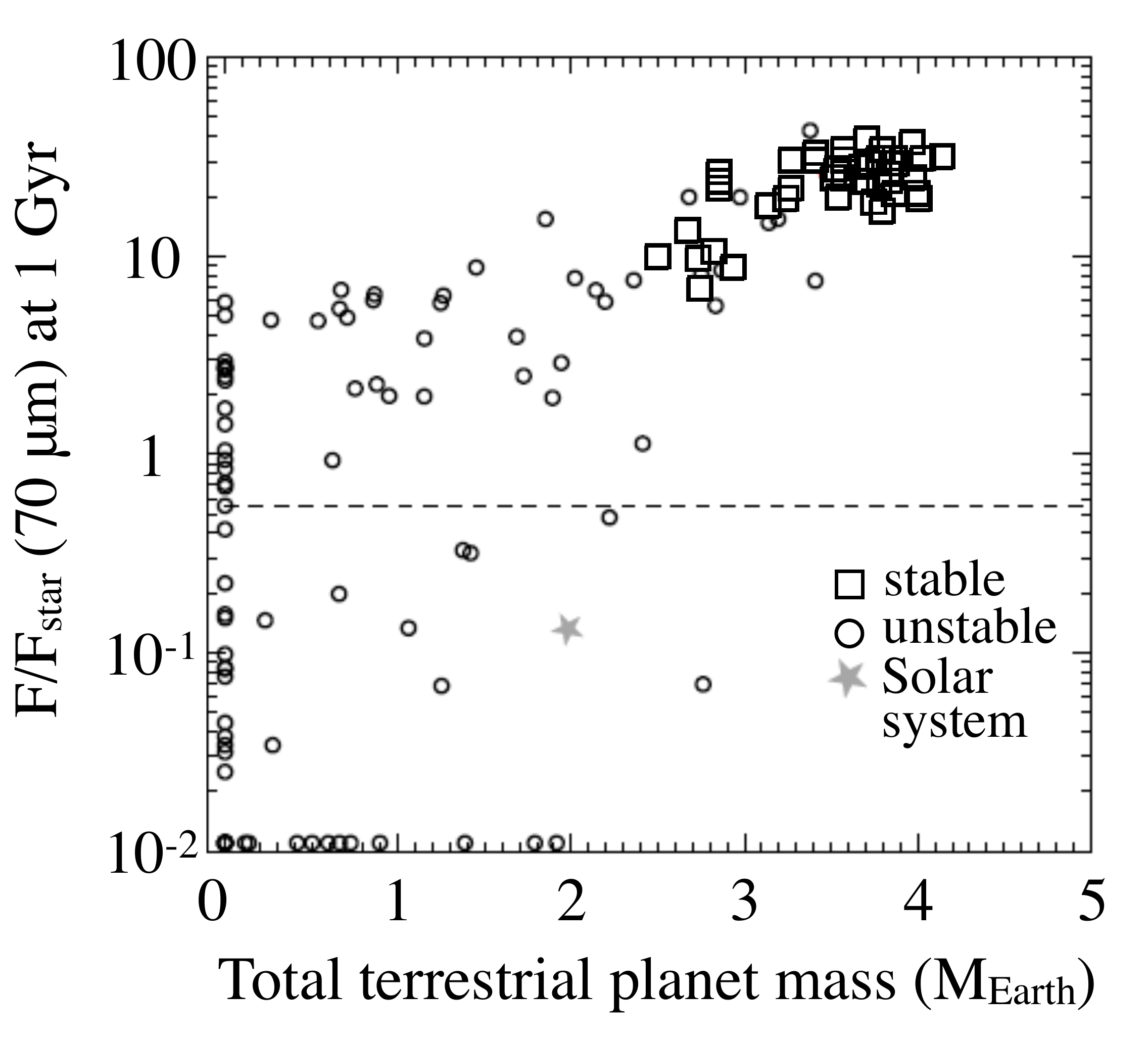}
\caption{The dust-to-stellar flux ratio at 70 $\mu$m versus the total mass in terrestrial planets from 400 dynamical simulations that include three giant planets, two belts of planetesimals  (inner and outer), and a population of Mars-sized embryos in the inner belt (that can grow into terrestrial planets). The circles correspond to unstable planetary systems, the squares to stable systems, and the grey star is an estimate of the Solar system's debris disk flux approximately 900 Myr after the LHB. The dashed line is the {\it Spitzer} detection limit.  The figure shows the correlation between bright cold dust and the efficiency of terrestrial planet formation and survival: debris disks trace terrestrial planets because dusty systems mean a calm dynamical evolution where terrestrial planets are able to grow and survive. From Raymond et al. (2011).
}
\label{sean}    
\end{center}
\end{figure}

Of particular interest is whether the presence of terrestrial planets might be correlated with debris dust. It is well known that the future detection and characterization of terrestrial planets in extra-solar planetary systems might be compromised by the emission from their inner debris disks (the equivalent to the zodiacal cloud). This  means that stars with evidence of warm dust are  not good candidates for terrestrial planet searches. On the other hand, a study by Raymond et al. (2011, 2012) indicates that stars with evidence of cold dust might turn out to be good targets. This study is based on numerical simulations of the dynamical evolution of about 400 different planetary systems, each consisting on three giant planets, two belts of planetesimals  (inner and outer), and a population of Mars-sized embryos in the inner belt (that can grow into terrestrial planets). The simulations are able to reproduce the observed distribution of giant planet eccentricities.  The ensemble of models by Raymond et al. (2011, 2012) show that: (1) 40--70\% of the systems that become unstable destroy their terrestrial planets; in these cases, the giant planet instability is too strong and the embryos are thrown into the star or ejected from the system. (2) The terrestrial planet outcome correlates with the eccentricity of the surviving giant planets; higher eccentricities imply that the instability was more violent and therefore the terrestrial planets were likely destroyed, while lower eccentricities indicate that more terrestrial planets survive. And (3) there is a strong correlation between the presence of bright cold dust and the occurrence of terrestrial planets (Figure \ref{sean}; Raymond et al 2011, 2012); cold debris disks trace terrestrial planets because dusty systems mean a calm dynamical evolution where terrestrial planets are able to grow and survive\footnote{In this context, the Solar system is an outlier because it harbors four terrestrial planets but very little dust. The explanation might be that the instability in the Solar system was sufficiently strong to clear out the planetesimals, but sufficiently weak not to affect the stability of the terrestrial planets.}.  

Finally, the recent release of the all-sky near to mid-infrared survey carried out by the {\it WISE} space telescope, together with ground-based observations, may soon allow to assess whether there is a correlation between the presence of planets and the presence of warm dust (that may hint dynamical activity in the terrestrial-planet region). 

\subsection{Debris disk composition}
\label{sec:composition}

{\it Spitzer/IRS} carried out spectral surveys of debris disks around star from A to K-type in the 5--35 $\mu$m wavelength range. Even though silicate emission features should be prominent around 10--20 $\mu$m, the spectra of most  debris disks do not show any  features, so little is known about the composition of the dust. In total, about two dozen disks show spectral features in this wavelength range; they are generally young stars ($<$ 50 Myr), and the majority of them show evidence of warm crystalline silicates and multiple planetesimal belts (with a warm component responsible for the spectral features, and a cold component that accounts for the emission at longer wavelengths -- Chen et al. 2006). It is found that for both FGK and A-type stars, there is a diversity of compositions and degree of crystallinity (related to the processing history) for stars of similar spectral types, even those of similar ages.  

The lack of spectral features in the majority of the disks indicates that the dust grains are either large and/or cold (T $<$ 110 K).  Grains with sizes $s > \lambda$/2$\pi $ tend to absorb and emit energy efficiently with a constant emissivity (i.e. with no spectral features), so the lack of features at $\lambda$ $<$ 35 $\mu$m implies grain sizes $s > $ 5.5 $\mu$m. For debris disks around A-type stars, radiation pressure can account for the lack of grain smaller than this size because the median blow-out size for the stars in the A star sample is $\sim$ 4.7 $\mu$m (see discussion on the blow-out size in Section \ref{sec:radiation_stellar_wind_effect_on_dynamics}). For solar-type stars, the blow out size is $\sim$ 1 $\mu$m, so  small grains are expected to be present and should produce spectral features if they are  warm. In fact, the two systems found with excess emission at $\lambda$ $<$ 25 $\mu$m, i.e. with warm dust, show spectral features at 7--20 $\mu$m.

One of these two systems is HD 69830 (Figure \ref{hd69830}). This mature K0 star, several Gyr old, harbors three close-in Neptune-like planets, and shows no excess emission at 70 $\mu$m (i.e. no evidence of cold dust),  a strong excess at 24 $\mu$m (corresponding to 1000 $\times$ the emission of the zodiacal cloud) and its spectra is dominated by strong silicate features that can be fitted with 80\% Mg-rich olivines (25\% of which is amorphous) and 20\% crystalline pyroxenes (Beichman et al. 2005).  At first sight, the spectrum is very similar to that of comet  Hale Bopp, but further analysis reveals it differs from cometary spectra (because there is no evidence for water gas, amorphous carbon, amorphous pyroxene, PAHs,  pyllosilicates and metallic sulfides); the best fit is to the dust produced by a highly processed low carbon P- or D-type asteroid (Lisse et al. 2007). These asteroids are the most common in the outer asteroid belt, and the break up of these type of objects gave rise to the Veritas and Karin asteroidal families in events that should have been important sources of dust (see discussion in Section \ref{zodiacaldust}).  The dust production in HD 69830 is also thought to be transient because the rate that would be needed too account for the observed amount of dust  is too high to be sustain for the age of the star (Wyatt et al. 2007 -- see discussion in Section \ref{ddevolution}). It is likely that the strong silicate features observed are associated with collision/disruption events, and in the case of HD 69830, the  spectra reveals that the composition of the parent body was similar to that of an asteroid. It is inferred that the dust around this planet-bearing star is located at 0.93--1 .16 AU, near the  2:1 and 5:2 MMRs of the outermost Neptune-like planet, raising the question whether the increased level of dust production is the  result of gravitational perturbations by the planets. 

The "colors" of the debris disks (obtained from scattered-light images taken at different wavelengths) can also provide some information about their composition. Debris disks tend to be red or neutral, with their redness commonly  explained by the presence of small ($\sim$ 0.4~$\mu$m) silicate grains. However, spatially resolved spectra have shown that debris disks do not generally contain large amounts of small silicate grains, in which case it must be the composition of the grains that makes them look intrinsically red. The red color of the debris disks around HR 4796A, an A0V star 8 Myr old, has been identified as a signature of tholins, the complex organic materials found in the surface of icy bodies and in the atmosphere of Titan; however, other fits to the data consist in porous grains made of amorphous silicates, amorphous carbon and water ice (Debes et al. 2008). Higher resolution spectroscopy (spatially resolved) is required to further constrain the models.  

The study of the atmospheric composition of some white dwarfs provide a unique opportunity to probe the elemental composition of planetesimals in these systems. About one fifth of white dwarfs expected to have pure hydrogen or pure helium atmospheres (due to quick sedimentation of the heavy elements under the effect of the strong gravitational field) show evidence for heavier elements, likely due to pollution from external sources. In the case of white dwarfs that show atmospheric pollutants and near infrared excesses, where the latter is inferred to be produced by dust located inside the tidal radius of the star, it is likely that the source of the pollutants is this circumstellar dust.  This is favored over an interstellar origin based on several observations: the spectra of some of these infrared excesses show a strong 10 $\mu$m silicate feature with a shape that resembles that observed in the zodiacal cloud and differs from the characteristic interstellar dust emission; and an observed depletion of carbon relative to iron in the atmospheric composition of the white dwarf also suggests that the infalling material is asteroid-like rather than interstellar  (Jura 2006). Detailed observations of the atmospheric composition of one of the white dwarfs (GD 362) revels that the relative enhancement of refractory elements and the relative depletion of volatiles is similar to that found in the Earth (Jura et al. 2007). All this favors a scenario in which tidally disrupted\footnote{During the stellar evolution leading to the white dwarf stage, there is a significant mass loss from the star. As a result,  the semimajor axes of orbiting bodies increase and this may lead to dynamical instabilities caused by resonance sweeping and other effects that may send asteroid like bodies into the tidal radius of the star.} asteroid-like bodies are responsible for the near infrared excess emission and the atmospheric pollution.   The advantage of using this method to assess the composition of these asteroid-like bodies is that, since the material has already been broken into its elemental composition, the results  do not depend on unknown dust grain properties. 

\section{Future prospects in debris disks studies}
\label{sec:future}

The discovery of debris disks in 1984, a decade before the detection of extra-solar planets around main sequence stars, provided the first evidence that a critical step in the process of planet formation (the formation of planetesimals) is taking place around other stars. Since then, our knowledge of debris disks has greatly improved and this chapter has described how it has shed light on the formation, evolution and diversity of planetary systems. Debris disks observations with {\it Herschel} (on-going), with upcoming observatories like {\it ALMA} and {\it JWST} (under development), and with future missions like {\it SPICA} (proposed), together with new developments in planet detection techniques, warrant that the field of debris disks studies will keep developing rapidly, enabled not only by the improved sensitivity and spatial and spectral resolution of the observations, but also by the interest of the astronomical community. The latter is reflected in the Astro2010 "Decadal Survey"\footnote{New Worlds, New Horizons in Astronomy and Astrophysics (2010--2020).} by the National Academies (2010--2010) and the "Cosmic Visions" by ESA (2015--2025), where questions intimately related to debris disks have been identified as priorities for the next decade, namely: How do circumstellar disks from and evolve into planetary systems? What is the diversity of planetary systems? How does it depends on stellar properties? How does the Solar system fit in the context of other planetary systems? Is the Solar system unique in its formation, characteristics and/or evolution? What is the composition of primitive planetesimals in the Solar system? Is there a radial gradient? What is the frequency of stars with terrestrial planets? Which stars are the best candidates for planet detection? Do they harbor debris disks bright enough to impede planet detection and characterization? To advance in answering some of these question, the Astro2010 "Decadal Suvey" made the following specific recommendations related to debris disks: 
(1) {\it To carry out debris disks surveys around stars of different ages and spectral types}; the goals are the characterization of the disk properties as a function of the stellar properties, the determination of the necessary conditions for the formation of planetesimals, and the study of the temporal evolution of the systems. 
(2) {\it Study of debris disk structure with high spatial resolution observations} in scattered light (optical--near infrared), in thermal emission (mid infrarred--submm), and at different epochs (to allow the detection of proper motions and to exclude features that might be background galaxies); the goals are to identify morphological features that can reveal the presence of planets (allowing to constrain their mass, eccentricity and period), and can shed light on the dynamical evolution of the system. 
(3) {\it Development of theoretical debris disks models} (including planet-disk interactions and collisions), with the goal of interpreting high spatial resolution observations. 
(4) {\it Direct detection of planets in protoplanetary and debris disks}, with the goal of studying the planet-disk interaction. 
(5) {\it To carry out KBOs surveys} to constrain their size distribution, physical properties (which depend on the size of the body) and their collisional state, with the goal of shedding light on the formation and dynamical evolution of the Solar system. 

Examples of these studies are the on-going {\it Hershel} surveys DUNES and DEBRIS designed to search for debris disks around AFGKM stars at 70, 100 and 160 $\mu$m, with follow-up at 350, 450 and 500 $\mu$m; these observations are already allowing to characterize a new population of cold disks (Eiroa et al. 2001) and to increase the number of spatially resolved observations (Booth et al. in preparation). {\it ALMA}'s unprecedented high spatial resolution will be able to advance in the study of debris disk structure, and to test the models of planet-disk interactions; its long wavelength observations will allow to better constrain the disks outer radii. Debris disks surveys with {\it JWST} in the near to mid-infrared will allow to characterize the warm dust component, setting constrains on the frequency of planetesimal formation in the terrestrial planet region, and identifying stars with low debris dust contamination that may be good targets for terrestrial planet detection. Deep debris disks surveys with {\it JWST} and {\it SPICA} (the approval of the latter is pending) will be able to study the debris disk evolution and the dust production rate as a function of stellar age; this will help identify systems undergoing LHB-type of events, and to assess whether the dynamical evolution of the Solar system was particularly benign (in the sense that it did not affect the orbital stability of the terrestrial planets and it happened early during the Solar system evolution). The SAFARI instrument planned for {\it SPICA} (a proposed telescope similar to {\it Herschel} but cool down to 5 K allowing a greatly improved sensitivity), covering the wavelength range of 30--210 $\mu$m, has debris disks at one of its focus. In fact, three out of its eight proposed core programs  are related to debris disks studies, namely: (1) the study of the occurrence and mass of debris disks (carrying out an unbiased survey of all stellar types out to a few hundred pc); (2) the study of the dust mineralogy in debris disks (doing an spectral survey of debris disks and, for nearby disks, spectral imaging); (3) the study of the composition of the KB (with an unbiased survey of KBOs). Of special interest is that, for the nearby debris disks, it will be possible to trace the variation in the dust mineral content as a function of disk radius that can be compared to the compositional gradient in the Solar system. Regarding the Solar system, advancements need to be made in the study of the dust-producing planetesimals, e.g. with KBO surveys with {\it Pan-STARRS}, {\it LSST} and {\it Subaru-HSC}, and in the study of the Solar system's dust (which properties in the outer Solar system are greatly unknown). For the latter, there are several space missions proposed, including sample return missions to an asteroid and a comet, dust detection experiments, and the study of the dust scattered light using a small telescope on board a spacecraft traveling into the outer Solar system.  Finally there are programs to detect planets in stars harboring debris disks using ground based telescopes (e.g. {\it Subaru/HiCIAO}, {\it Gemini/GPI} and {\it VLT/SPHERE}) that  will allow to study the planet-disk interaction.  

These are some of the future research lines in debris disk studies that will help us understand our Solar system in the context of the wide diversity of planetary systems: debris disks allow us to  "see worlds in grains of sand"\footnote{''To see the world in a grain of sand, and to see heaven in a wild flower, hold infinity in the palm of your hands, and eternity in an hour" (William Blake).}. 
	

\begin{center}
\Large
{\bf Part II}\\
\huge 
{\bf Physical Processes acting on Dust}
\end{center}
\normalsize

{\it Part II} reviews the dominant physical processes acting on dust particles and their effect of the particle size and spatial distribution, focussing on radiation and stellar wind forces (Section \ref{sec:radiation_stellar_wind_forces}), gravitational forces  in the presence of planets (Section \ref{sec:gravitational_forces}) and collisions (Section \ref{sec:coll}). The discussion applies the gas-free environment of the Solar system's interplanetary space and extra-solar debris disks.  There are many reviews on this topic, e.g. Mukai et al. (2001), Gustafson et al. (2001) and Dermott et al. (2001), Wyatt (2008a), Krivov (2010). 

\section{Radiation and Stellar Wind Forces}
\label{sec:radiation_stellar_wind_forces}

\subsection{Radiation Pressure}
\label{sec:radiation_pressure}

If the circumstellar dust particle is at rest, the radiation pressure force exerted by the stellar photons on the particle is given by $F_{\rm rad} = \frac{S}{h\nu} \frac{h \nu}{c} Q_{\rm pr} A$, where $\frac{S}{h\nu}$ is the flux of incoming photons,  $\frac{h \nu}{c}$ is the momentum per photon and  $Q_{\rm pr} A$ is the particle cross-section  for radiation pressure; $A$ is the particle geometric cross-section ($A = \pi s^2$) and $Q_{\rm pr}$ is the dimensionless radiation pressure factor averaged over the stellar spectrum; $Q_{\rm pr}$ is a function of the grain optical properties, size, shape and chemical composition and a measure of the fractional amount of energy scattered and/or absorbed by the grain. Substituting the energy flux density $S = \frac{L_{*}}{4 \pi r^2}$, one gets that  $F_{\rm rad} = \frac{L_{*} Q_{\rm pr} s^2}{4 r^2c}$,  where $L_{*}$ is the stellar luminosity, $s$ is the particle radius and $r$ is the heliocentric distance (Burns, Lamy and Sotter 1979). Because the radiation pressure force has the same dependency on $r$ as the gravitational force, $F_{\rm grav} = \frac{G M_{*} m}{r^2}$, it is useful to define the dimensionless parameter $\beta$, given by 

\begin{equation}
\beta = \frac{F_{\rm rad}} {F_{\rm grav}} = \left(\frac{3L_{*} }{16 \pi G M_{*} c}\right) \left( \frac{Q_{\rm pr}}{\rho s} \right), 
\label{beta}
\end{equation}

\noindent where $Q_{\rm pr}$ is the radiation pressure factor averaged over the stellar spectrum. 

\begin{figure}[]
\begin{center}
\includegraphics[width=0.6\textwidth]{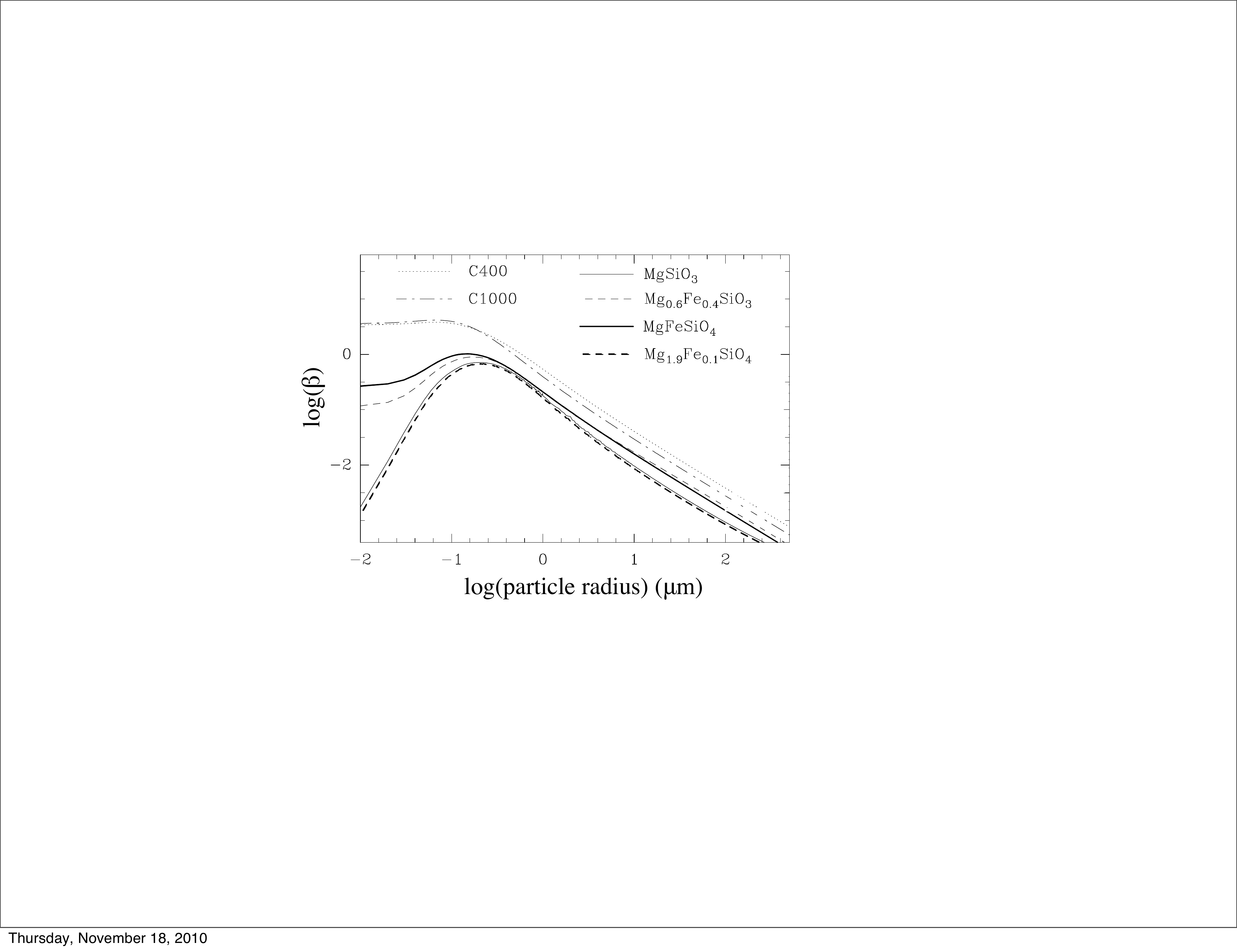}
\caption{Correspondence between $\beta = \frac{F_{\rm rad}} {F_{\rm grav}}$ and particle size when the central star is solar-type, under the assumption of spherical grains and for the following chemical compositions (based on spectroscopic observations of debris disks and evolved stars): MgSiO$_3$ and Mg$_{0.6}$Fe$_{0.4}$SiO$_3$ (Fe-poor and Fe-rich pyroxene), MgFeSiO$_4$ and Mg$_{1.9}$Fe$_{0.1}$SiO$_4$ (amorphous and crystalline olivine), and C400 and C1000 (graphite-poor and graphite-rich carbon). 
From Moro-Mart{\'{\i}}n, Wolf \& Malhotra (2005).
}
\label{fig_beta_size2}    
\end{center}
\end{figure}

\begin{deluxetable}{lcccccc}
\tablewidth{0pt}
\tabletypesize{\scriptsize}
\tablecaption{Correspondence between $\beta$ and radius ($\mu$m) for dust particles orbiting a solar-type star}
\tablehead{
\colhead{$\beta$} & 
\colhead{MgSiO$_3$} & 
\colhead{Mg$_{0.6}$Fe$_{0.4}$SiO$_3$} & 
\colhead{MgFeSiO$_4$} &
\colhead{Mg$_{1.9}$Fe$_{0.1}$SiO$_4$}& 
\colhead{C 400} & 
\colhead{C 1000}}
\startdata
0.4     		& 0.53 & 0.59  & 0.58 & 0.50 & 1.3   & 0.99 \\
0.2     		& 0.93 & 1.0   & 1.1  & 0.86 & 2.3   & 1.8  \\
0.1      		& 1.5  & 1.8   & 1.8  & 1.3  & 4.3   & 3.2  \\
0.05     	& 2.5  & 3.4   & 3.4  & 2.3  & 8.2   & 6.1  \\
0.025    	& 4.4  & 6.7   & 6.4  & 4.0  & 15.9  & 11.7 \\
0.0125   	& 8.0  & 13.7  & 12.4 & 7.1  & 31.2  & 22.7 \\
0.00625  	& 14.8 & 27.8  & 24.3 & 13.3 & 61.6  & 44.8 \\
0.00312  	& 28.6 & 54.8  & 48.1 & 25.7 & 122.7 & 89.0 \\
0.00156  	& 57.3 & 113.5 & 95.5 & 51.4 & 244.5 & 177.2 \\
\enddata
\tablenotetext{~}{Grain radii are given in $\mu$m. Particles are assume to be spherical, with bulk densities in g cm$^{-3}$ of 2.71 (MgSiO$_3$), 3.1 (Mg$_{0.6}$Fe$_{0.4}$SiO$_3$), 3.71 (MgFeSiO$_4$), 3.3 (Mg$_{1.9}$Fe$_{0.1}$SiO$_4$ - crystalline olivine), 1.435 (C 400) and 1.988 (C1000). From Moro-Mart{\'{\i}}n, Wolf \& Malhotra (2005).}
\label{tab_beta_size2}
\end{deluxetable}

\noindent  For interplanetary dust particles in the Solar system, $\beta = 5.7 \cdot 10^{-5} \left( \frac{Q_{\rm pr}}{\rho s} \right)$, where $\rho$ is the grain density in g cm$^{-3}$ and $s$ is the grain radius in cm (Burns, Lamy and Sotter 1979).  Figure \ref{fig_beta_size2} and Table  \ref{tab_beta_size2} show $\beta$ as a function of the particle radius for a range of representative dust compositions and assuming spherical grains. 

If the grains are highly non-spherical or have significant porosity (as might be expected for aggregates of smaller grains that result naturally from grain growth via coagulation - see e.g. Figure \ref{idp}), their increased surface and decreased bulk density will result on an increased value of $\beta$ (with respect to that of spherical grains). 

\subsection{Poynting-Robertson Drag}
\label{sec:pr_drag}
If the circumstellar particle is moving with respect to the central star, it experiences a force given by 

\begin{equation}
{\mu{d^2{\bf{r}} \over dt^2} = {S A Q_{\rm pr} \over c}\left[\left(1-{\dot{r}\over c}\right)
{\bf{\hat{S}}} - {{\bf{v}} \over c}\right],}
\label{mua}
\end{equation}

\noindent (to terms of order $\it{v}$/c), where $\bf{r}$ and $\bf{v}$ are the particle  position and velocity with respect to the central star, $\hat{S}$ is the unit vector in the direction of the incident radiation ($\bf{\hat{S}}$=$\bf{r}$/$\it{r}$) and $\mu$ is the particle mass. The radial term, ${S A Q_{\rm pr} \over c}\left(1-{\dot{r}\over c}\right){\bf{\hat{S}}}$ is the radiation pressure force with the added factor ($1-\frac{\dot{r}}{c}$) to account for the Doppler effect, while the velocity-dependent term, ${S A Q_{\rm pr} \over c} {{\bf{v}} \over c}$,  is known as the Poynting-Robertson (P-R) drag. The latter is a relativistic effect that can be intuitively explained in the following way:  in the reference frame of the particle, the stellar radiation appears to come at a small angle forward from the radial direction (due to the aberration of light) that results in a force with a component against the direction of motion; in the reference frame of the star, the radiation appears to come from the radial direction, but the particle reemits more momentum into the forward direction due to the photons blueshifted by the Doppler effect, resulting in a drag force (Burns, Lamy and Sotter 1979). Using  $\beta$ in Equation \eqref{beta}, the equation of motion becomes

\begin{equation}
{{d^2{\bf{r}} \over dt^2} = {-G\it{M}_*(1-\beta) \over \it{r}^{3}} {\it{\bf{r}}} 
- {\beta \over c} {G \it{M}_* \over \it{r}^{2} }\left[\left({\dot{r}\over r}\right)
{\bf{r}} + {\bf{v}}\right].
}
\label{d2rdt2}
\end{equation}

\subsection{Stellar Wind Forces}
\label{sec:stellar_wind_forces}

Radiation pressure and P-R drag arise from the transfer of momentum between the photons and the dust particle. Similarly, the dust grain interacts with the stellar wind particles giving rise to a corpuscular pressure force and a corpuscular drag force; these forces depend on the stellar wind properties (mass-loss, relative velocity between the stellar wind and the dust particle, and molecular weight), and on how efficiently the stellar wind particles interact with the dust grain. In the Solar system, the solar wind carries a momentum  flux  that is on average about 2 $\cdot$10$^{-4}$ times the momentum flux carried by radiation, and therefore the corpuscular pressure force can be neglected. On the contrary, the corpuscular drag force is about 35\% of the P-R drag force (Gustafson 1994); its increased significance in this case is due to the slower velocity of the solar wind compared to the speed of light which, in the frame of the particle,  increases the aberration angle and therefore the  component of the force against the direction of motion; the aberration angle of the stellar wind particles is arctan$(v/v_{\rm sw})$ compared to arctan$(v/c)$) for the stellar photos, where $v$ and $v_{\rm sw}$ are the velocity of the particle and stellar wind, respectively (Burns, Lamy and Sotter 1979). Defining the  ratio of the  solar wind drag to the P-R drag as $\it{sw}$, the equation of motion becomes  
\begin{equation}
{{d^2{\bf{r}} \over dt^2} = {-G\it{M}_*(1-\beta) \over \it{r}^{3}} {\it{\bf{r}}} 
- {(1+sw)\beta \over c} {G \it{M}_* \over \it{r}^{2} }\left[\left({\dot{r}\over r}\right)
{\bf{r}} + {\bf{v}}\right].
}
\label{d2rdt2sw}
\end{equation}

\noindent For the Solar system, $\it{sw}$ = 0.35 (Gustafson 1994); for M-type stars, the contribution of the corpuscular drag force could be significantly more important because of the increased mass-loss rate and low stellar luminosities (Plavchan, Jura \& Lipscy 2005). 

\subsection{Effect of radiation forces on the dust dynamics}
\label{sec:radiation_stellar_wind_effect_on_dynamics}

As soon as the dust particle is released from its parent body and begins to be subject to radiation forces, its equation of motion changes from that of the parent body, ${d^2{\bf{r}} \over dt^2} = {-G\it{M}_* \over \it{r}^{3}} {\it{\bf{r}}}$,  to  Equation \ref{d2rdt2} (or Equation \ref{d2rdt2sw} if considering stellar wind forces), resulting in a change of the particle orbital elements (Burns, Lamy and Sotter 1979). The degree of change will depend on the  $\beta$-value of the particle (the ratio of the radiation force to the gravitational  force acting on the particle). Figure \ref{fig_beta_size2}  shows that for very large and very small grains $\beta \rightarrow 0$; therefore, particles in this size range are unaffected by radiation forces. Intermediate-sized particles might be blown-out from the system if their specific orbital energy becomes positive, i.e. 

\begin{equation}
\frac{E}{m} = \frac{v^2}{2}  - \frac{GM_*(1 - \beta)}{r}  \equiv -\frac{GM_*(1 - \beta)}{2a} \ge 0, 
\label{Em}
\end{equation}

\noindent where $v$ is the particle velocity and $a$ its semimajor axis. If the particle is released at perihelion,  $r = a (1 - e)$, $v^2 = \frac{\mu}{a}\frac{1 + e}{1-e}$ and ejection occurs for $\beta \ge \frac{(1 - e)}{2}$; if the particles are released at aphelion, ejection occurs for $\beta \ge \frac{(1 + e)}{2}$. Radiation pressure blow-out is very fast, with a timescale similar to the orbital period, $t_{\rm blow} = \frac{1}{2}\Big(\frac{(r/AU)^3}{M_*/M_{\odot}}\Big)^{1/2}~~~\rm{yrs}$. Because $\beta$ depends on the particle size, e.g.  for the Solar system $\beta = 5.7 \cdot 10^{-5} \left( \frac{Q_{\rm pr}}{\rho s} \right)$, the condition above sets up a lower limit for the size of a particle on a bound orbit. In the Solar system,  the particles smaller than the blow-out size are known as  "$\beta$-meteoroids'' (Zook \& Berg 1975); these small grains, of asteroidal or cometary origin, are escaping from the Solar system on hyperbolic orbits as the result of radiation pressure; they have been inferred to exist from the lunar micro-crater record and from  {\it in situ}  detections  on board spacecraft (Gr\"un et al. 1994).  

The orbital energy of dust particles with $\beta \le \frac{(1 \pm e)}{2}$, i.e. with sizes larger than the blow-out size, will stay negative after release and therefore these particles will remain on bound orbits;  because the dust particle and its parent body are  effectively moving under different gravitational potentials, their orbital elements will differ (this is because the particle "feels" a stellar mass reduced by the factor (1-$\beta$) (Burns, Lamy and Sotter 1979). The position and velocity  of the parent body and the dust particle are the same at release, 

\begin{equation}
v = \Big[ GM \Big(\frac{2}{r}-\frac{1}{a}\Big)\Big]^{1/2} = \Big[GM(1-\beta) \Big(\frac{2}{r}-\frac{1}{a'}\Big)\Big]^{1/2}, 
\label{velocity}
\end{equation}

\noindent from which one gets that the particle semi-major axis right after release ($a'$) is
\begin{equation}
a'=a{1-\beta \over 1-2a\beta/r}, 
\label{a_after_release}
\end{equation}

\noindent  and its eccentricity ($e'$) is
\begin{equation}
e'={\Big|1 - { (1-2a\beta /r)(1-e^2) \over (1-\beta^2)}\Big|}^{1/2}
\label{e_after_release}
\end{equation}

\noindent (where $a$ and $e$ are the parent body semi-major axis and eccentricity). The particle inclination remains the same as that of its parent body because radiation pressure is a radial force.  

Radiation and stellar wind forces not only change the orbital elements of the dust particles upon release, but also affect their evolution with time. The drag forces make the dust particles lose orbital energy and spiral toward the central star, with
\begin{equation}
{\left<{ da \over dt}\right>_{\rm pr}=-{\beta GM_* \over c}{2+3e^2 \over a(1-e^2)^{3/2}}}
\label{dadt}
\end{equation}

\noindent (averaged over an orbit); for a particle in a circular orbit, $e = 0$, the timescale for orbital collapse can be found from $\int\limits_{a}^{0} a da = \int\limits_{0}^{t_{\rm pr}} - \frac{\beta 2GM_*}{c} dt$, resulting in ${t_{\rm pr} = \frac{a^2 c}{4 G M_* \beta} = \frac{4 \pi \rho s a^2 c^2}{3 L_* Q_{pr}}}$ ${\approx 690 \Big(\frac{\rho}{g/cm^3}\Big) \Big(\frac{s}{\mu m}\Big) \Big(\frac{a}{AU}\Big)^2 \Big(\frac{L_{\odot}}{L_*}\Big) \frac{1}{Q_{pr}} \rm{yr}}$, where $\rho$ is the particle bulk density, $s$ is the particle size, $a$ is the initial heliocentric distance, $L_*$ is the stellar luminostiy and $Q_{pr}$ is the radiation pressure factor; in the Solar system, $t_{\rm pr} \approx 400 \Big(\frac{a}{AU}\Big)^2 \frac{1}{\beta} $yr $\approx 2000 \Big(\frac{s}{\mu m}\Big) \Big(\frac{a}{AU}\Big)^2$ yr (Burns, Lamy and Sotter 1979).  A micron-sized dust particle at 40 AU (at the distance of the KB) will spiral into the Sun in $\sim$ 3 Myr, a timescale much smaller than the age of the Sun; this means that the dust observed in the  Solar system (and around other mature stars) cannot be primordial but must be replenished by planetesimals. 

Because P-R drag is of order $\frac {v}{c}$ and $v$ is highest at perihelion, the orbit not only shrinks but also circularizes, with 

\begin{equation}
{\left<{ de \over dt}\right>_{\rm pr}=-{5\beta GM_* \over 2c}{e \over a^2(1-e^2)^{1/2}}}
\label{dedt}
\end{equation}

\begin{figure}[]
\begin{center}
\includegraphics[width=0.5\textwidth]{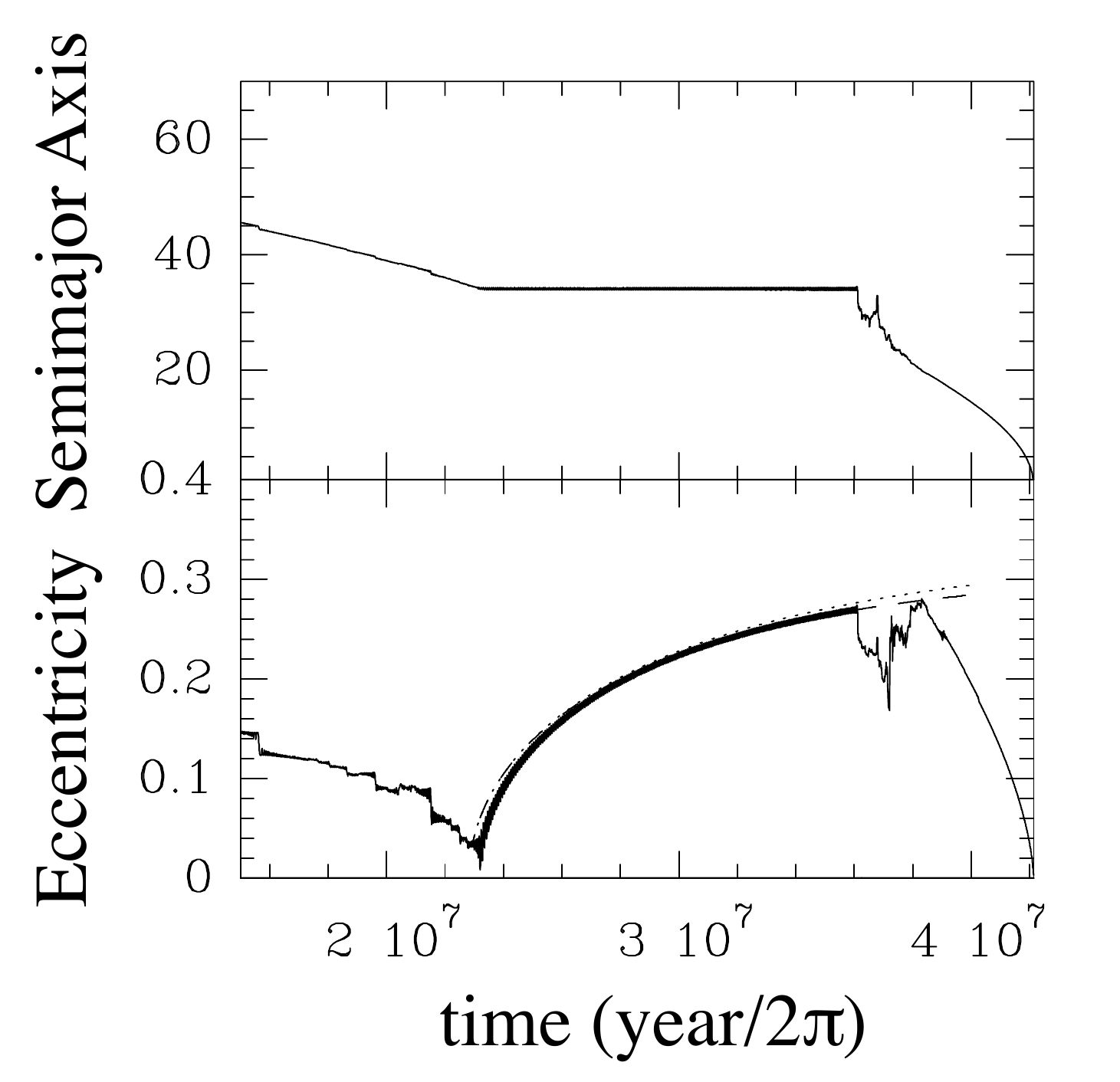}
\caption{Evolution of semimajor axis and eccentricity for a Kuiper belt dust particle with $\beta$=0.17  in a planetary system with a solar-type star (with $sw$ = 0.35) and a Neptune mass planet in a circular orbit. The solid line is the numerical result and the dotted and dashed lines are two analytical results. At first, the particle drifts inward due to P-R and corpuscular drag; during that time its semimajor axis and eccentricity decrease. Then the particle is trapped for 14 Myr in the exterior 4:3 MMR with Neptune; during this time the eccentricity of the particle increases until is sufficiently high to leave the resonance, after which the particle keeps migrating inward. From Moro-Mart{\'{\i}}n \& Malhotra (2002).}
\label{tp_evolution}      
\end{center}
\end{figure}

\noindent (averaged over an orbit). The particle inclination does not evolve with time because radiation pressure is a radial force (Burns, Lamy and Sotter 1979).  An example of the evolution of a dust particle under P-R drag can bee seen in Figure \ref{tp_evolution}. 

\subsection{Effect of radiation forces on the dust spatial distribution}
\label{sec:radiation_stellar_wind_effect_on_distribution}

In steady state, dust production and dust loss is balanced and  the amount of mass that crosses a given radius $r$ per unit time is a constant, i.e.  $\dot{M} = 2 \pi v r \sigma(r) =$ const., where $v$ is the dust particles velocity and $\sigma(r)$ is the dust disk surface density.  For unbound grains being blown out by radiation pressure, $v \sim$ const., $\sigma(r) \propto \frac{1}{r}$ and $n(r) \propto \frac {1}{r^2}$ (where the latter is the disk number density). For bound grains drifting inward under P-R drag, $v = v_{\rm pr} = \frac{r}{t_{\rm pr}}   \propto \frac{1}{r}$, which results in a dust disk with constant surface density, $\sigma(r) \propto \dot{M} = {\rm const.}$, and a number density that is inversely proportional to the distance to the central star, $n(r) \propto \frac {1}{r}$.

\section{Gravitational Forces  in the presence of Planets}
\label{sec:gravitational_forces}
If the dust particles are constantly being released by a planetesimal belt, P-R drag would create a dust disk of wide radial extent and uniform surface density. However, if one or more planets are present in the system,  on their journey toward the central star, the dust grains will be affected by gravitational perturbations that will modify the spatial distribution of the dust. The following describes the effect of the giant planets on the distribution of Kuiper belt dust in the Solar system. 

\subsection{Resonant perturbations}
\label{sec:mmrs}

Figure \ref{tp_evolution} shows the dynamical evolution of a typical dust particle from the Kuiper belt.  As the particle drifts inward due to P-R and solar wind drag, its semi-major and eccentricity decrease (following Equations \eqref{dadt} and \eqref{dedt}). Then the particle might get trapped in a mean motion resonance (MMR) with one of the giant planets - most commonly with Neptune because it is the outermost planet. Mean motion resonances are located where the orbital period of the dust particle is $\frac{p+q}{p}$ times that of the planet, $T_{\rm dust} = \frac{p+q}{p} T_{\rm pl}$, where $p$ and $q$ are integers, $p > 0$ and $p+q \geqslant 1$. While the planet orbital period is $T_{\rm pl} = 2\pi\Big(\frac{a_{\rm pl}^3}{GM_*}\Big)^{1/2}$, the effect of radiation pressure results in a dust particle orbital period of $T_{\rm dust} = 2\pi\Big(\frac{a_{\rm dust}^3}{GM_*(1-\beta)}\Big)^{1/2}$  (because the particle "feels" a less massive Sun by a factor of 1-$\beta$). Therefore, mean motion resonances take place when $a_{\rm dust~(p+q):p} = a_{\rm pl} (1-\beta)^{\frac{1}{3}}\Big(\frac{p+q}{p}\Big)^\frac{2}{3}$. To account for corpuscular drag forces, one would  substitute $\beta$  by $\beta(1+sw)$. A particle is  trapped when in the reference frame co-rotating with the planet, the closest approach  between the particle and the planet is always at the same point(s) along the particle orbit (one point if $q$ = 1, two points if $q$ = 2); at these location(s), the particle receives repetitive "kicks" from the perturbing planet that are always in the same direction and can balance the energy loss due to P-R drag, halting the particle migration. While trapped, the particle semi-major axis stays constant, while its eccentricity slowly increases (as it can be seen in Figure \ref{tp_evolution}). The amount of time the particle stays trapped in the MMR is highly variable. The particle escapes from the resonance when its eccentricity becomes sufficiently high ($\sim$ 0.3 for the KB dust particle shown in Figure \ref{tp_evolution}). After escaping, the particle keeps spiraling inward under P-R and solar wind drag  (following again Equations \eqref{dadt} and \eqref{dedt}).

\begin{figure}[]
\begin{center}
\includegraphics[width=0.75\textwidth]{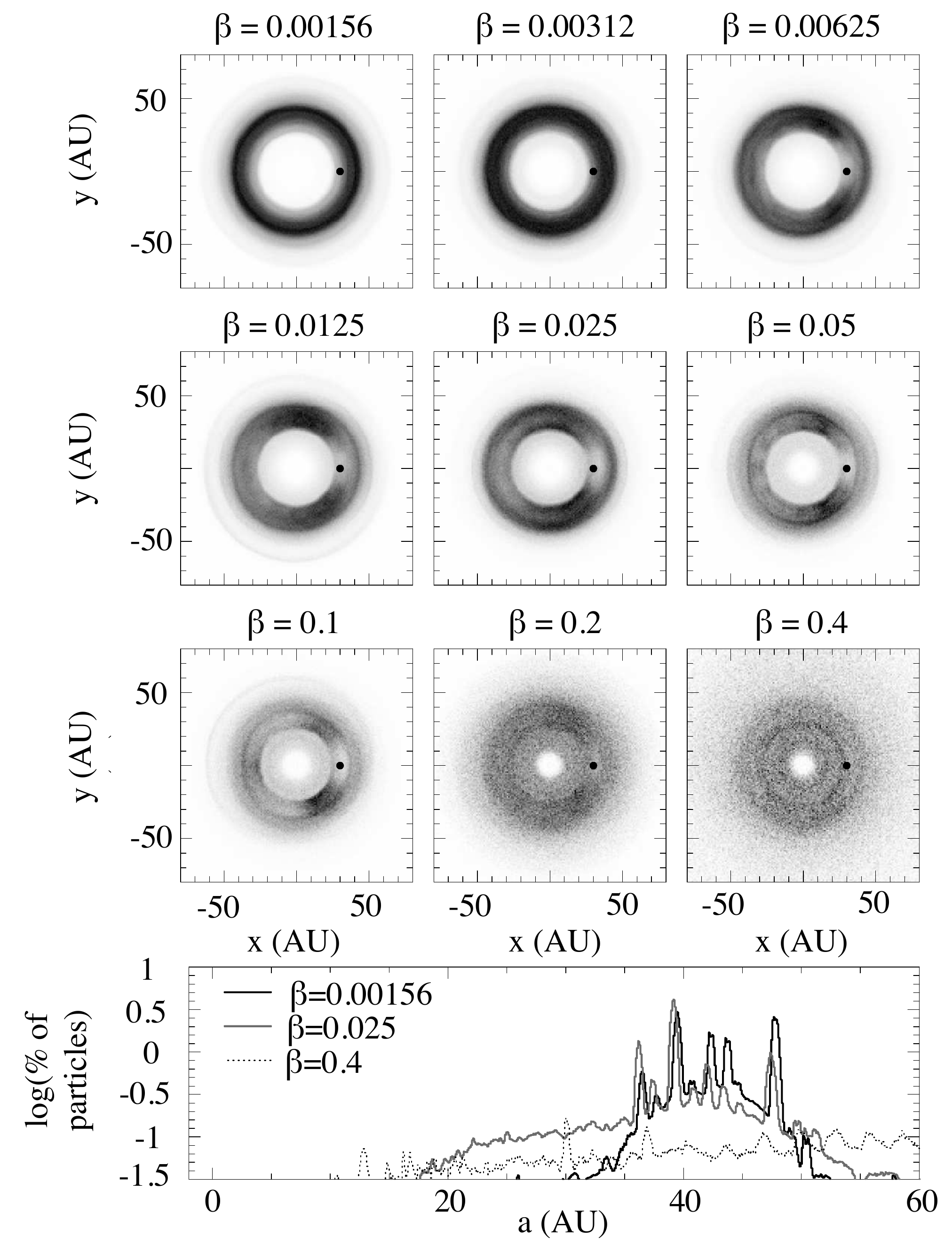}
\caption{Expected number density distribution of the Kuiper belt dust disk for nine different particle sizes (or $\beta$ values).  Assuming that the grains are composed of spherical
astronomical silicates,  $\beta$ values of 0.4, 0.2, 0.1, 0.05, 0.025, 0.0125, 0.00625, 0.00312, 0.00156 correspond to grain radii of 0.7, 1.3, 2.3, 4.5, 8.8, 17.0, 33.3, 65.9, 134.7
$\mu$m, respectively. The trapping of particles in MMRs with Neptune is responsible for the ring-like structure, the asymmetric clumps along the orbit of Neptune, and the clearing of dust at Neptune's location (indicated with a black dot). The disk structure is more prominent for larger particles (smaller $\beta$ values) because the P-R drift rate is slower and the trapping is more efficient. The disk is more extended in the case of small grains (large $\beta$ values) because small particles are more strongly affected by radiation pressure. The histogram shows the relative occurrence of the different MMRs for different sized grains, where the large majority of the peaks correspond to MMRs with Neptune. The inner depleted region inside $\sim$ 10 AU is created by gravitational scattering of dust grains with Jupiter and Saturn. 
From Moro-Mart{\'{\i}}n \& Malhotra (2002). }
\label{KBdust}       
\end{center}
\end{figure}

The histogram in Figure \ref{KBdust} shows the trapping of Kuiper belt dust particles in MMRs.  Even though the widths of the resonant regions are finite,  these narrow regions of parameter space can become densely populated with dust because they are constantly being fed by the inward migration of dust particles under P-R and stellar wind drag.  The features in the histogram are more pronounced for the larger grains (small $\beta$) than for the smaller grains (large $\beta$) because the larger grains have a higher trapping probability due to their slower migration velocity ($v_{\rm pr} \propto \frac{1}{t_{\rm pr}}  \propto {\beta} \propto \frac{1}{s}$, where $s$ is the radius of the particle).  Generally, more massive planets exert a stronger perturbing force  and allows the trapping of dust particles at resonances located further away from the planet. However, as if can be seen in the histogram, Neptune dominates the trapping of Kuiper belt dust because, even though is not the most massive planet,  it is the outermost planet and its exterior resonances are not affected by the interior MMRs (located where the orbital period of the planet is $\frac{p+q}{p}$ times that of the particle -- cf. Moro-Mart\'in et al. 2008, Wyatt 2008a). 
 
\subsection{Gravitational scattering}
\label{sec:gravitational_scattering}

\begin{figure}[]
\begin{center}
\includegraphics[width=0.65\textwidth]{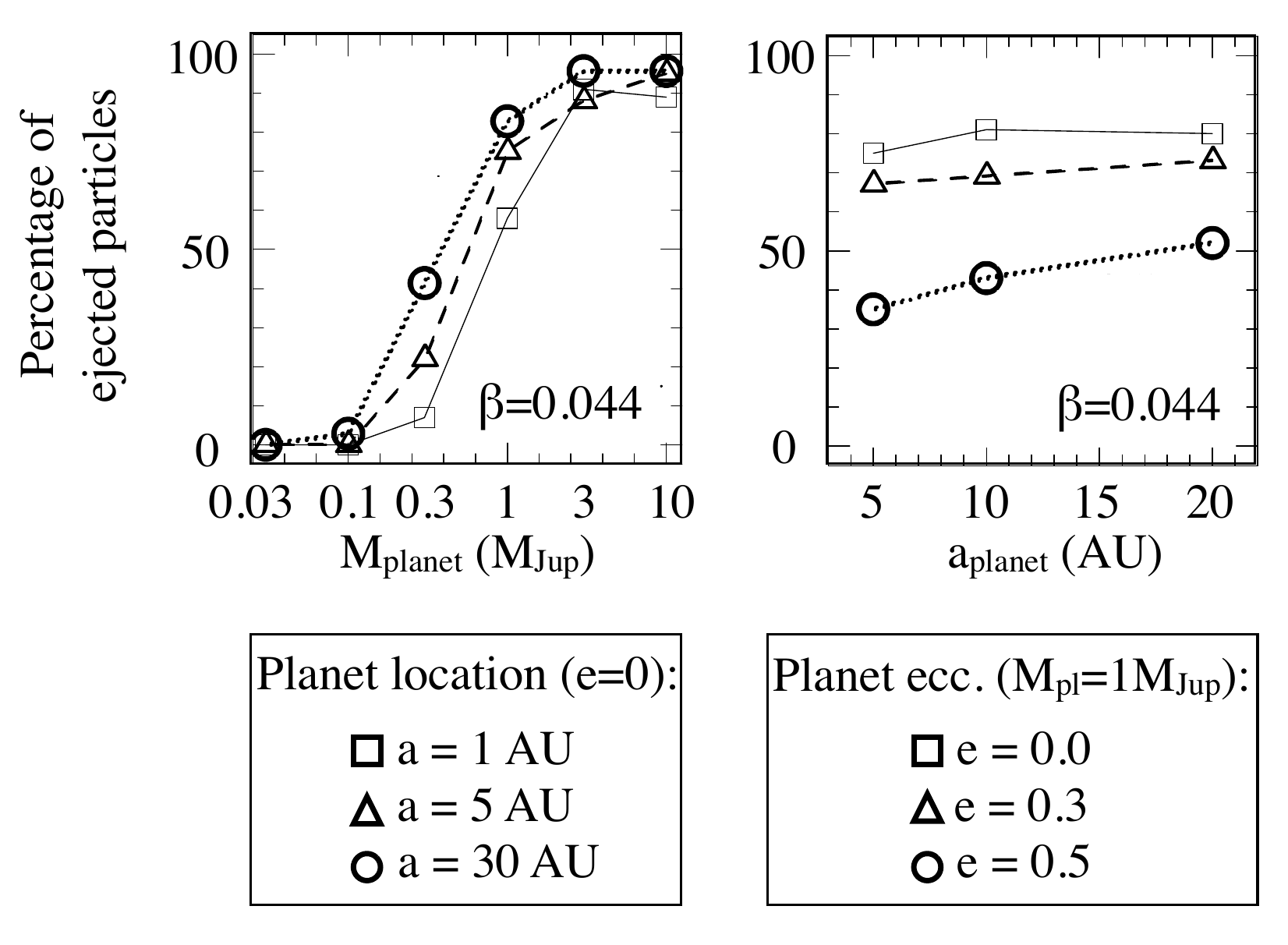}
\caption{Percentage of dust particles ejected from the system by gravitational scattering with a planet. {\it Left}: dependency on planet mass
(x-axis) and the planet semimajor axis (indicated by the different symbols). {\it Right}: dependency on planet semi-major axis (x-axis) and eccentricity (corresponding to the different symbols). The particle size is fixed, corresponding to $\beta$ = 0.044. In the case of Kuiper Belt dust,  ejection efficiencies for particles with $\beta \le$ 0.4 are approximately 5--20\% for Uranus or Neptune, and  25--40\% for Saturn or Jupiter (Moro-Mart{\'{\i}}n \& Malhotra 2003, 2005).}
\label{efficiency} 
\end{center}
\end{figure}

Figure \ref{tp_evolution} shows that after the Kuiper belt dust particle leaves the resonance, it will continue spiraling inward under the effects of P-R and corpuscular drag. This inward journey will take the particle near the orbit of a giant planet, where neighboring MMRs overlap and make the orbit of the particle chaotic and subject to gravitational ejection. This chaotic region extends from  $a_{\rm pl} - \Delta a < a < a_{\rm pl} + \Delta a$, where $m_{\rm pl}$ and $a_{\rm pl}$ are the planet mass and semi-major axis and $\Delta a \simeq \pm 1.5 a_{\rm pl} \Big(\frac{m_{\rm pl}}{m_\star}\Big)^{2/7}$ (Wisdom 1980). Dynamical models show that most of the Kuiper Belt dust grains get ejected from the Solar system by gravitational scattering with the giant planets: for particles with $\beta \le$ 0.4, the percentage of particles ejected by Uranus and Neptune is 5--20\%, while for Saturn and Jupiter is 25--40\% (where the efficiency of ejection refer to each planet); only 10--20\% of the Kuiper belt dust particles are able to pass these gravitational barriers and drift into the inner Solar system. Figure \ref{efficiency} shows the dependency of the efficiency of gravitational ejection on the planet mass, semi-major axis and eccentricity: planets with masses of 3 M$_{\rm Jup}$--10 M$_{\rm Jup}$ located between 1 AU--30 AU in a circular orbit around a solar-type star eject $>$90\% of the dust grains that go past their orbits; a 1 M$_{\rm Jup}$ planet at 30 AU ejects $>$80\% of the grains and about 60\% if located at 1 AU; while a 0.3 M$_{\rm Jup}$ planet  ejects about  40\% if located at 30 AU and $<$ 10\% if it is at 1 AU. The efficiency of ejection decreases significantly as the planetÕs eccentricity increases, e.g. for a 1 M$_{\rm Jup}$ planet at 5 AU the efficiency of ejection decreases from 80\% for $e$ = 0 to 30\% for $e$ = 0.5. For eccentric planets, the particle ejection preferentially takes place along the major axis of the planet's orbit, with the number of particles ejected in the apoastron direction exceeding that in the periastron direction by a factor of 5 for $e$ = 0.5 (because the planet spends more time near
apoastron and therefore the probability of encounter with a dust particle is higher near that location). 

\subsection{Secular perturbations}
\label{sec:secular_resonances}

\begin{figure}[]
\begin{center}
\includegraphics[width=0.70\textwidth]{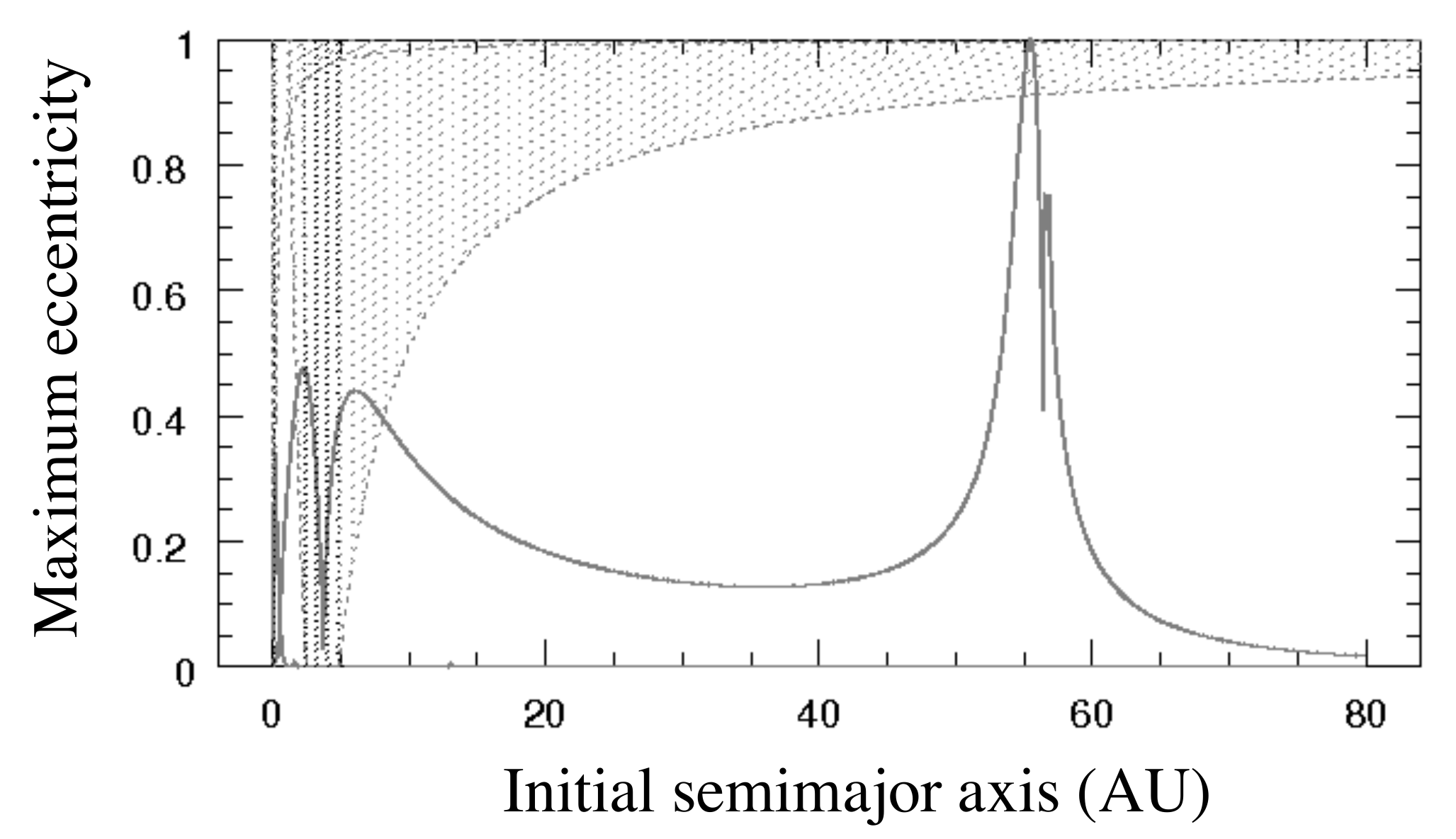}
\caption{
Effect of the secular perturbations created by the two planets in the HD 38529 system. 
The planetary masses, semi-major axes and eccentricities are the following: $M_{\rm b}({\rm sin}i)$ = 0.8 M$_{\rm Jup}$, $a_{\rm b}$ = 0.13 AU, $e_{\rm b}$ = 0.25, for planet b, and $M_{\rm c}({\rm sin}i)$ = 12.2 M$_{\rm Jup}$, $a_{\rm c}$ = 3.74 AU, $e_{\rm c}$ = 0.35, for planet c.  The shaded zones denote areas that are strongly unstable due to planet-crossing orbits and overlapping 1st order mean motion resonances.  The y-axis is the maximum eccentricity imposed on the test particles (initially on circular orbits).   The secular modes of the two planets excite the eccentricities of the test particles. The effect of these perturbations can be felt at a wide range of distances from the star: secular eccentricity excitation exceeds 0.1 to nearly 60 AU; the sharp peak at 55 AU is due to a resonance with the slow mode. The dust observed in the system  is likely produced by planetesimals located in the regions of low eccentricity from $\sim$ 20--50 AU. (From Moro-Mart\'in et al. 2007b).
}
\label{hd38529} 
\end{center}
\end{figure}

The gravitational forces on the dust particle exerted by planetary companions are described by a sum of many terms, known as the perturbing function.  Secular perturbations are the long-term average of these forces; they result from the terms of the perturbing function that are independent of the mean longitude of the planets and the dust particles. Unlike the resonant perturbations described above, secular perturbations are non-periodic in nature and act on longer timescales ($>$0.1 Myr). They can be thought of as the perturbations that would arise if the mass of the perturbing planet were to be spread out along its orbit (like a wire), weighting the mass density to reflect how much time the planet spends in each region.  A planet  on an  eccentric orbit can force an eccentricity on the test particles;  if the planet and the test particle are not co-planar, the secular perturbations will tend to align their orbits; secular perturbations do not affect the semi-major axis of the particles. If there is only one perturbing planet, the strength of the perturbation is independent of its mass, but the smaller the mass the longer the secular timescale.  If there are more than one perturbing planets, particles with precession rates that coincide with the eigenfrequencies of the planetary system (resulting from secular perturbations between the planets) will be strongly affected by secular perturbations (see e.g. Figure \ref{hd38529}), resulting in the ejection of these particles from the secular resonant region (cf. Murray \& Dermott 1999, Wyatt et al. 1999,  Wyatt 2008a). 

\subsection{Effect of gravitational forces on the dust spatial distribution}
\label{sec:grav_effect_on_dynamics}

\subsubsection{Resonant perturbations}
Figure \ref{KBdust} shows that the trapping of Kuiper belt dust in MMRs results in the formation of structure in the dust disk consisting on resonant rings outside the orbit of the perturbing planet, asymmetric clumps and a clearing of dust at the planet position. The rings are created when the dust particles are on nearly circular orbits and because they spend a significant part of their lifetime trapped at certain semi-major axis, corresponding to the most favorable MMRs. For Kuiper Belt dust, with a low eccentricity perturbing planet and low eccentricity dust producing parent bodies, the most favorable MMRs are the first order resonances 2:1, 3:2, 4:3... The clumps appear when the particles are on eccentric orbits: when trapped in a resonance, the particles orbits tend to be oriented always in the same direction with respect to the location of the planet, and clumps are created near apocenter, where the particles spend more time; because the clumps are fixed with respect to the planet position, they should follow the orbit of the planet (rotating in the reference frame of the star), and this can be used as an observational test to assess if the proper motion of the clump is consistent with dust particles trapped in an MMR. 
The clearing of dust near the planet position is created because when trapped in a resonance, the particle avoids being close to the perturbing planet. 

The resonant features described above for the Kuiper belt dust disk (Figure \ref{KBdust}) have yet to be observed because the foreground thermal emission from the dust in the inner Solar system (of asteroidal and cometary origin) overwhelms the background emission from the colder Kuiper belt dust. However, resonant features have been observed in the zodiacal cloud itself: a  ring of asteroidal dust particles trapped in the 1:1 co-rotating resonance with the Earth at around 1 AU, with a 10\% number density enhancement on the Earth's wake that results from the resonance geometry (Dermott et al. 1994).  Modeling and observations of the Kuiper belt dust and of the zodiacal cloud  indicate that planets with a wide range of masses (down to Neptune and Earth-masses) can create high-contrast features via resonant trapping.

Figure \ref{KBdust} shows that the resulting dust disk structure depends on the particle size under consideration. This is because the dynamical evolution of the particle depends on its size: large particles interact weakly with the stellar radiation field, migrate slowly, and can get easily trapped in MMRs; small grains on the other hand interact strongly with radiation, their inward migration is faster, and this gives rise to more extended and more uniform disks (Liou \& Zook 1999, Moro-Mart\'in \& Malhotra 2002). 
 
Mean motion resonances can also be populated if the perturbing planet migrates outward; in this case, the resonant trapping probability depends on the planet  migration rate and the extent of the migration.  In the Solar system, the early migration of Neptune resulted in the trapping of Kuiper belt objects in MMRs, primarily in the 4:3, 3:2, 5:3 and 2:1 (Malhotra 1993). This means that the dust-producing planetesimals may have an asymmetric distribution (note that the disk models in Figure \ref{KBdust} assumed a uniform distribution of the dust-producing bodies, with angular elements uniformly distributed between 0 and 2$\pi$). If the parent bodies themselves are trapped in a resonance,  the largest dust grains released  from them would remain trapped in the same MMR, the smallest grains (with $\beta > 0.5$) would escape the system producing a spiral structure, and the intermediate size grains (with $\beta < 0.5$)  would leave the resonance but remain on bound orbits producing an axisymmetric distribution (Wyatt 2008a). 

\subsubsection{Gravitational scattering}
\begin{figure}[]
\begin{center}
\includegraphics[width=0.7\textwidth]{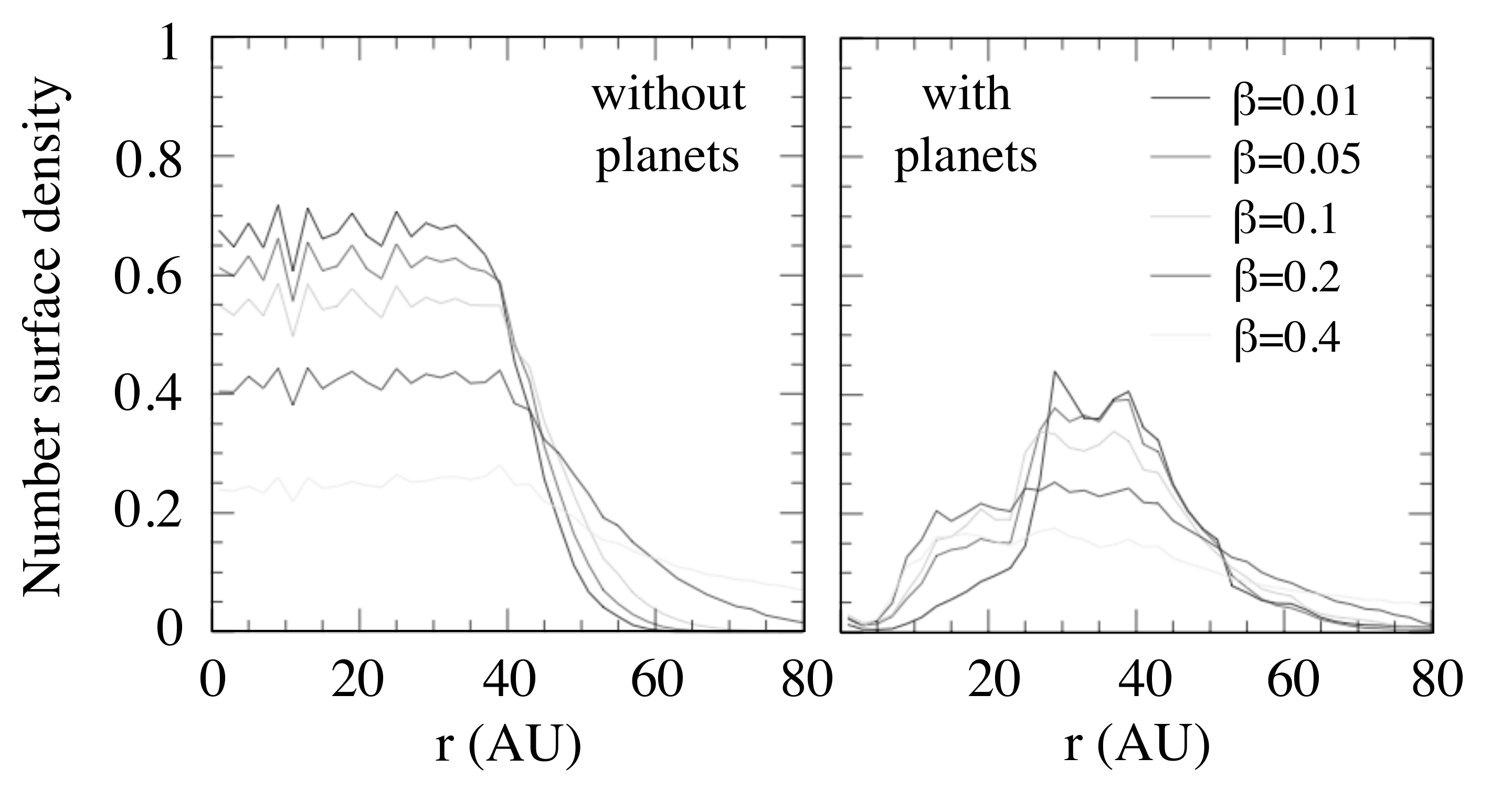}
\caption{({\it Right}): Expected radial distribution of the number surface density for the Kuiper belt dust disk, taking into account the gravitational perturbations from all the planets in the Solar system (excluding Mercury). ({\it Left}): Same but assuming there are no planets in the system, leading to a uniform surface density. The main features from the comparison of the two panels are the depletion of particles in the inner 10 AU, due to scattering by Jupiter and Neptune, and the enhancement of particles from 30 to 50 AU, due to trapping in MMRs with Neptune. From Moro-Mart\'in and Malhotra (2002).}
\label{radialprofile}    
\end{center}
\end{figure}

High-contrast features can also be produced by gravitational scattering:   for planetary systems in which the source of the dust is outside the orbit of a massive planet, gravitational ejection can result in the formation of a dust-depleted region inside the planet orbit, where the depletion factor depends on the planet's mass and orbit  (Figure \ref{efficiency}). Models indicate that a dust depleted region of  $\sim$ 10 AU in radius is expected to be present in the Kuiper belt dust disk due to gravitational scattering by Jupiter and Saturn (Figures \ref{KBdust} and \ref{radialprofile}). There are also indications that large inner cavities are very common in extra-solar debris disks (see Figure \ref{carpenter2} and the discussion in Section \ref{innergaps}).

\subsubsection{Secular perturbations}
Secular perturbations can also produce warps, spirals and brightness asymmetries.  A warp is created when the planet and the test particles are on non-coplanar orbits, and because the secular perturbations will tend to align their orbits on a shorter timescale closer to the star. With time, the inflection point of the warp (marking the boundary between perturbed and unperturbed orbits) will move outward at a rate that depends on the planet mass and semi major axis and is proportional to $M_{pl}$ $a_{pl}^2$ (Mouillet et al. 1997), which in principle could be used to constrain the planet mass if the age of the planet were known. With time, the warp will end up disappearing, unless multiple planets on non-coplanar orbits are present in the system. Spirals are created if the planet is in an eccentric orbit; in this case, the secular perturbations can force an eccentricity on the test particles creating two spiral structures that with time propagate away from the planet. This also creates a brightness asymmetry because, after all the test particles have been affected, there is an offset in the dust disk center with respect to the star. 

Some of the structure observed in the zodiacal cloud is the result of secular perturbations, namely the inner edge of the cloud around 2 AU (due to a secular resonance with Saturn that also explains the inner edge of the main asteroid belt), the offset of the cloud center with respect to the Sun, the inclination of the cloud with respect to the ecliptic, and the cloud warp (cf. Murray \& Dermott 1999, Wyatt et al. 1999, Wyatt 2008a). And as it was discussed in {\it Part I},  these features can also be seen in extrasolar dust disks, e.g. the warps  in AU Mic and $\beta$ Pic, the offsets with respect to the central star in $\epsilon$-Eri and Fomalhaut, the brightness asymmetries in Fomalhaut and HD 32297, and the spiral structure in HD 141569, to name a few (see Figure \ref{structure_planets} and discussion in Section \ref{sec:ddstructure}). 

\subsubsection{Debris disk structure can unveil the presence of planets}
Most of the structural features discussed above depend on the mass and orbit of the planet, and as the case of the Solar system illustrates, the structure is sensitive to small planets (like the Earth) and to planets located far from the star (like Neptune). As discussed in Section \ref{sec:ddstructure}, this opens the possibility of using the study of the dust disk structure as a  detection technique of planets of a wide range of masses and semi-major axes. The discovery of the planets around Fomalhaut and $\beta$-Pic, that were previously predicted to exist based on the structure of both debris disks, illustrates this idea (see discussion in Section \ref{sec:ddstructure} and Figure \ref{structure_planets}). Particularly interesting is that this method  is complementary to radial velocity and transit surveys (which are limited to planets relatively close to the star), and to direct imaging (which is limited to young and massive planets). 

\section{Collisions}
\label{sec:coll}

\subsection{Collisional lifetimes}
\label{sec:coll_lifetime}

The timescale for a collision between two equal-sized grains of radius $s$ is $t_c = \frac{1}{\pi(s+s)^2 F}$, where $F$ is the particle flux. The flux is given by $F = \frac{N}{V}\Delta v$, where $N$ is the total number of particles in the disk, $V$ is the total volume they occupy and $\Delta v$ is their velocity dispersion. If the particles in the dust disk have non-zero eccentricities and inclinations ($e$ and $i$, with $i$ in radians), they will occupy a volume $V$= $2 \pi a \cdot 2ae \cdot 2 i a$ = $8 \pi a^3 e i$, where the factor $2 \pi a$ is the disk circumference,  $2ae$ is its width (from pericenter, $a (1-e)$, to apocenter,  $a(1+e)$), and $2 i a$ is its thickness; the velocity dispersion is $\Delta v = v_K(e^2 + i^2)^{1/2}$, where $V_K$ is the Kepler velocity. This leads to $t_c = \frac{8 \pi a^3 e i}{\pi 4 s^2 N v_K(e^2 + i^2)^{1/2}}$.  If there are $N$ grains with a characteristic size $s$,  the optical depth is $\tau = \frac{N \pi s^2}{4 \pi a^2 e}$. The collisional timescale in terms of the optical depth and orbital period (for $i$ = $e$) is 

\begin{equation}
t_c = \frac{P}{\tau} \frac{i}{4\pi(e^2 + i^2)^{1/2}} \sim \frac{1}{9} \frac{P}{\tau} \sim \frac{1}{9\tau} \Big(\frac{r}{AU}\Big)^{3/2} \sim \frac {1}{\tau\Omega},
\label{tcol}
\end{equation}
where $\Omega$ is the angular velocity. The collisional velocity above can also be approximated by  

\begin{equation}
t_c  \sim 0.1 \Big(\frac{a}{AU}\Big)^{3/2}  \Big(\frac{M_{\odot}}{M_*}\Big)^{1/2} \frac{1}{\tau} \sim 1.1\cdot10^4 \Big(\frac{a}{AU}\Big)^{3/2}  \Big(\frac{M_{\odot}}{M_*}\Big)^{1/2} \Big(\frac{10^5}{L_{dust}/L_*}\Big),
\label{tcol2}
\end{equation}
where $L_{dust}/L_*$ is the fractional luminosity of the debris disk. All these estimates assume equal-sized grains (the following subsection considers the more realistic case of a distribution of particle sizes). Comparing this collision timescale to the PR timescale, 

\begin{equation}
\frac{t_c}{t_{\rm pr}} \sim \frac{\frac{1}{9\tau} \Big(\frac{r}{AU}\Big)^{3/2}}{\frac{400}{\beta} \Big(\frac{r}{AU}\Big)^2} \sim \frac{\beta}{3600\tau\Big(\frac{r}{AU}\Big)^{1/2}}. 
\label{tcol_tpr}
\end{equation}

In the Kuiper Belt, $r \sim 40$ AU and it is estimated that $\tau \sim 10^{-7}$  so that $\frac {t_c}{t_{\rm pr}} \sim 440 \beta$ (with $t_c \sim$ 280 Myr and $t_{\rm pr} \sim \frac {0.6}{\beta}$ Myr), suggesting that collisional destruction is unimportant and removes only a small fraction of KB dust grain as they drift into the inner Solar system (this is referred to as the P-R-dominated regime).  Compared to interstellar collisions, mutual collisions in the present day Kuiper belt are less significant because the relative velocity of Kuiper belt dust grains ($\sim$1.6 km s$^{-1}$) is significantly smaller than the impact velocity of interstellar grains ($\sim$25 km s$^{-1}$), making shattering less likely, and because the optical depth of the Kuiper Belt is very small so that collisions are infrequent. Due to limitations in the sensitivity of the detectors, the majority of the extra-solar debris disks observed with {\it Spitzer} have optical depths $\gtrsim$ 100 times that of the Kuiper belt; unlike in the Solar system, in these systems collisional destruction plays a  significant role in the dust dynamics and resulting disk structure (this is referred to as the collision-dominated regime, Wyatt 2005, 2008b). The improved sensitivity of observatories like {\it ALMA}, {\it JWST} and {\it SPICA} (and to some degree {\it Herschel}) will enable the study of debris disks that are in the P-R dominated regime (i.e. KB dust disk analogs). 

\subsection{Effect of collisions on the dust  size distribution}
\label{sec:coll_effect_on_dust_size}

The size distribution most commonly adopted for both interstellar and debris dust is $n(s){\rm d}s \propto s^{-3.5}{\rm d}s$ (Dohnanyi 1969; Mathis,  Rumpl, \& Nordsieck 1977). It results from a catastrophic quasi-steady state collisional cascade in which the dust is derived from the grinding down of larger bodies, assuming that: (1) the strength of the particle does not depend on the target size (where the strength is measured as the energy per unit volume needed to break and disperse the target); and (2) the system is in quasi-steady state, i.e. the same amount of mass that enters one size bin as larger particles break down, leaves the size bin as the particles continue breaking in smaller pieces. Obviously, the bins at the two extremes would not be in steady state and  the reservoir of large particles would get depleted with time. For  $q = -3.5$, the mass is dominated by the large grains and the cross-section is dominated by the small grains. For comparison, in a size distribution with a power-law index $q$ = -3, each size bin will contain an equal amount of cross-sectional area, while for $q$ = -4, each size bin will contain an equal amount of mass. 

\begin{figure}
\begin{center}
\includegraphics[width=0.8\textwidth]{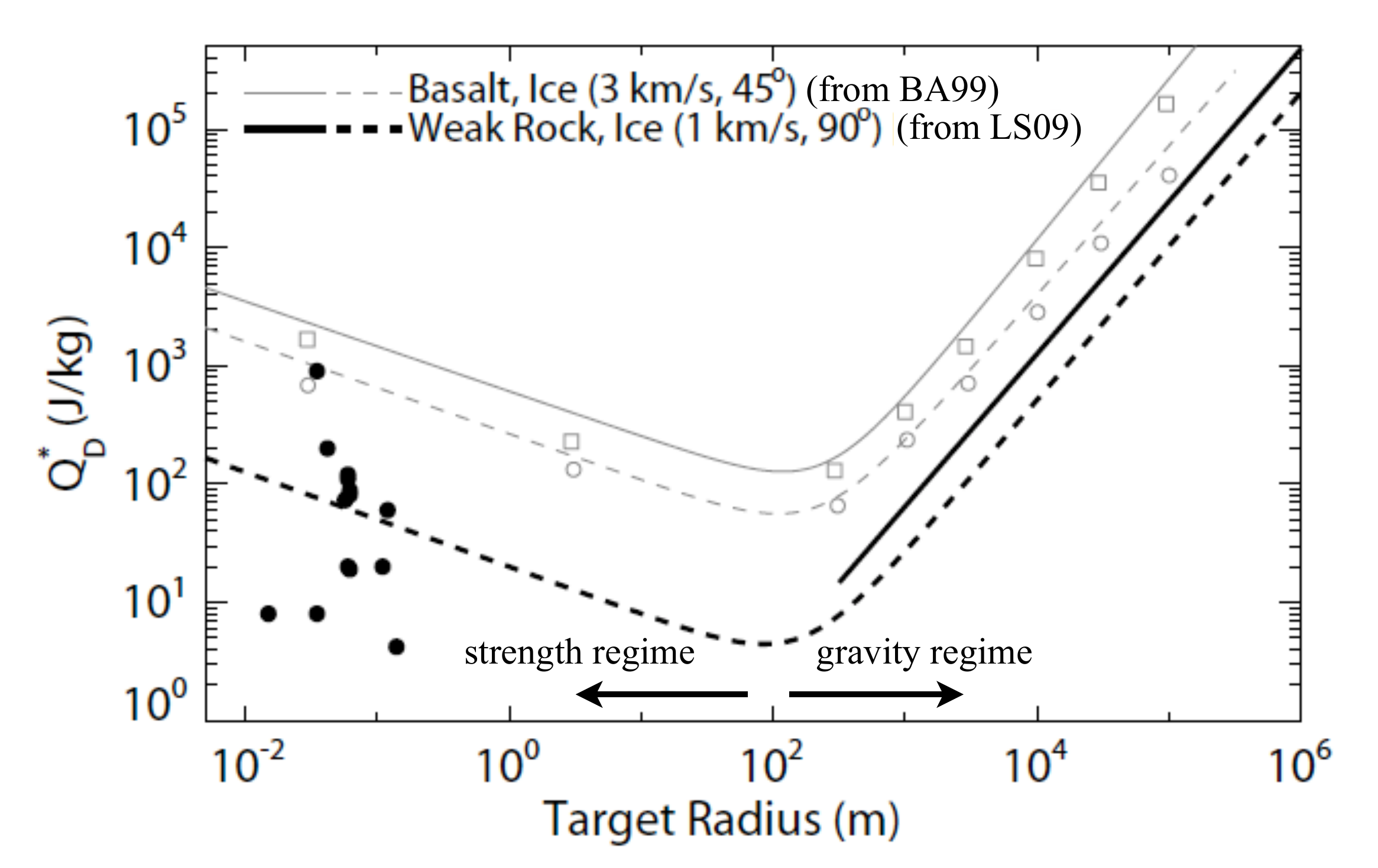}
\caption{Energy per unit mass necessary for a collision to result in the largest fragment to have half the mass as the original target, with target properties that could be similar to those of KBOs. The lines correspond to models (solid for rocks and dashed for ice) and the symbols to laboratory experiments (squares for rocks and circles for ice). BA09 is Benz \& Asphaug (1999) and LS09 is  Leinhardt \& Stewart (2009). Figure from Leinhardt \& Stewart (2009).}
\label{yield}
\end{center}
\end{figure}

However, the particle strength is in fact a function of the target size (unlike assumed above).  For small objects, this critical specific energy for disruption and dispersal, $Q_D^*$, is dominated by the material strength, while for large objects is dominated by self-gravity, with a turnover is around 100 m (even for relatively small particles, gravity can play an important role because the transfer of momentum is inefficient and more energy than that required to shatter the targets is needed to dispersed the fragments). The turn over results in a sum of two power-laws (see Figure \ref{yield}), 

\begin{equation}
Q_D^* = Q_s \Big(\frac{s}{{\rm 1~m}}\Big)^{-b_s} + Q_g \Big(\frac{s}{{\rm 1~km}}\Big)^{b_g},  
\label{qd}
\end{equation}

\noindent with $Q_s$ and $Q_g$ in the range 10$^5$--10$^7$ erg g$^{-1}$, $b_s$ $\sim$ 0--0.5, and $b_g$ $\sim$ 1--2 ($s$ is for the scattering regime and $g$ for the gravitational regime -- Benz \& Asphaug, 1999; Leinhardt \& Stewart, 2009).  Krivov et al. (2005) estimate that two colliders of mass $m_t$ and $m_p$ are disrupted if their relative velocity is larger than the critical value of $\Big(\frac{2(m_t+m_p)^2}{m_t m_p}Q_D^*\Big)^{1/2}$, which for equal-sized particles would be $(8Q_D^*)^{1/2}$. Given that $Q_D^* \sim 10^8$ erg g$^{-1}$, dust grains would get destroyed if their relative velocity exceeds several hundreds m s$^{-1}$, typical of dust particles with $e \sim 1$ at 10s of AU from the central star (Krivov 2010). 

\begin{figure}[]
\begin{center}
\includegraphics[width=0.95\textwidth]{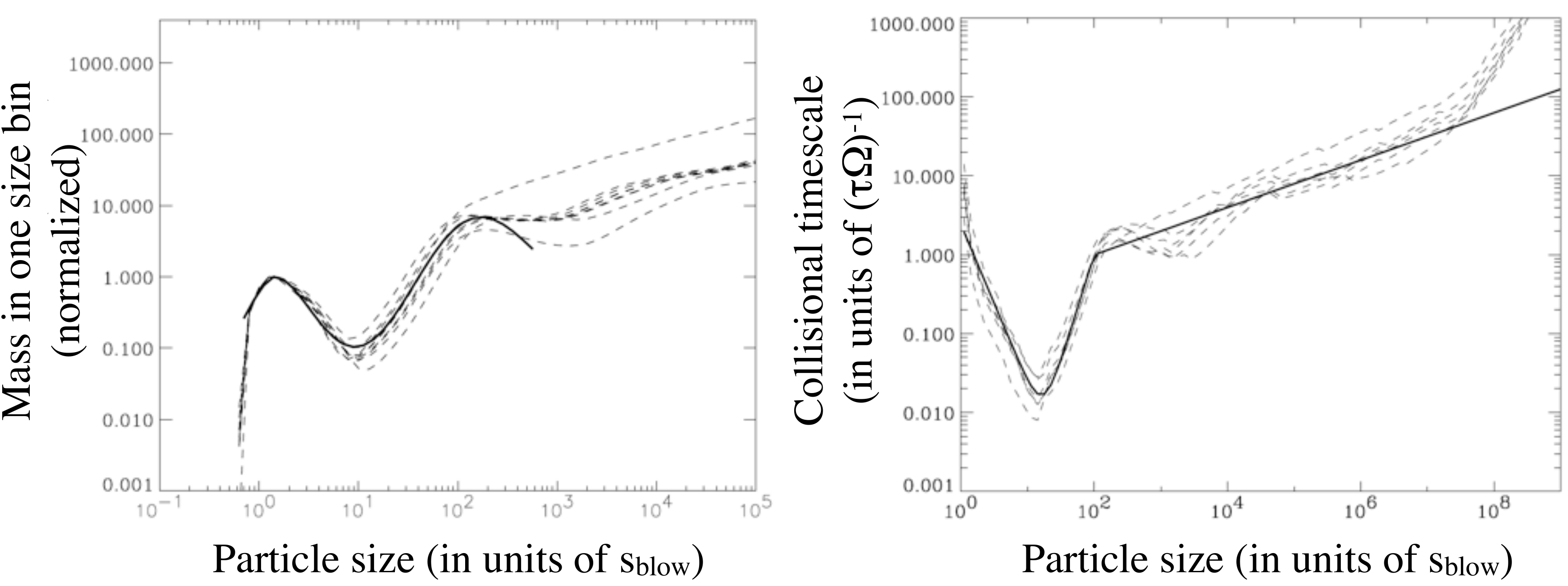}
\caption{
({\it Left}): Particle size distribution integrated over the entire disk. The different dashed lines  correspond to models with different initial disk masses, surface density profiles and dynamical excitation. The y-axis is normalized to the peak at $\sim1.5 s_{\rm blow}$. 
({\it Right}): Collisional lifetime at $r = 55$ AU. 
In both panels, the solid lines correspond to the empirical approximations described in the text. 
(From Thebault \& Auregearu 2007). 
}
\label{thebault_18_19} 
\end{center}
\end{figure}

The particle-in-the-box scenario described above that gives rise to the $q$ = -3.5 power-law index can be less restrictive  by taking into account a size-dependent particle strength, the removal of the smallest particles by radiation pressure, and the effect of collisions with grains coming from the inner regions on highly eccentric orbits. Numerical models that take these factors into account find that the size distribution changes from a strict single power-law of index -3.5 to a wavy distribution; the resulting size distribution depends on the  distance from the star, with the waves being more pronounced in the outer regions because the particles there are more affected by collisions with high-velocity grains (just below the blow-out size) coming from the inner disk. The following results from numerical simulations by Th\'ebault \& Augereau (2007)  illustrate the effect of collisions. Integrated over the entire disk, these models find an overdensity of particles with sizes $\sim 2 s_{blow}$, and a depletion of particles with sizes in the range (10--50)$s_{\rm blow}$, where $s_{\rm blow}$ is the blow-out size (see Figure \ref{thebault_18_19}).  For sizes $ < 100 s_{\rm blow}$, these results are found to be weakly independent of the initial disk mass, the initial surface density profile and the dynamical excitation, while they depend significantly on the particle strength. An empirical estimate for $s < 100 s_{\rm blow}$ based on these numerical simulations give
${\rm d} N \propto G(s)s^{-3.59}{\rm d}s$, for $\frac{2}{3}s_{\rm blow} < s \lesssim 100 s_{\rm blow}$, with
${\rm log}_{10} (G(s)) = \frac{2}{3}\big[{\rm cos}\big(2\pi\big[\vert\frac{1}{2}{\rm log}_{10}\big(\frac{s}{1.5s_{\rm blow}}\big)\vert\big]^{0.85}\big)-1\big]$. 
For $s > 100 s_{\rm blow}$, a rough extrapolation is  ${\rm d} N \propto s^{-3.7}{\rm d}s$ (Th\'ebault \& Augereau 2007). 

Regarding the changes in the collisional lifetime, the  expression in \eqref{tcol}, ${t_c} \sim \frac{1}{9} \frac{P}{\tau} \sim \frac {1}{\tau\Omega}$, assumed that all impactors are equal-sized and all the collisions are destructive, and ignored both  the effect of collisions with grains coming from the inner regions ($< r$), and the dynamics of the smallest grains affected by radiation pressure. Figure \ref{thebault_18_19} shows that when taking these factors in to account, the collisional lifetime depends strongly on the particle size (with a wavy  pattern), and in some cases (for particles $<$ 100 $\mu$m) can be  2 orders of magnitude smaller than ${t_c} \sim \frac{1}{9} \frac{P}{\tau} \sim \frac {1}{\tau\Omega}$; the main features are a sharp increase near the blow-out size ($s_{blow}$), a sharp minimum at $\sim 10 s_{blow}$, a sharp increase between between $\sim 10 s_{blow}$ and $\sim 100 s_{blow}$, and a slow increase for larger sizes. An empirical estimate of the collisional timescale derived from numerical simulations give 
$t_c (s,r)= \frac {1}{\tau\Omega} \big[\big(\frac{s}{s_1}\big)^{-2}+\big(\frac{s}{s_2}\big)^{2.7}\big]$, for $s < s_2$, and $t_c (s,r)= \frac {1}{\tau\Omega} \big(\frac{s}{s_2}\big)^{0.3}$, for $s > s_2$, where $s$ is the particle size, $s_1 = 1.2 s_{\rm blow}$, $s_2 = 100 s_{\rm blow}$,  $\tau$ is the geometrical vertical optical depth, $r$ is the distance to the star, and $\Omega$ is the angular velocity at $r$ (Th\'ebault \& Augereau 2007). 

The size distribution of the collisional debris is expected to evolve with time because the collisional timescale depends on the particle size. Small dust-size particles have shorter lifetimes and they reach collisional equilibrium faster, i.e. their size distribution changes quickly from the primordial one to that of reprocessed material, while larger particles will retain their primordial size distribution for a longer period, creating a "knee" in the size distribution. The critical size at which the transition from a primordial  to a collisional equilibrium population takes place increases with time. The largest bodies may not achieve collisional equilibrium within the age of the system. This will results in a combination  of power-laws for the size distribution, with waves in the smaller end triggered by the dust loss processes. 

\begin{figure}[]
\begin{center}
\includegraphics[width=0.7\textwidth]{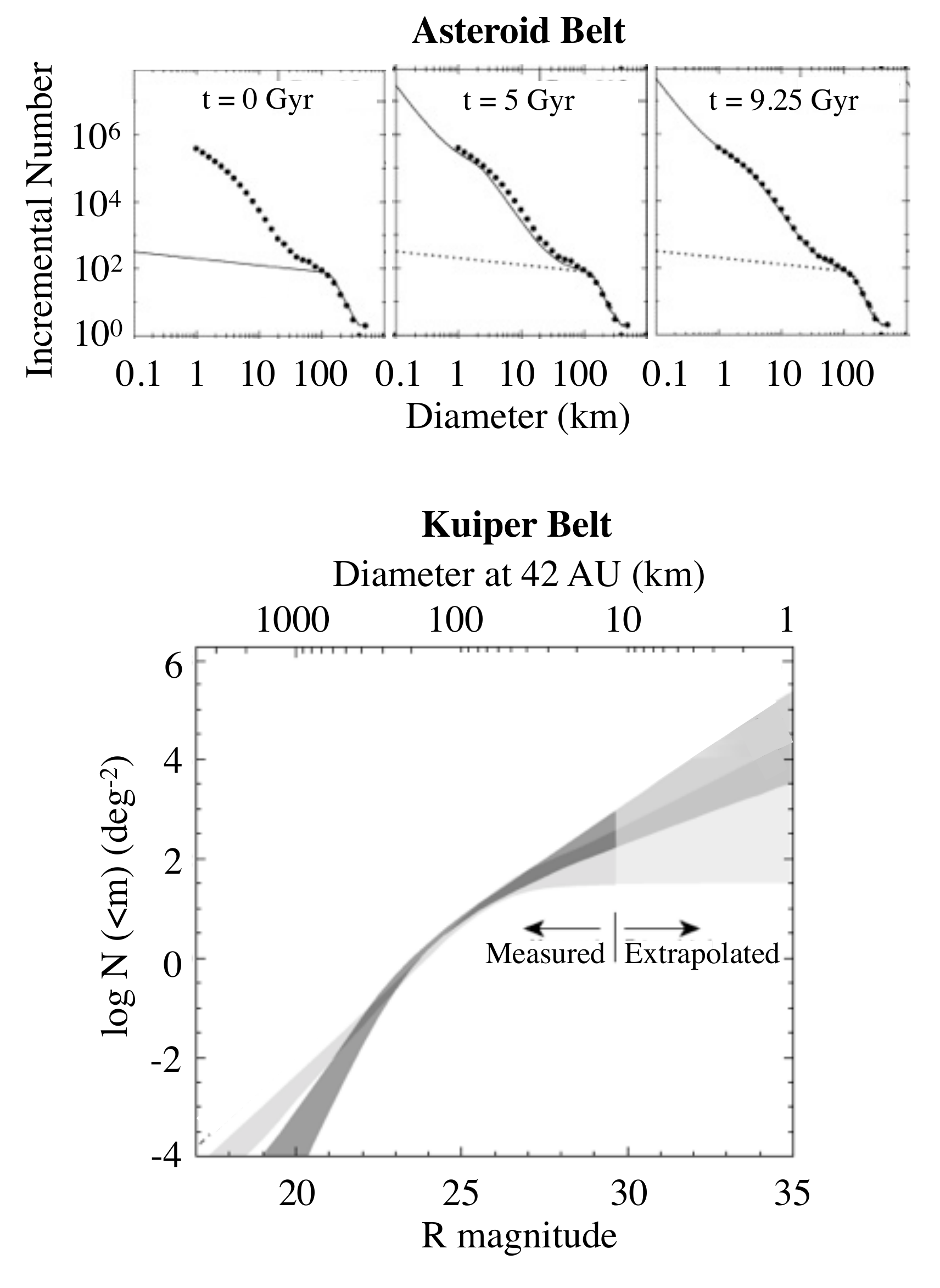}
\caption {(Top): observations (thick dots) and collisional evolution models (thin solid line) of the asteroid size distribution from Bottke et al. (2005). Note that the best model is for $t$ = 9.5 Gyr;  $t$ is a pseudo-time that measures the degree of collisional evolution; the fact that $t$ is greater than the age of the Solar system indicates that the asteroid belt was significantly more massive in the past and that dynamical clearing must have played an important role. (Bottom): cumulative surface density of the KBOs in the classical disk (dark grey) and in the scattered disk (light grey) with 95\% confidence upper and lower bounds (from Bernstein et al. 2004). 
}
\label{bottke_bernstein}
\end{center}
\end{figure}

The size distributions of the small body population in the Solar system (asteroids and KBOs) are partly the result of these processes.  The top panel of Figure \ref{bottke_bernstein} (from Bottke et al. 2005) shows the results of the collisional evolution models of the asteroid belt: as time goes by, the initial size distribution changes from a power-law (solid straight line in the left panel) to the observed wavy distribution (thick dotted line); the observed distribution has two main features, one at $D$$\sim$120 km, leftover from the planetesimal accretion process, and another one at $D$$\sim$200 m, marking the transition between the regime where particles are held by strength forces and the regime where particles are held by self-gravity. The  bottom panel of Figure \ref{bottke_bernstein} (from Bernstein et al. 2004) shows the size distribution of the KBOs; even though this size distribution is more difficult to constrain due to the greater distance  (in particular at the small size end), observations show a strong break to a shallower size distribution at $D$$<$100 km, the boundary where the particles become more susceptible to collisional destruction.  The size distributions of the asteroid and the Kuiper belts indicate that collisions have played an important role in the evolution of the Solar system. 

\subsection{Effect of collisions on the dust spatial distribution}
\label{sec:coll_effect_on_spatial_dist}

When the optical depth of the disk, $\tau$, is large enough so that collisions dominate the dynamics, i.e. when
$t_c \sim \frac{1}{9\tau} \Big(\frac{r}{AU}\Big)^{3/2} < t_{pr} \sim \frac{r^2 c}{4 G M_* \beta}$,
the particles will not be able to migrate far from their parent bodies under P-R drag. This is not the case in the Solar system, but it is the case in many extrasolar debris disks observed to date (Wyatt 2005). Under these conditions, small particles will not be able to spiral in and sweep a wide range of semimajor axes that may allow them to get trapped in MMRs or get ejected by gravitational scattering; on the contrary, these grains will get quickly ground down into smaller particles that radiation pressure sets on eccentric or hyperbolic orbits. On the other hand, large bodies with longer collisional lifetimes will closely trace the spatial distribution of their parent bodies (which could show structure due to the effect of gravitational perturbations).  

Models in Thebault \& Auregearu (2007) show the effect of collisions on a planetesimal belt with an initial radial distribution $\propto r^{-1.5}$, characteristic of the minimum mass solar nebula. They found that the radial distribution of sub-mm size particles (which dominate the disk optical depth) gets significantly flatter because of two reasons: the smallest grains just over the blow-out size are set on eccentric orbits and spend most of their time outside the orbit of their parent planetesimals, and the erosion of the larger 0.05 to 1 mm grains (the source of the smallest particles) proceeds faster in the inner regions. 

\subsection{Effect of collisions on the dust disk evolution}
\label{sec:coll_effect_on_dust_evolution}

The steady state scenario described in Section \ref{sec:coll_effect_on_dust_size} cannot be sustained for an indefinite period of time because the reservoir of large particles that feeds the collisional cascade gets depleted, resulting in a decay of the amount of dust in the system. In the simplest scenario, the dust is derived from the grinding down of planetesimals, the planetesimals are destroyed after one collision, and the number of collisions is proportional to the square of the number of planetesimals, $N$; in this case, $\frac{dN}{dt} \propto N^2$ and $N \propto \frac{1}{t}$. 
In a collisional cascade, the dust production rate, $R_{\rm prod}$, would be proportional to the loss rate of planetesimals times a proportionality constant,  $R_{\rm prod} \propto \frac{dN}{dt}(\frac{s_{\rm dust}}{s_{\rm parent}})^{-3.5}$, where $s_{\rm dust}$ and $s_{\rm parent}$ are the sizes of the dust and the parent bodies, respectively. In this scenario, $\frac{s_{\rm dust}}{s_{\rm parent}}$  is independent of time and one gets $R_{\rm prod} \propto \frac{dN}{dt} \propto  N^2 \propto \frac{1}{t^{2}}$.  One can solve for the amount of dust in the disk in steady state by equating the dust production rate to the dust loss rate, $\it{R}$$_{\rm loss}$. Depending on the number density of dust, $n$, 
there are two different solutions:
(1) If the number density of dust particles is low, the disk is in the P-R drag-dominated regime where the dust loss rate is determined by P-R drag ($t_{c}$ $>$ $t_{\rm pr}$); in this case, the dust loss rate is proportional to the number density of particles, $R_{\rm loss} \propto n$, and from $R_{\rm prod} = R_{\rm loss}$ one gets ${n} \propto \frac{1}{t^{2}}$. (2) If the number density of dust is high, the disk is in the collision-dominated regime, where the main dust removal process is grain-grain collisions ($t_{c}$ $<$ $t_{\rm pr}$); in this case, the dust loss rate is given by  $R_{\rm loss} \propto n^2$, and from $R_{\rm prod} = R_{\rm loss}$ one gets ${n} \propto \frac{1}{t}$, i.e.  both the dust mass and the number of parent bodies (and therefore the total disk mass) decay as $\frac{1}{t}$, with a characteristic timescale that is inversely proportional to the initial disk mass (Dominik \& Decin 2003).  

The Solar system is in the P-R drag-dominated regime, while, due to limited sensitivity of the observations, the majority of the extra-solar debris disks known to date are in the collision-dominated regime; this explains why the intensity of the dust emission of the entire sample of debris disks systems (for similar-type stars, and including stars with a wide range of ages)  roughly follows a $t^{-1}$ decay (see Figure \ref{diskevolution}). More specifically, the decay of the dust mass, the fractional luminosity $L_{dust}$/$L_{star}$ and the thermal excess is better approached by  ${t^{\xi}}$ with $\xi$ = -0.3 to -0.4 (L\"ohne, Krivov \& Rodmann  2008); it deviates from $t^{-1}$ because this corresponds to the assumption that all the bodies have achieved collisional equilibrium, however, the largest planetesimals of some of the debris disks observed (with ages ranging from 10 Myr to 10 Gyr), would not have had enough time to achieve collisional equilibrium within the age of the system. The index depends on the particle strength as a function of particle size and the properties of the largest planetesimals (primordial size distribution, eccentricities and inclinations - the latter two determining their rate of collisions). 

\section{Other Physical Processes}
\label{sec:otherforces}

\subsection{Dust sublimation} 
\label{sec:sublimation}

Silicates dust grains sublimate at $\sim$1500 K, while for icy grains 1 to 100 $\mu$m in size the destruction temperature is $\sim$ 120 K  (the timescale of sublimation increases dramatically above 100 K).  The following is an estimate of the corresponding sublimation distance. If the grain is larger than the peak wavelengths of both the absorbed and emitted radiation, the grain is in the blackbody regime where it emits and absorbs efficiently and its temperature is given by 

\begin{equation}
T_{\rm dust} = 278 \Big(\frac{L_*}{L_\odot}\Big)^{1/4}\Big(\frac{r}{AU}\Big)^{-1/2}~~~{\rm K},
\label{tgrainbb}
\end{equation}

\noindent where $r$ is the heliocentric distance; if the grain is larger than the peak wavelength of the absorbed radiation but smaller than the peak wavelength of the emitted radiation, the grain is in the intermediate size regime and 

\begin{equation}
T_{\rm dust} = 468 \Big(\frac{L_*}{L_\odot}\Big)^{1/5}\Big(\frac{r}{AU}\Big)^{-2/5} (\xi s_{\mu m})^{-1/5}~~~{\rm K},
\label{tgrainis}
\end{equation}

\noindent where $s_{\mu m}$ is the grain radius and $\xi$ $\sim$ 2$\pi$, $\frac{1}{2\pi}$ and 1 for strongly, weakly and moderately absorbing material, respectively (Backman \& Paresce 1993). From Wien's law, the peak wavelength is given by  $\lambda_{\rm peak} = \frac{3669}{T}~~~{\rm \mu m}$. The incident solar radiation ($T_{\odot}$ = 5778 K) peaks at 0.6 $\mu$m, while the emitted radiation would peak at  2.5 $\mu$m if $T_{\rm dust}$ = 1500 K or 30 $\mu$m if $T_{\rm dust}$ = 120 K. In the former case, silicate grains larger than $\sim$ 2.5 $\mu$m would be in the blackbody regime in which case the sublimation temperature of 1500 K would be achieved at a distance of $r \sim$ 0.03 AU. In the latter case, icy grains smaller than 30 $\mu$m would be in the intermediate size regime achieving the destruction temperature of 120 K at 4 AU (for 1 $\mu$m grains) and 2 AU (for 30 $\mu$m  grains). 

\subsection{Lorentz force}
\label{sec:lorentz}

Dust grains are expected to be electrically charged due to the accretion of stellar wind ions and electrons, ionization due to impacts with stellar wind particles, and the ejection of electrons due to UV radiation (the latter process dominating the charging of Solar system dust).  Once charged, the dust grains will be subject to the Lorentz force,  $F_L = q v B$, where $q$ is the electric charge, $v$ is the velocity of the grain with respect to the field, and $B$ is the magnetic field flux density. 

In the case of the Solar system, the grains inside the heliosphere ($<$ 150 AU) are subject to the interplanetary magnetic field. This field has a dipole component that changes polarity every 11 years with the 22 year Solar cycle; near the ecliptic, these sign reversals take place more rapidly because of the presence of the heliospheric current sheet (the extension of the Sun's magnetic equator into interplanetary space, separating regions of opposite polarity). At solar minimum, the current sheet extends from approximately $-25^\circ$ to $25^\circ$ from the solar equator; particles within this latitude range will cross the current sheet at least twice every solar rotation  ($\sim$27 days) or four or even six times if the current sheet is wrapped because of higher order terms in the magnetic field; particles at higher ecliptic latitudes will cross the current sheet at  least twice as they orbit the Sun. These sign reversals will cause a random walk in the semi-major axis of the particles; for particles smaller than a few microns, this random walk will dominate over the P-R drift on timescales from a few orbital periods in the inner Solar system to a few tens of orbital periods in the outer Solar system. In addition to the dipole, the interplanetary magnetic field has a dominant component that is perpendicular to the radial solar wind vector, with a magnitude $\sim \frac {3\cdot10^{-5}}{r (AU)}$ Gauss (for heliocentric distances {\it r} exceeding a few AU); the Lorentz force in this case will tend to scatter the smallest dust particles out of the ecliptic plane by perturbing the particle inclinations while keeping the energy of the orbit unchanged. 

Because, as described above, the circumstellar magnetic field can have a complex structure and time behavior,  it is difficult to include the effect of the Lorentz force in the study of the dynamics and spatial distribution of dust particles in extra-solar debris disks for which the magnetic field properties are unknown. 

\subsection{Sputtering} 
\label{sec:sputtering}

Dust grains erode with time due to the impact of energetic stellar wind particles (a process known as sputtering). The study of exposed lunar rocks allows to estimate the rate of erosion, resulting in values that differ by two orders of magnitude, where a rate of $\sim$ 0.2 \AA~yr$^{-1}$ at 1 AU would be on the high-end; this rate scales with distance as $r^{-2}$.  Dynamical models indicate that a dust particle with Kuiper belt origin has a typical lifetime of $\sim10^7 $ yr, most of which is spent at $a>$20 AU. Assuming the particle spends 10$^7$ yr at 20 AU from the Sun, the fraction of mass loss due to erosion (at the highest measured rate) would be $\sim$ 50\% for a 3 \mum~ particle, scaling as {\it s}$^{-1}$, where $\it{s}$ is the particle radius; if the erosion rate is 100 times smaller (which is within the present uncertainties), the mass loss would be negligible. The erosion rate is uncertain because sputtering can cause chemical alteration on the dust grain surface (via the implantation of stellar wind ions) that can create molecular bondings between layers of dissimilar materials making them erosion-resistant. These chemical alterations may also change the optical properties of the grain (e.g. by producing a blackened highly carbonized and refractory surface layer from organic and volatile grain mantles), which can affect the particle's response to radiation forces and therefore its lifetime. 

\section{Open questions}
\label{sec:open}

There are many open questions in debris disk modeling; some are related to the model input parameters, while others are related to the physics involved and the modeling approach. Regarding the input parameters, the lack of spectral features in most of debris disk spectra make it difficult to constrain the dust properties -- like particle size distribution, composition, shape and porosity -- which determine the particle emission properties and dynamics (because they affect how the particle interacts with radiation forces).  The properties of the parent planetesimals are also unknown, with the particle strength playing a key role  in the dust-producing process via collisions. The origin of the debris dust itself is also uncertain: while a component may originate from steady-state erosion, as the models generally assume, it has now became clear that stochastic collisions are required to explain some of the debris disks observations, however, the origin and ubiquity of stochastic collisions remain unknown; in addition, in some cases the debris dust could also be due to cometary activity rather than planetesimal collisions.  Another caveat is that, in most systems, little is known about the presence of planets, in particular long-period planets and planets with low masses (and even when planets are identified, their masses are generally not well determined); this makes it difficult to constrain the collisional state of the dust-producing planetesimals and the effect of the planets on the dust disk structure. 

But even if the input parameters were known, there are open questions regarding the modeling procedures and the physics involved. Regarding the physics, collisions are the most difficult to account for because they take place across 12 orders of magnitude and involve a wide range of relative velocities and incoming angles, from head-on to grazing collisions, with their outcome ranging from particle growth, to cratering, to complete disruption.  

Regarding the modeling, while the N-body approach (like that in Figures \ref{tp_evolution} and \ref{KBdust}) follows the trajectory of individual particles and can take into account the effect of planetary perturbations, radiation forces, gas drag and the interstellar medium, the CPU power is limited and such models cannot treat a number of particles sufficiently large to cover a wide range of particle sizes; therefore, the N-body approach cannot model the particle size distribution. On the contrary, statistical methods (like those in Figures \ref{lohne} and \ref{thebault_18_19}), where the particles are replace with packages with given distributions, allow to study the outcome of collisions and the size distributions but, due to the averaging over angular orbital elements, they lose accuracy in the dynamical modeling and cannot study in detail the spatial distribution of the dust.

  



\begin{thebibliography}{99.}

\bibitem{} 
A'Hearn, M. F. \ 2008: 
Deep Impact and the Origin and Evolution of Cometary Nuclei.\ 
Space Sci. Rev., 138, 237

\bibitem{2007JGRA..11207105A} Altobelli, N., Dikarev, V., Kempf, S., Srama, 
R., Helfert, S., Moragas-Klostermeyer, G., Roy, M., 
\& Gr{\"u}n, E.\ 2007: Cassini/Cosmic Dust Analyzer in situ dust measurements between Jupiter and Saturn.\ Journal of Geophysical Research (Space Physics), 112, 7105 

\bibitem{} 
Aumann H. H., Beichman C. A., Gillett F. C., de Jong T., Houck J. R. et al. \ 1984: 
Discovery of a shell around Alpha Lyrae.\
AJ, 278, L23

\bibitem{1993prpl.conf.1253B} 
Backman, D.~E., \& Paresce, F.\ 1993: 
Main-sequence stars with circumstellar solid material - The VEGA phenomenon.\ 
In {\it Protostars and Planets III}  (E. H. Levy and J. I. Lunine, eds.), University of Arizona Press, Tucson, p. 1253

\bibitem{1995ApJ...450L..35B} 
Backman, D.~E., Dasgupta, A., \& Stencel, R.~E.\ 1995: 
Model of a Kuiper Belt Small Grain Population and Resulting Far-Infrared Emission. \ 
ApJL, 450, L35 

\bibitem{2005ApJ...626.1061B} 
Beichman, C. A., Bryden, G., Gautier, T. N., Stapelfeldt, K. R.; Werner, M. et al.\ 2005: 
An Excess Due to Small Grains around the Nearby K0 V Star HD 69830: Asteroid or Cometary Debris?.\ 
ApJ, 626, 1061 

\bibitem{2006ApJ...639.1166B}
Beichman C. A., Tanner, A., Bryden G., Stapelfeldt K. R., Werner, M. W. et al. \ 2006:
IRS Spectra of Solar-Type Stars: A Search for Asteroid Belt Analogs.\
ApJ, 639, 1166

\bibitem{} Benz, W., \& Asphaug, E.\ 1999:
Catastrophic Disruptions Revisited. 
Icarus, 142, 5 

\bibitem{2004AJ....128.1364B} 
Bernstein, G.~M., Trilling, D.~E., Allen, R.~L., Brown, M.~E., Holman, M., \& Malhotra, R.\ 2004: 
The Size Distribution of Trans-Neptunian Bodies.\ 
AJ, 128, 1364 

\bibitem{2009MNRAS.399..385B} 
Booth, M., Wyatt, M.~C., Morbidelli, A., Moro-Mart{\'{\i}}n, A., \& Levison, H.~F.\ 2009: 
The history of the Solar system's debris disc: observable properties of the Kuiper belt.\ 
MNRAS, 399, 385 

\bibitem{2005Icar..175..111B} 
Bottke, W.~F., Durda, D.~D., Nesvorn{\'y}, D., Jedicke, R., Morbidelli, A., Vokrouhlick{\'y}, D., \& Levison, H.\ 2005:
The fossilized size distribution of the main asteroid belt.\ 
Icarus, 175, 111 

\bibitem{2006Sci...314.1711B} 
Brownlee, D., Tsou, P., Al\'eon, J., Alexander, C.,  Araki, T. et al.\ 2006: 
Comet 81P/Wild 2 Under a Microscope.\ 
Science, 314, 1711 

\bibitem{2006ApJ...636.1098B} 
Bryden, G., Beichman, C. A., Trilling, D. E., Rieke, G. H., Holmes, E. K. et al.\ 2006: 
Frequency of Debris Disks around Solar-Type Stars: First Results from a Spitzer MIPS Survey.\ 
ApJ, 636, 1098 

\bibitem{2009ApJ...705.1226B} 
Bryden, G., Beichman, C. A., Carpenter, J. M., Rieke, G. H., Stapelfeldt, K. R et al.\ 2009: 
Planets and Debris Disks: Results from a Spitzer/MIPS Search for Infrared Excess.\ 
ApJ, 705, 1226 

\bibitem{1979Icar...40....1B} 
Burns, J.~A., Lamy, P.~L., \& Soter, S.\ 1979: 
Radiation forces on small particles in the solar system.\ 
Icarus, 40, 1 

\bibitem{2009ApJS..181..197C} 
Carpenter, J. M., Bouwman, J., Mamajek, E. E., Meyer, M. R., Hillenbrand, L. A. et al.\ 2009: 
Formation and Evolution of Planetary Systems: Properties of Debris Dust Around Solar-Type Stars.\ 
ApJS, 181, 197 

\bibitem{2007Icar..189..233C} 
Chapman, C.~R., Cohen, B.~A., \& Grinspoon, D.~H.\ 2007: 
What are the real constraints on the existence and magnitude of the late heavy bombardment? \  
Icarus, 189, 233

\bibitem{2006ApJS..166..351C} 
Chen, C. H., Sargent, B. A., Bohac, C., Kim, K. H., Leibensperger, E. et al.\ 2006: 
Spitzer IRS Spectroscopy of IRAS-discovered Debris Disks.\ 
ApJS, 166, 351 

\bibitem{2009ApJ...693..734C} 
Chiang, E., Kite, E., Kalas, P., Graham, J.~R., \& Clampin, M.\ 2009: 
Fomalhaut's Debris Disk and Planet: Constraining the Mass of Fomalhaut b from disk Morphology.\ 
ApJ, 693, 734 

\bibitem{1990Natur.343..129C} 
Chyba, C.~F.\ 1990: 
Impact delivery and erosion of planetary oceans in the early inner solar system.\ 
Nature, 343, 129 

\bibitem{} 
Clampin M., Krist J. E., Ardila D. R., Golimowski D. A., Hartig G. F. et al. \ 2003:
Hubble Space Telescope ACS Coronagraphic Imaging of the Circumstellar Disk around HD 141569A.\ 
Astron. J., 126, 385 

\bibitem{2008ApJ...673L.191D} 
Debes, J.~H., Weinberger, A.~J., \& Schneider, G.\ 2008: 
Complex Organic Materials in the Circumstellar Disk of HR 4796A.\ 
ApJL, 673, L191 

\bibitem{} 
Decin, G., Dominik, C., Waters, L.~B.~F.~M., \& Waelkens, C.\ 2003:
Age Dependence of the Vega Phenomenon: Observations.\
ApJ, 598, 636 

\bibitem{} 
Dermott, S. F., Durda, D.D., Gustafson, B. A. S., Jayaraman, S., Liou, J. C., Xu, Y. L.,\ 1994: 
Zodiacal dust bands. In: Asteroids, Comets, Meteors 1993 (Milani, A., et al., eds.), pp. 127-142.

\bibitem{1994Natur.369..719D} 
Dermott, S.~F., Jayaraman, S., Xu, Y.~L.,  Gustafson, B.~{\AA}.~S.,  \& Liou, J.~C.\ 1994: 
A circumsolar ring of asteroidal dust in resonant lock with the Earth.\ 
Nature, 369, 719 

\bibitem{} 
Dermott, S. F., Grogan, K., Durda, D. D., Jayaraman, S., Kehoe, T. J. J. et al. \ 2001:
Orbital Evolution of Interplanetary Dust:
In {\it Interplanetary Dust} (E. Gr\"un,  B. A. S. Gustafson, S. F. Dermott, H. Fechtig, eds.), Springer A\&A Library, p. 295

\bibitem{2002ESASP.500..319D} 
Dermott, S.~F., Kehoe, T.~J.~J., Durda, D.~D., Grogan, K., \& Nesvorn{\'y}, D.\ 2002: 
Recent rubble-pile origin of asteroidal solar system dust bands and asteroidal interplanetary dust particles.\ 
Asteroids, Comets, and Meteors: ACM 2002, 500, 319 

\bibitem{1969JGR....74.2431D} 
Dohnanyi, J. S. \ 1969: Collisional Model of Asteroids and Their Debris.\ 
J. Geophys. R., 74, 2431

\bibitem{2003ApJ...598..626D} 
Dominik C. and Decin G. \ 2003:  
Age Dependence of the Vega Phenomenon: Theory. \ 
Astrophys. J., 598, 626

\bibitem{} 
Eiroa, C., Marshall, J.~P., Mora, A., et al.\ 2011:
Herschel discovery of a new class of cold, faint debris discs.
A\&A, 536, L4 

\bibitem{}
Farley, K. A. \ 1995:\
Cenozoic variations in the flux of interplanetary dust recorded by 3He in deep sea sediments.\
Nature, 376, 153

\bibitem{}
Fischer D. A. \& Valenti J. 2005: \\
The Planet-Metallicity Correlation. 
ApJ, 622, 1102

\bibitem{1999ApJ...525..492F} 
Frisch, P.~C., et al.\ 1999: 
Dust in the  Local Interstellar Wind.\ 
ApJ, 525, 492 

\bibitem{2007ApJ...667..527G} 
Gautier, T. N., III,  Rieke, G. H., Stansberry, J., Bryden, G. C., Stapelfeldt, Karl R. et al.\ 2007: 
Far-Infrared Properties of M Dwarfs.\ 
ApJ, 667, 527 

\bibitem{2006AJ....131.3109G}	
Golimowski, D. A., Ardila, D. R., Krist, J. E., Clampin, M., Ford, H. C. et al. \ 2006:	
Hubble Space Telescope ACS Multiband Coronagraphic Imaging of the Debris Disk around $\beta$-Pictoris. \
AJ, 131, 3109

\bibitem{2005Natur.435..466G} 
Gomes, R., Levison, H.~F., Tsiganis, K., \& Morbidelli, A.\ 2005: 
Origin of the cataclysmic Late Heavy Bombardment period of the terrestrial planets.\ 
Nature, 435, 466 

\bibitem{1997GeoRL..24.3125G} 
Gurnett, D.~A., Ansher, J.~A., Kurth, W.~S. Granroth, L. \ 1997: 
Micron-sized dust particles detected in the outer solar system by the Voyager 1 and 2 plasma wave instruments.\ 
Geoph. R. Lett., 24, 3125 

\bibitem{} 
Greaves J. S., Holland, W. S., Moriarty-Schieven G., Jenness, T., Dent, W. R. F. et al. \ 1998:  
A Dust Ring around $\epsilon$-Eridani: Analog to the Young Solar System.\  
ApJ, 506, L133

\bibitem{} 
Greaves, J.~S., Holland, W.~S., Jayawardhana, R., Wyatt, M.~C., 
\& Dent, W.~R.~F.\ 2004:
A search for debris discs around stars with giant planets.
MNRAS, 348, 1097 

\bibitem{2005ApJ...619L.187G} 
Greaves, J.~S., et al.\ 2005: 
Structure in the {$\epsilon$} Eridani Debris Disk.\ 
ApJ, 619, L187 

\bibitem{}
Greaves J. S., Fischer D. A. \& Wyatt M. C. 2006: \
Metallicity, debris discs and planets. 
MNRAS, 366, 283

\bibitem{1994A&A...286..915G} 
Gr\"un, E., Gustafson, B., Mann, I., Baguhl, M., Morfill, G.~E., Staubach, P., Taylor, A., \& Zook, H.~A.\ 1994: 
Interstellar dust in the heliosphere.\ 
A\&A, 286, 915 

\bibitem{} 
Gr\"un, Baguhl, M., Svedhem, H. \& Zook, H. A. \ 2001: 
In situ measurements of cosmic dust. \ 
In {\it Interplanetary Dust} (E. Gr\"un,  B. A. S. Gustafson, S. F. Dermott, H. Fechtig, eds.), Springer A\&A Library, p. 295

\bibitem{} 
Gustafson, B. A. S. \ 1994: 
Physics of Zodiacal Dust. \ 
Ann. Rev. Earth Planet. Sci., 22, 553

\bibitem{} 
Gustafson, B. A. S., Greenbert, J. M., Kolokolova, L., u, Y., Stognienko, R. \ 2001:
Interactions with Electromagnetic Radiation: Theory and Laboratory Simulations. 
In {\it Interplanetary Dust} (E. Gr\"un,  B. A. S. Gustafson, S. F. Dermott, H. Fechtig, eds.), Springer A\&A Library, p. 57

\bibitem{} 
Habing, H.~J., Dominik, C., Jourdain de Muizon, M., et al.\ 2001:
Incidence and survival of remnant disks around main-sequence stars. 
A\&A, 365, 545 

\bibitem{2000ApJ...539..435H} 
Heap, S.~R., Lindler, D.~J., Lanz, T.~M.,  Cornett, R.~H., Hubeny, I., Maran, S.~P., \& Woodgate, B.\ 2000: 
Space Telescope Imaging Spectrograph Coronagraphic Observations of {$\beta$} Pictoris.\ 
ApJ, 539, 435 

\bibitem{2008ApJ...677..630H} 
Hillenbrand, L. A., Carpenter, J. M., Kim, J. S. Meyer, M. R., Backman, D. E. et al.\ 2008: 
The Complete Census of 70 {$\mu$}m-bright Debris Disks within ``the Formation and Evolution of Planetary Systems'' Spitzer Legacy Survey of Sun-like Stars.\ 
ApJ, 677, 630 

\bibitem{1998Natur.392..788H} 
Holland, W. S., Greaves, J. S., Zuckerman, B., Webb, R. A., McCarthy, C. et al.\ 1998: 
Submillimetre images of dusty debris around nearby stars.\ 
Nature, 392, 788 

\bibitem{2004ApJS..154...18H} 
Houck, J. R., Roellig, T. L., van Cleve, J., Forrest, W. J., Herter, T. et al.\ 2004: 
The Infrared Spectrograph (IRS) on the Spitzer Space Telescope.\ 
ApJS, 154, 18 

\bibitem{} 
Humes, D. \ 1980: Results of Pioneer 10 and 11 meteoroid experiments - Interplanetary and near-Saturn.\  
J. Geophys. R.,  85 (A/II), 5841

\bibitem{}
Jessberger, E. K., Stephan, T., Rost, D., Arndt, P., Maetz, M. et al. \ 2001:
Properties of Interplanetary Dust: Information from Collected Samples. \
In {\it Interplanetary Dust} (E. Gr\"un,  B. A. S. Gustafson, S. F. Dermott, H. Fechtig, eds.), Springer A\&A Library, p. 253

\bibitem{1993Natur.362..730J} 
Jewitt, D., \& Luu, J.\ 1993: Discovery of the candidate Kuiper belt object 1992 QB1.\ 
Nature, 362, 730 

\bibitem{2000prpl.conf.1201J} 
Jewitt, D.~C., \& Luu, J.~X.\ 2000: 
Physical Nature of the Kuiper Belt.\ 
Protostars and Planets IV, 1201 

\bibitem{2010Natur.467..817J} 
Jewitt, D., Weaver, H., Agarwal, J., Mutchler, M.,  \& Drahus, M.\ 2010: 
A recent disruption of the main-belt asteroid P/2010A2.\ 
Nature, 467, 817 

\bibitem{2006ApJ...653..613J} 
Jura, M.\ 2006: 
Carbon Deficiency in Externally Polluted White Dwarfs: Evidence for Accretion of Asteroids. \ 
ApJ, 653, 613

\bibitem{2007AJ....133.1927J} 
Jura, M., Farihi, J., Zuckerman, B., \& Becklin, E.~E.\ 2007: 
Infrared Emission from the Dusty Disk Orbiting GD 362, an Externally Polluted White Dwarf.\ 
AJ, 133, 1927 

\bibitem{2005Natur.435.1067K} 
Kalas, P., Graham, J.~R.,  \& Clampin, M.\ 2005: 
A planetary system as the origin of structure in Fomalhaut's dust belt.\ 
Nature, 435, 1067 

\bibitem{2006ApJ...637L..57K}	
Kalas, P., Graham, J. R., Clampin, M. C. \& Fitzgerald, M. P. 2006: \	
First Scattered Light Images of Debris Disks around HD 53143 and HD 139664.\
Astrophys. J., 637, 57

\bibitem{2008Sci...322.1345K} 
Kalas, P., et al.\ 2008: 
Optical Images of an Exosolar Planet 25 Light-Years from Earth.\ 
Science, 322, 1345 

\bibitem{2005AJ....130..269K} 
Kenyon, S.~J., \& Bromley, B.~C.\ 2005: 
Prospects for Detection of Catastrophic Collisions in Debris Disks.\ 
AJ, 130, 269 

\bibitem{}
K{\'o}sp{\'a}l, {\'A}., Ardila, D.~R., Mo{\'o}r, A., \& {\'A}brah{\'a}m, P.\ 2009: 
On the Relationship Between Debris Disks and Planets.\
ApJ, 700, L73 

\bibitem{} 
Krist J. E., Ardila D. R., Golimowski D. A., Clampin M. \& Ford H. C. \ 2005: 
Hubble Space Telescope Advanced Camera for Surveys Coronagraphic Imaging of the AU Microscopii Debris Disk.
Astron. J., 129, 1008

\bibitem{} 
Krivov, A.~V., Srem{\v c}evi{\'c}, M., \& Spahn, F.\ 2005: 
Evolution of a Keplerian disk of colliding and fragmenting particles: a kinetic model with application to the Edgeworth Kuiper belt. 
Icarus, 174, 105 

\bibitem{2010RAA....10..383K} 
Krivov, A.~V.\ 2010: 
Debris disks: seeing  dust, thinking of planetesimals and planets.\ 
Research in Astronomy and Astrophysics, 10, 383 

\bibitem{2010Sci...329...57L} 
Lagrange, A.-M., et al.\ 2010: 
A Giant Planet Imaged in the Disk of the Young Star {$\beta$} Pictoris.\ 
Science, 329, 57 

\bibitem{2002AJ....123.2857L} 
Landgraf, M., Liou, J.-C., Zook, H.~A., \& Gr{\"u}n, E.\ 2002: 
Origins of Solar System Dust beyond Jupiter.\ 
AJ, 123, 2857 

\bibitem{} Leinhardt, Z. M. \& Stewart, S. T. \ 2009:
Full numerical simulations of catastrophic small body collisions. \
Icarus, 199, 542

\bibitem{} 
Levasseur-Regourd, A. C., Mann, I., Dumont, R. \& Hanner, M.  \ 2001: 
Optical and Thermal Properties of Interplanetary Dust. \
In {\it Interplanetary Dust} (E. Gr\"un,  B. A. S. Gustafson, S. F. Dermott, H. Fechtig, eds.), Springer A\&A Library, p. 57

\bibitem{2010arXiv1012.1570L} 
Li, J., \& Jewitt, D., Clover, J. M. \& Jackson, B. V. \ 2010: 
Outburst of Comet 17P/Holmes Observed With The Solar Mass Ejection Imager.\ 
arXiv:1012.1570

\bibitem{1999AJ....118..580L} 
Liou, J.-C.,  \& Zook, H.~A.\ 1999: 
Signatures of the Giant Planets Imprinted on the Edgeworth-Kuiper Belt Dust Disk.\ 
AJ, 118, 580 

\bibitem{2007ApJ...658..584L} 
Lisse, C.~M., Beichman, C.~A., Bryden, G., \& Wyatt, M.~C.\ 2007: 
On the Nature of the Dust in the Debris Disk around HD 69830.\ 
ApJ, 658, 584 

\bibitem{2008ApJ...673.1123L} 
L\"ohne, T., Krivov, A. V. \& Rodmann, J. \ 2008:  
Long-Term Collisional Evolution of Debris Disks.\ Astrophys. J., 673, 1123

\bibitem{2006Natur.441..305L} 	
Lovis, C., Mayor, M., Pepe, F., Alibert, Y., Benz, W.  et al.\ 2006: 
An extrasolar planetary system with three Neptune-mass planets.\ 
Nature, 441, 305 

\bibitem{1993Natur.365..819M} 
Malhotra, R.\ 1993: 
The origin of Pluto's peculiar orbit.\ 
Nature, 365, 819 

\bibitem{2005PThPS.158...24M} 
Marcy, G., Butler, R.~P., Fischer, D., Vogt, S., Wright, J.~T. et al.\ 2005: 
Observed Properties of Exoplanets: Masses, Orbits, and Metallicities.\ 
Progress of Theoretical Physics Supplement, 158, 24 

\bibitem{2008Sci...322.1348M} 
Marois, C., Macintosh, B., Barman, T.,  Zuckerman, B., Song, I. et al.\ 2008: 
Direct Imaging of Multiple Planets Orbiting the Star HR 8799.
Science, 322, 1348 

\bibitem{1977ApJ...217..425M} 
Mathis, J.~S., Rumpl, W., \& Nordsieck, K.~H. \ 1977:  
The size distribution of interstellar grains.\  
Astrophys. J., 217, 425

\bibitem{}
Maurette, M., Immel, G., Hammer, C., Harvey, R., Kurat, G. \& Taylor, S. \ 1994:
Collection and curation of IDPs from the Greenland and antarctic ice Sheets. \ 
In {\it Analysis of Interplanetary Dust} (M. E. Zolensky, T. L. Wilson, F. J. M. Rietmeijer and G. J. Flynn, eds.), Amer. Inst. Physics, pp. 277-289

\bibitem{}
Maurette, M., Engrand, C. \& Kurat, G. \ 1996:
Collection and Microanalysis of Antarctic Micrometeorites. 
In {\it Physics, Chemistry and Dynamics of Interplanetary Dust} (B. A. S. Gustafson and M. S. Hanner), ASP Conf. Ser., Vol. 104, pp. 265-273

\bibitem{} 
Mayor, M., Marmier, M.,  Lovis, C., Udry, S. SŽgransan, D., Pepe, F., Benz, W., Bertaux, J. -L., Bouchy, F., Dumusque, X., Lo Curto, G., Mordasini, C., Queloz, D., Santos, N. C.\ 2011: 
The HARPS search for southern extra-solar planets XXXIV. Occurrence, mass distribution and orbital properties of super-Earths and Neptune-mass planets. 
arXiv:1109.2497 (A\&A in press)

\bibitem{} 
McDonnell, T., McBride, N., Green, S. F., Ratcliff, P. R. et al. \ 2001:
Near Earth Environment. \
In {\it Interplanetary Dust} (E. Gr\"un,  B. A. S. Gustafson, S. F. Dermott, H. Fechtig, eds.), Springer A\&A Library, p. 163

\bibitem{2006Sci...314.1724M}
McKeegan, , K. D., Al\'eon, J. B., Bradley, J., Brownlee, D., Busemann, H. A et al.\ 2006:
Isotopic Compositions of Cometary Matter Returned by Stardust.\
Science, 314, 1724

\bibitem{2008ApJ...673L.181M} 
Meyer, M. R., Carpenter, J. M., Mamajek, E. E., Hillenbrand, L. A., Hollenbach, D. et al.\ 2008: 
Evolution of Mid-Infrared Excess around Sun-like Stars: Constraints on Models of Terrestrial Planet Formation.\ 
ApJL, 673, L181 

\bibitem{2009ApJ...699.1067M} 
Morales, F. Y., Werner, M. W., Bryden, G., Plavchan, P., Stapelfeldt, K. et al.\ 2009: 
Spitzer Mid-IR Spectra of Dust Debris Around A and Late B Type Stars: Asteroid Belt Analogs and Power-Law Dust Distributions.\ 
ApJ, 699, 1067 

\bibitem{2002AJ....124.2305M} 
Moro-Mart{\'{\i}}n, A.,  \& Malhotra, R.\ 2002: 
A Study of the Dynamics of Dust from the Kuiper Belt: Spatial Distribution and Spectral Energy Distribution.\ 
AJ, 124, 2305 

\bibitem{2003AJ....125.2255M} 
Moro-Mart{\'{\i}}n, A.,  \& Malhotra, R.\ 2003: 
Dynamical Models of Kuiper Belt Dust in the Inner and Outer Solar System.\ 
AJ, 125, 2255 

\bibitem{2005ApJ...621.1079M} 
Moro-Mart{\'{\i}}n, A., Wolf, S.,  \& Malhotra, R.\ 2005: 
Signatures of Planets in Spatially Unresolved Debris Disks.\ 
ApJ, 621, 1079 

\bibitem{2005ApJ...633.1150M} 
Moro-Mart{\'{\i}}n, A.,  \& Malhotra, R.\ 2005: 
Dust Outflows and Inner Gaps Generated by Massive Planets in Debris Disks.\ 
ApJ, 633, 1150 

\bibitem{2007ApJ...658.1312M} 
Moro-Mart{\'{\i}}n, A., Carpenter, J. M., Meyer, M. R., Hillenbrand, L. A., Malhotra, R. et al.\ 2007a: 
Are Debris Disks and Massive Planets Correlated?.\ 
ApJ, 658, 1312 

\bibitem{2007ApJ...668.1165M} 
Moro-Mart{\'{\i}}n, A., et al.\ 2007b: 
The Dust, Planetesimals, and Planets of HD 38529.\ 
ApJ, 668, 1165 

\bibitem{} 
Moro-Mart{\'{\i}}n, A., Wyatt, M.~C., Malhotra, R., \& Trilling, D.~E.\ 2008: 
Extra-Solar Kuiper Belt Dust Disks. 
In {\it The Solar System Beyond Neptune} (A. Barucci, H. Boehnhardt, D. Cruikshank and A. Morbidelli, eds.), University of Arizona Press,  Tucson, pp. 465--482 (arXiv:astro-ph/0703383)


\bibitem{2010ApJ...717.1123M} 
Moro-Mart{\'{\i}}n, A., Malhotra, R., Bryden, G., Rieke, G.~H., Su, K.~Y.~L., Beichman, C.~A., \& Lawler, S.~M.\ 2010: 
Locating Planetesimal Belts in the Multiple-planet Systems HD 128311, HD 202206, HD 82943, and HR 8799.\ 
ApJ, 717, 1123 

\bibitem{} 
Mouillet, D., Larwood, J.~D., Papaloizou, J.~C.~B., \& Lagrange, A.~M.\ 1997: 
A planet on an inclined orbit as an explanation of the warp in the Beta Pictoris disc.
MNRAS, 292, 896 

\bibitem{}
Mukai, T., Blum, J., Nakamura, A. M., Johnson R. E., Havnes, O. 2001:
Physical Processes on Interplanetary Dust, 
In {\it Interplanetary Dust} (E. Gr\"un,  B. A. S. Gustafson, S. F. Dermott, H. Fechtig, eds.), Springer A\&A Library, p. 445

\bibitem{2010ApJ...708.1728M} 
M{\"u}ller, S., L{\"o}hne, T., 
\& Krivov, A.~V.\ 2010: The Debris Disk of Vega: A Steady-state Collisional Cascade, Naturally.\
ApJ, 708, 1728 

\bibitem{}
Murray, C. D. \& Dermott, S. F.  1999, 
Solar System Dynamics 
(Cambridge: Cambridge Univ. Press)

\bibitem{2010ApJ...713..816N} 
Nesvorn{\'y}, D., Jenniskens, P., Levison, H.~F., Bottke, W.~F., Vokrouhlick{\'y}, D., \& Gounelle, M.\ 2010: 
Cometary Origin of the Zodiacal Cloud and Carbonaceous Micrometeorites. Implications for Hot Debris Disks.\ 
ApJ, 713, 816 

\bibitem{2005ApJ...631.1161P} 
Plavchan, P., Jura, M., \& Lipscy, S.~J.\ 2005: 
Where Are the M Dwarf Disks Older Than 10 Million Years?.\ 
ApJ, 631, 1161 

\bibitem{2006MNRAS.372L..14Q} 
Quillen, A.~C.\ 2006: 
Predictions for a  planet just inside Fomalhaut's eccentric ring.\ 
MNRAS, 372, L14 

\bibitem{} 
Raymond, S.~N., Armitage, P.~J., Moro-Mart{\'{\i}}n, A., Booth, M., Wyatt, M. C., Armstrong, J. C., Mandell, A. M., Selsis, F., West, A. A. 2011:\ 
Debris disks as signposts of terrestrial planet formation.\
A\&A, 530, A62 

\bibitem{} 
Raymond, S.~N., Armitage, P.~J., Moro-Mart{\'{\i}}n, A., Booth, M., Wyatt, M. C., Armstrong, J. C., Mandell, A. M., Selsis, F., West, A. A. 2012:\ 
Debris disks as signposts of terrestrial planet formation. II Dependence of exoplanet architectures on giant planet and disk properties.
arXiv:1201.3622 (A\&A in press)

\bibitem{1995Natur.374..521R} 
Reach, W. T., Franz, B. A., Weiland, J. L., Hauser, M. G., Kelsall, T. N.  et al.\ 1995: 
Observational confirmation of a circumsolar dust ring by the COBE satellite.\ 
Nature, 374, 521 

\bibitem{2003Icar..164..384R} 
Reach, W.~T., Morris, P., Boulanger, F., \& Okumura, K.\ 2003: 
The mid-infrared spectrum of the zodiacal and exozodiacal light. \ 
 Icarus, 164, 384 

\bibitem{2004ApJS..154...25R} 
Rieke, G. H., Young, E. T., Engelbracht, C. W., Kelly, D. M., Low, F. J. et al.\ 2004: 
The Multiband Imaging Photometer for Spitzer (MIPS).\ 
ApJS, 154, 25 

\bibitem{2006Sci...314.1720S}
Sandford, S. A.; Al\'eon, J., Alexander, C.,  Araki, T., Bajt, S. et al.\ 2006:
Organics Captured from Comet 81P/Wild 2 by the Stardust Spacecraft.\
Science, 314, 1720

\bibitem{2005ApJ...629L.117S} 
Schneider, G., Silverstone, M.~D., \& Hines, D.~C.\ 2005: 
Discovery of a Nearly Edge-on Disk around HD 32297.\
ApJ, 629, L117 

\bibitem{2007ApJ...654..580S} 
Siegler, N., Muzerolle, J., Young, E.~T., Rieke, G.~H., Mamajek, E.~E., Trilling, D.~E., Gorlova, N., \& Su, K.~Y.~L.\ 2007: 
Spitzer 24 {$\mu$}m Observations of Open Cluster IC 2391 and Debris Disk Evolution of FGK Stars.\ 
ApJ, 654, 580 

\bibitem{2004ApJS..154..458S} 
Stapelfeldt, K. R., Holmes E. K., Chen C., Rieke G. H. and Su K. Y. L. et al. \ 2004: 
First Look at the Fomalhaut Debris Disk with the Spitzer Space Telescope.\ 
ApJS, 154, 458 

\bibitem{2009ApJ...707..543S} 
Stark, C.~C., \& Kuchner, M.~J.\ 2009: 
A New Algorithm for Self-consistent Three-dimensional Modeling of Collisions in Dusty Debris Disks.\ 
ApJ, 707, 543 

\bibitem{1996A&A...310..999S}
Stern, S.~A.\ 1996:
Signatures of collisions in the Kuiper Disk.\
A\&A, 310, 999

\bibitem{2005Sci...309.1847S} 
Strom, R.~G., Malhotra, R., Ito, T., Yoshida, F., \& Kring, D.~A.\ 2005: 
The Origin of Planetary Impactors in the Inner Solar System.\ 
Science, 309, 1847 

\bibitem{2005ApJ...628..487S} 
Su, K. Y. L., Rieke, G. H., Misselt, K. A., Stansberry, J. A., Moro-Mart\'in, A. et al.\ 2005: 
The Vega Debris Disk: A Surprise from Spitzer.\ 
ApJ, 628, 487 

\bibitem{2006ApJ...653..675S} 
Su, K. Y. L., Rieke, G. H., Stansberry, J. A., Bryden, G., Stapelfeldt, K. R.  et al.\ 2006: 
Debris Disk Evolution around A Stars.\ 
ApJ, 653, 675 

\bibitem{2009ApJ...705..314S} 
Su, K. Y. L., Rieke, G. H., Stapelfeldt, K. R., Malhotra, R., Bryden, G. et al.\ 2009: 
The Debris Disk Around HR 8799.\ 
ApJ, 705, 314 

\bibitem{1986Icar...65...51S} 
Sykes, M.~V., \& Greenberg, R.\ 1986: 
The formation and origin of the IRAS zodiacal dust bands as a consequence of single collisions between asteroids.\ 
Icarus, 65, 51 

\bibitem{2007A&A...472..169T} 
Th{\'e}bault, P., \& Augereau, J.-C.\ 2007: 
Collisional processes and size distribution in spatially extended debris discs.\ 
A\&A, 472, 169 

\bibitem{2008ApJ...674.1086T} 
Trilling, D.~E., et al.\ 2008: 
Debris Disks around Sun-like Stars.\ 
ApJ, 674, 1086 

\bibitem{} 
Vitense, C., Krivov, A.~V., Kobayashi, H., {\ L\"o}hne, T.\ 2012:
An improved model of the Edgeworth-Kuiper debris disk.\
A\&A, 540, A30 

\bibitem{} 
Weingartner J. C. and Draine B. T. \ 2001:  
Dust Grain-Size Distributions and Extinction in the Milky Way, Large Magellanic Cloud, and Small Magellanic Cloud. \
ApJ, 548, 296

\bibitem{1980AJ.....85.1122W} 
Wisdom, J.\ 1980: 
The resonance overlap criterion and the onset of stochastic behavior in the restricted three-body problem.\ 
AJ, 85, 1122 

\bibitem{1999ApJ...527..918W} 
Wyatt, M.~C., Dermott, S.~F., Telesco, C.~M.,  Fisher, R.~S., Grogan, K., Holmes, E.~K.,  \& Pi{\~n}a, R.~K.\ 1999:\
How Observations of Circumstellar Disk Asymmetries Can Reveal Hidden Planets: Pericenter Glow and Its Application to the HR 4796 Disk.\ 
ApJ, 527, 918 

\bibitem{}
Wyatt M. C. 2005: \
The insignificance of P-R drag in detectable extrasolar planetesimal belts. 
A\&A, 433, 1007

\bibitem{2007ApJ...658..569W} 
Wyatt, M.~C., Smith, R., Greaves, J.~S., Beichman, C.~A., Bryden, G., \& Lisse, C.~M.\ 2007: 
Transience of Hot Dust around Sun-like Stars.\ 
ApJ, 658, 569 

\bibitem{}
Wyatt, M.~C. 2008a: 
Dynamics of small bodies in planetary systems. 
In {\it Small Bodies in Planetary Systems}, Lect. Notes Phys. 758, Springer (arXiv:astro-ph/0807.1272) 

\bibitem{2008ARA&A..46..339W} 
Wyatt, M.~C.\ 2008b: Evolution of Debris Disks.\ 
ARA\&A, 46, 339 


\bibitem{1998A&A...329..785Y}
Yamamoto, S., \& Mukai, T.\ 1998:
Dust production by impacts of interstellar dust on Edgeworth-Kuiper Belt objects. \
A\&A, 329, 785

\bibitem{2006Sci...314.1735Z}
Zolensky, M. E.; Zega, T. J., Yano, H., Wirick, S., Westphal, A. et al.\ 2006a:
Mineralogy and Petrology of Comet 81P/Wild 2 Nucleus Samples.\
Science, 314, 1735

\bibitem{} 
Zolensky, M., Bland, P., Brown, P. \& Halliday, I. \ 2006b: \ 
Flux of Extraterrestrial Materials.\
In {\it Meteorites and the Early Solar System II} (D. S. Lauretta \& H. Y. McSween Jr., eds), University of Arizona Press, Tucson, p. 869 
 
\bibitem{1975P&SS...23..183Z} 
Zook, H.~A., \& Berg, O.~E.\ 1975:\ 
A source for hyperbolic cosmic dust particles.\ 
P\&SS, 23, 183 

\end{thebibliography}
\end{document}